\documentclass[preprint]{emulateapj}
\slugcomment{{\sc Accepted to ApJ:} May 20, 2013} 
\usepackage{epstopdf}
\usepackage{amsmath,amsopn,amsxtra,txfonts}
\usepackage{comment} 
\usepackage{natbib} 

\newcommand{\ha}{H$\alpha$}
\newcommand{\hb}{H$\beta$}
\newcommand{\kms}{ \ifmmode{\rm km\thinspace s^{-1}}\else km\thinspace s$^{-1}$\fi}

\newcommand{\pc}{\ensuremath{ \, \mathrm{pc}}}
\newcommand{\kpc}{\ensuremath{\, \mathrm{kpc}}}
\newcommand{\cm}{\ensuremath{ \, \mathrm{cm}}}

\newcommand{\s}{\ensuremath{ \, \mathrm{s}}}

\newcommand{\R}{\ensuremath{ \, \mathrm{R}}}
\newcommand{\sr}{\ensuremath{ \, \mathrm{sr}}}

\newcommand{\dg}{\ifmmode{^{\circ}}\else $^{\circ}$\fi}

\newcommand{\hi}{\ion{H}{1}}
\newcommand{\hii}{\ion{H}{2}}

\newcommand{\lb}{\ifmmode{(\ell,b)} \else $(\ell,b)$\fi}
\newcommand{\nii}{[\ion{N}{2}]}

\newcommand{\sii}{[{\ion{S}{2}}]}
\newcommand{\oiii}{[\ion{O}{3}]}

\newcommand{\oi}{[\ion{O}{1}]}

\newcommand{\vlsr}{\ifmmode{v_{\rm{LSR}}}\else $v_{\rm{LSR}}$\fi}
\newcommand{\av}{\ifmmode{A(V)}\else $A(V)$\fi}
\newcommand{\ebv}{\ifmmode{E(B-V)}\else $E(B-V)$\fi}
\newcommand{\iha}{\ifmmode{I_{\rm{H}\alpha}} \else $I_{\rm H \alpha}$\fi}
\newcommand{\ihb}{\ifmmode{I_{\rm{H}\beta}} \else $I_{\rm H \beta}$\fi}
\newcommand{\isii}{\ifmmode{I_{\ion{\rm{S}}{2}}} \else $I_{\rm [S \textsc{ ii}]}$\fi}
\newcommand{\inii}{\ifmmode{I_{\ion{\rm{n}}{2}}} \else $I_{\rm [N \textsc{ ii}]}$\fi}
\newcommand{\ioi}{\ifmmode{I_{\ion{\rm{o}}{1}}} \else $I_{\rm [O \textsc{ i}]}$\fi}

\newcommand{\mha}{\ensuremath{\mathrm{H} \alpha}}

\newcommand{\nhi}{$N_{\rm{H}\textsc{ i}}$}

\newcommand{\lhi}{$L_{\rm{H}\textsc{ i}}$}

\newcommand{\vgeo}{\ifmmode{v_{\mathrm{geo}}} \else $v_{\mathrm{geo}}$\fi}

\shorttitle{Warm Ionized Gas Revealed in the Magellanic Bridge}
\shortauthors{Barger et al.}

\altaffiltext{1}{Now an NSF Astronomy and Astrophysical Postdoctoral Fellow at the Department of Physics, University of Notre Dame, South Bend, IN 46556, USA.}

\begin{document}

\author{K. A. Barger\altaffilmark{1}, L. M. Haffner}
\affil{Department of Astronomy, University of Wisconsin-Madison, Madison, WI 53706, USA}
\email{kbargers@nd.edu,haffner@astro.wisc.edu}
\author{J. Bland-Hawthorn}
\affil{Sydney Institute for Astronomy, School of Physics A28, University of Sydney, NSW 2006}
\email{jbh@physics.usyd.edu.au}

\title{Warm Ionized Gas Revealed in the Magellanic Bridge Tidal Remnant: Constraining the Baryon Content and the Escaping Ionizing Photons around Dwarf Galaxies}

\begin{abstract}

The Magellanic System includes some of the nearest examples of galaxies disturbed by galaxy interactions. These interactions have redistributed much of their gas into the halos of the Milky Way and the Magellanic Clouds. We present Wisconsin \ha\ Mapper kinematically resolved observations of the warm ionized gas in the Magellanic Bridge over the velocity range of +100 to +300~\kms\ in the local standard-of-rest reference frame. These observations include the first full \ha\ intensity map and the corresponding intensity-weighted mean velocity map of the Magellanic Bridge across $(\mathit{l, b}) = (281\fdg5, -30\fdg0)$ to $(302\fdg5, -46\fdg7)$. Using the \ha\ emission from the SMC-Tail and the Bridge we estimate that the mass of the ionized material is between $\left(0.7-1.7\right)\times10^8~M_\odot$, compared to $3.3\times10^8~M_\odot$ for the neutral mass over the same region. The diffuse Bridge is significantly more ionized than the SMC-Tail, with an ionization fraction of $36-52\%$ compared to $5-24\%$ for the Tail. The \ha\ emission has a complex multiple-component structure with a velocity distribution that could trace the sources of ionization or distinct ionized structures. We find that incident radiation from the extragalactic background and the Milky Way alone are insufficient to produced the observed ionization in the Magellanic Bridge and present a model for the escape fraction of the ionizing photons from both the Small and Large Magellanic Clouds. With this model, we place an upper limit of $4.0\%$ for the average escape fraction of ionizing photons from the LMC and an upper limit of $5.5\%$ for the SMC. These results, combined with the findings of a half a dozen results for dwarf galaxies in different environments, provide compelling evidence that only a small percentage of the ionizing photons escape from dwarf galaxies in the present epoch to influence their surroundings.

\end{abstract}

\keywords{galaxies: Magellanic Clouds - galaxies: dwarf - Galaxy: evolution - Galaxy: halo - ISM: individual (Magellanic Bridge)}

\maketitle

\section{Introduction}

Galaxy interactions can lead to the formation of bridges and tails and to the triggering of star formation in the individual systems (e.g., \citealt{1992ARA&A..30..705B}). Bridges, material that links two galaxies, have been detected in many systems and are often the signature of recent interactions. An \ha\ bridge connects M86, a giant elliptical galaxy, to NGC~4438, a disturbed spiral galaxy \citep{2008ApJ...687L..69K}. \hi\ bridges connect the Magellanic Irregular galaxy pairs NGC~4027-4027A \citep{2007ApJ...659L.115C} and NGC~3664-3995 \citep{2004AJ....127.1900W}. In the Magellanic System, the galaxy interactions have made the removed material vulnerable to influence of the gravitational potential of the Milky Way and to the exchange of material between the Magellanic Clouds; the Magellanic Stream funnels roughly $0.4~{\rm M}_{\odot}\ {\rm yr}^{-1}$ in \hi\ gas \citep{2004ASSL..312..195V}---and as least as much in ionized gas \citep{2007ApJ...670L.109B, 2010ApJ...718.1046F}---to the Milky Way. Many of the other dwarf galaxies surrounding the Milky Way are gas poor. A combination of ram pressure and tidal shocks likely stripped material from these galaxies (e.g., \citealt{2007Natur.445..738M}). Galaxy interactions may play an important role in replenishing the star-formation reservoirs of $L_*$ galaxies (see \citealt{2008A&ARv..15..189S} for a detailed discussion on sources of replenishment). 

The nearby Magellanic System is an exquisite example of how galaxy interactions can affect galaxy evolution. Galaxy interactions have greatly altered the morphology of the Magellanic Clouds. Large-scale mapping of the 21-cm emission reveals signatures of these interactions with several large, gaseous structures originating from the Magellanic Clouds, including the Leading Arm, the Magellanic Stream, and the Magellanic Bridge (e.g., \citealt{2003ApJ...586..170P, 2005A&A...432...45B, 2009ApJS..181..398M}). These circumgalactic gas features contain roughly $\sim37\%$ of the \hi\ gas in the Magellanic System \citep{2005A&A...432...45B}. The Magellanic Bridge, which links the Large and Small Magellanic Clouds (LMC, SMC), and SMC-Tail contain almost 40\% of all the neutral gas surrounding the Magellanic System with an \hi\ mass totaling $1.8\times10^8~M_\odot$ \citep{2005A&A...432...45B}. A recent encounter between the Magellanic Clouds likely created this bridge $200~\rm{Myr}$ ago \citep{1996MNRAS.278..191G}. With the LMC only $50~\kpc$ and the SMC only $60\kpc$ away (see \citealt{1999ASSL..237..125W} and references therein), they provide a closeup view of galaxy interactions. Studying the extended gas structures in this system aids in understanding of the future evolution of these galaxies and other other tidally disturbed galaxies.

\begin{figure}
\begin{center}
\includegraphics[scale=0.425,angle=90]{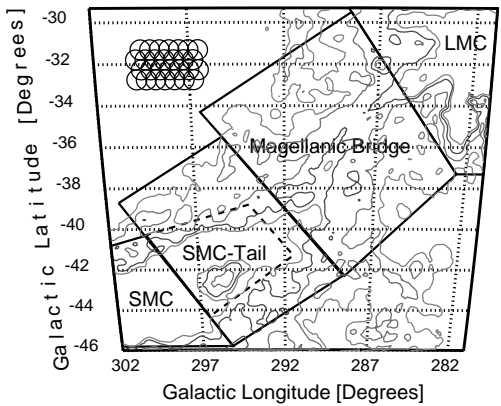}
\end{center}
\figcaption{A schematic view of the gas connecting the Magellanic Clouds. The SMC-Tail bounded by a solid black line indicates the \hi\ tail; the SMC-Tail region bounded by a dashed line indicates the \ha\ tail. The contours depict the $10^{19}~\cm^{-2}$ \hi\ column density at 10, 20, 35, and 50 increments. The grid of $1\arcdeg$ circles in the top left region of the map represents the Nyquist sampling, with pointings separated by $0\fdg5$ beam steps, used to map the \ha\ emission.
\label{figure:cartoon_map}}
\end{figure}

Observing the extended, faint emission from the diffuse ionized gas in the Magellanic Bridge, also known as the intercloud region, requires a high-sensitivity instrument as the emission scales with the density squared. \cite{1982MNRAS.198..985J} published an \ha\ map of  the Magellanic System on photographic plates using the SRC Schmidt telescope. This image indicated the presence of faint \ha\ nebulosities between the LMC and SMC. Unfortunately, these photographic plates only reveal relative \ha\ fluxes because of the difficulties involved with removing the atmospheric background, especially the bright geocoronal line and a bright OH line. The low signal-to-noise ratio of this image makes determining the morphology of the \ha\ emission difficult. Since then, \ha\ emission has only been observed towards dense \hii\ regions in the SMC-Tail \citep{1985Natur.316..705M, 2003ApJ...597..948P, 2007PASA...24...69M}---a prominent tidal feature connecting the SMC and the Magellanic Bridge---where the detections were sensitive down to 0.5--2 R.\footnote{$1~\rm{Rayleigh} = 10^6 / 4 \pi \textrm{ photons} \cm^{-2} \sr^{-1} \s^{-1}$, which is $\sim1.7\times10^{-6}\,{\rm erg} \cm^{-2}\s^{-1}\sr^{-1}$ at \ha.}  Throughout this paper, we separate the SMC-Tail from the diffuse bridge (or intercloud region) when referring to the Magellanic Bridge as shown in Figure \ref{figure:cartoon_map}.

Absorption-line studies have also revealed that the central region of the Magellanic Bridge contains ionized gas. Studies conducted by \citet{2002ApJ...578..126L} and \citet{2008ApJ...678..219L} using the Far Ultraviolet Spectroscopic Explorer (FUSE) and the Space Telescope Imaging Spectrograph (STIS) instrument on the Hubble Space Telescope (HST) confirmed the presence of ionized gas towards two early-type stars and a background quasar. These observations revealed multiple components with different ionization fractions, many lacking \oi\ absorption. The high level of ionization observed towards the background quasar could be explained if that sightline is serendipitously near an early-type star, which would make the sightline have an uncharacteristically high ionization fraction when compared to the rest of the Magellanic Bridge. 

The ionized gas observations of the Magellanic Bridge can constrain the source of the ionization. If the Lyman continuum from the Magellanic Clouds produces much of this ionization, then the strength of the \ha\ emission limits the fraction of ionizing photons that escape ($f_{esc}$) from both of the Magellanic Clouds. This quantity is of cosmological importance because the ionizing radiation from galaxies might be the dominant source of the reionization of the universe (e.g., \citealt{1999ApJ...514..648M, 2005MNRAS.357.1178B}). This reionization altered the structure and shape of the universe by reducing gas accretion onto galaxies, especially dwarf galaxies, \citep{1992MNRAS.256P..43E, 1996ApJ...465..608T, 2004ApJ...601..666D} and subsequent galaxy formation \citep{1999ApJ...523...54B, 2003astro.ph..5527S, 2004MNRAS.348..753S} due to the heating of the intergalactic medium that surrounds galaxies. High-mass galaxies alone are unable to reionize the universe \citep{2011ApJ...731...20F}, while the contribution from low-mass galaxies is uncertain. The $f_{esc}$ from galaxies at both the present epoch and the epoch of reionization is poorly constrained. 

To determine if the Magellanic Bridge and the SMC-Tail contain small pockets of high ionization or if they are ionized throughout, we present an \ha\ emission survey of these structures using the Wisconsin \ha\ Mapper (WHAM) observatory. WHAM is optimized to detect faint, optical emission from diffuse ionized sources with a sensitivity of a few hundredths of a Rayleigh. The spectrometer, described in detail by \citet{2003ApJS..149..405H}, consists of a dual-etalon Fabry-Perot spectrometer that produces a $200~\kms$\ wide spectrum with $12~\kms$\ velocity resolution from light integrated over a $1\arcdeg$ beam. Section \ref{section:obs} includes a description of the \ha\ observations. We detail the data reduction process in Section \ref{section:reduce}, which includes the velocity calibration, the removal of atmospheric lines, the merging of spectra taken over different velocity ranges, and the technique used to correct \ha\ observations for extinction. We present the non-extinction corrected \ha\ intensity map of the Magellanic Bridge in Section \ref{section:intensity_map} and discuss the differences and similarities of the \ha\ and \hi\ emission. In Section \ref{section: compare}, we compare the global behaviors of the \ha\ and \hi\ gas, including their emission levels and velocity distributions. We investigate the total mass of the Magellanic Bridge by addressing the distribution of neutral and ionized gas in Section \ref{section:mass}. In Section \ref{section:ionization}, we explore the source of the ionization and the escape fraction of ionizing photons from the LMC and SMC. Finally, we discuss the implications of these observations in Section \ref{section:implications} and list our major conclusions in Section \ref{section:summary}.

\section{Observations}\label{section:obs}

To survey the baryons cycling in and out of the Magellanic Clouds, we fully sampled the \ha\ emission of the Magellanic Bridge with WHAM at an angular resolution of $1\arcdeg$ and a velocity resolution of $12~\kms$ over the local standard-of-rest (LSR) velocity range $0$ to $+315~\kms$ from $(\mathit{l, b}) = (281\fdg5, -30\fdg0)$ to $(302\fdg5, -46\fdg7)$. This region was chosen to include the SMC-Tail. The high-throughput, dual Fabry-Perot spectrograph of WHAM---combined with a 1\arcdeg\ angular provides---enables an unprecedented sensitivity to faint \ha\ emission over large scales; however, these beams span almost a kiloparsec in diameter at distance of 55\kpc, the median distance between the Magellanic Clouds. As a result, this survey is less sensitive to emission from individual \hii\ regions---which often span only only a few hundred parsecs or less---since they are diluted by the contribution of diffuse emission within the beam. 

We grouped our observations into ``blocks'' of 30--50 Nyquist sampled pointings of the entire Magellanic Bridge at $0\fdg5$ spacings, as displayed in Figure \ref{figure:cartoon_map}. Each pointing in a block was observed sequentially in time such that subsequent rows of pointings alternated in direction. These observations were taken at Cerro Tololo Inter-American Observatory (CTIO), where Magellanic Bridge never ascends above 1.4 airmass. We took the large majority of the observations while the Magellanic Bridge had an airmass between 1.4 and 1.6; however, we also incorporated additional observations sampled at $1\arcdeg$ spacing with airmass greater than 1.6 to increase the total integration time of the map. 

\begin{figure}
\begin{center}
\includegraphics[scale=0.35,angle=90]{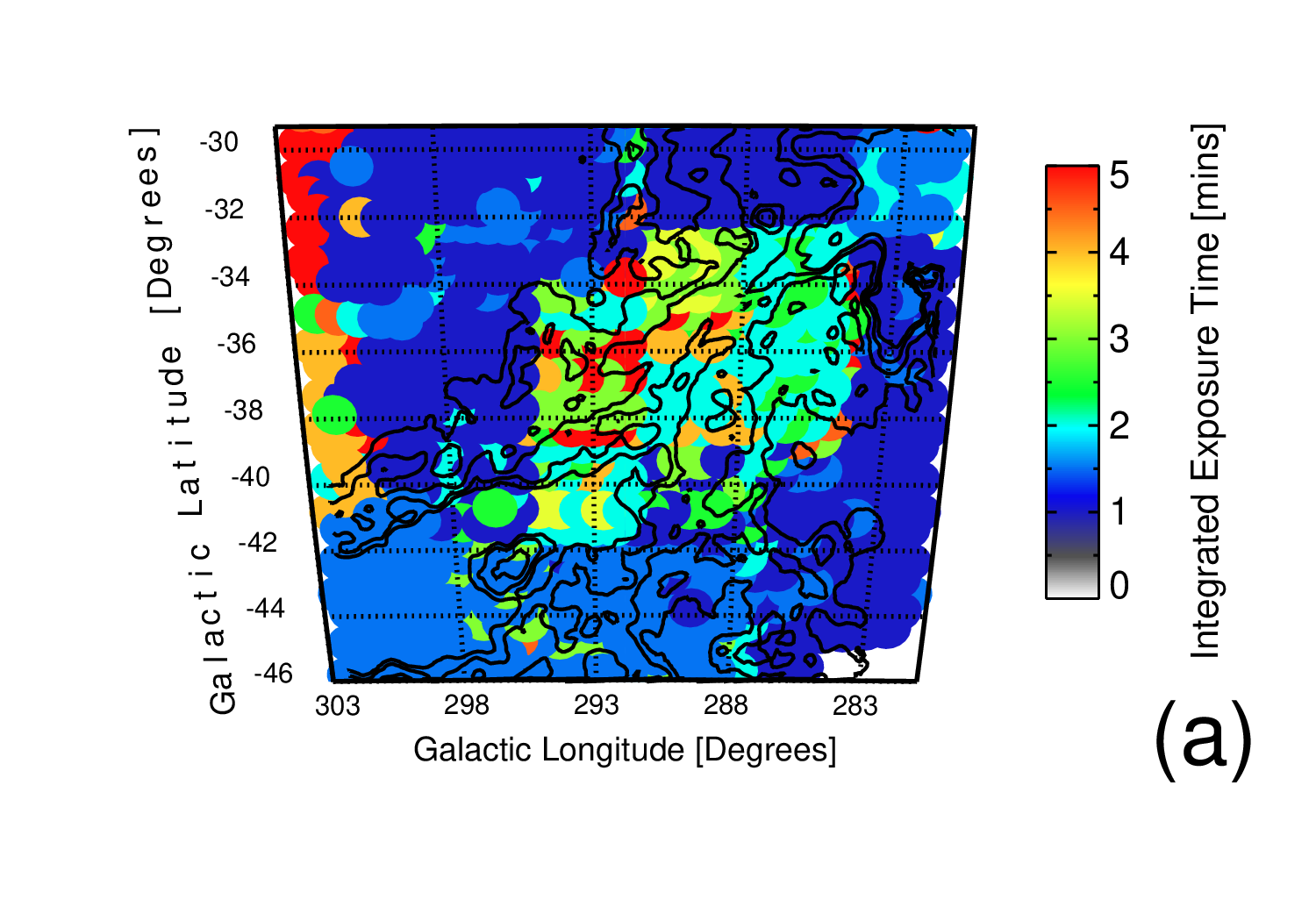} \\
\includegraphics[scale=0.35,angle=90]{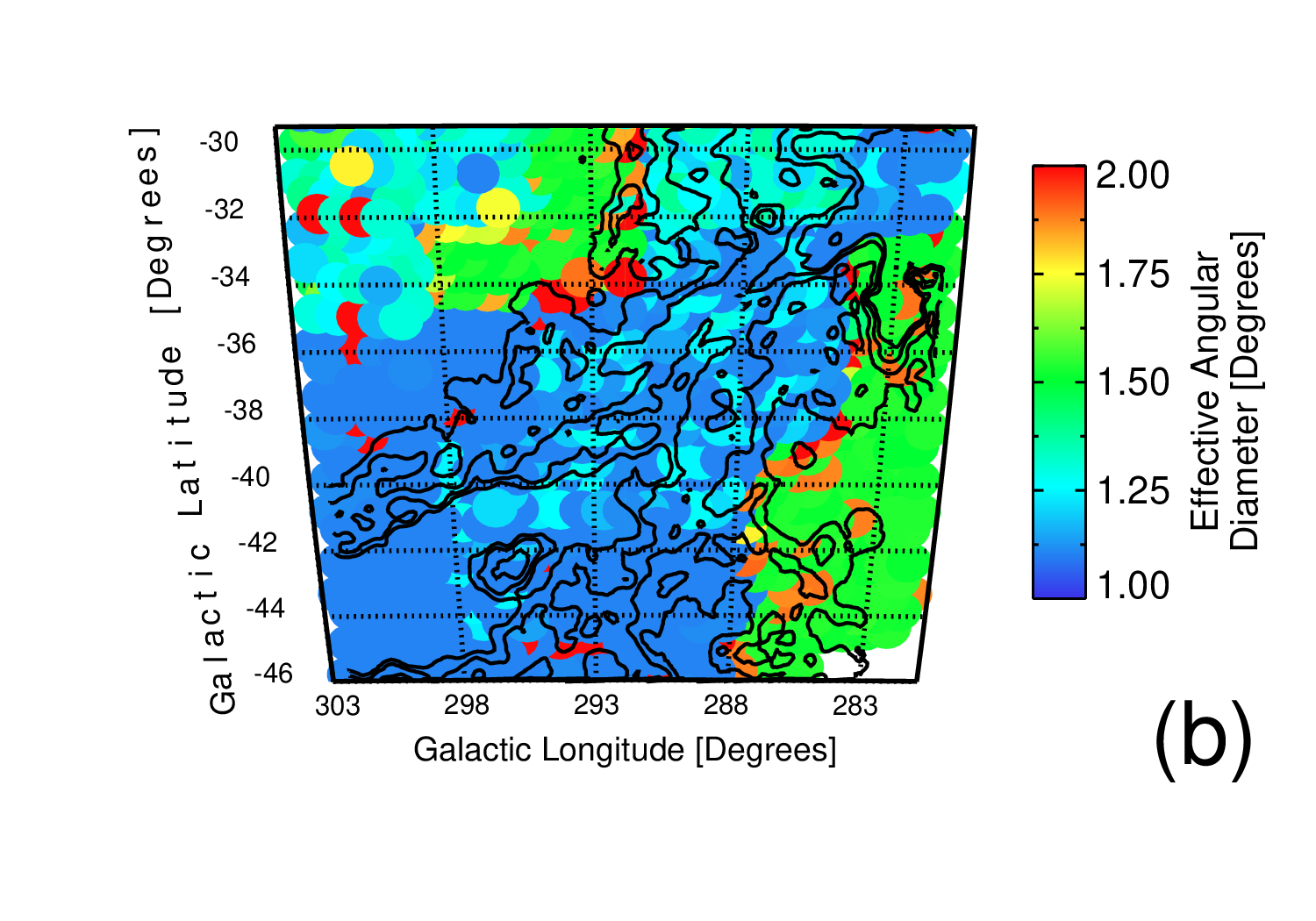}
\end{center}
\figcaption{Total integrated exposure time \textit{(a)} and smoothed angular diameter of each sightline \textit{(b)} of survey. The total expire time consists of multiple 30-second individual observations at each location of the \ha\ Magellanic Bridge survey. Each observation span $200~\kms$, which is only part of the $0$ to $+315~\kms$ velocity coverage. The majority of the observations were centered at either $+175$ or $+210~\kms$. The smoothed angular diameter is the effective diameter of the sightline after binning the non-nyquist sampled observations such that they conform to the nyquist grid spacing. The contours depict the $10^{19}~\cm^{-2}$ \hi\ column density at 10, 20, 35, and 50 increments. 
\label{figure:num_observations}}
\end{figure}

We kept single observations short to minimize subtle changes in the spectra caused by variations in atmospheric lines and observed each block multiple times over September 2011 to January 2012 to increase our sensitivity. Each single observation had an exposure time of 30-seconds, while the total integrated exposure time at each sightline ranges from 1.5 to 6.0 minutes. We sampled the high \hi\ column density regions $(N_{\rm H{~\textsc{i}}}>10^{19}~\cm^{-2})$ that connect the LMC and SMC the most. Figure \ref{figure:num_observations}(a) shows the total integrated exposure time for each location in this survey. Many of the repeated observations towards the same regions are separated by few weeks to allow the atmospheric lines to shift relative to the LSR velocity reference frame. Combining observations acquired over multiple months reduces the residuals from the faint atmospheric-line subtraction while reinforcing the astronomical emission at a specific LSR velocity. 

The observations taken with different spacing from the Nyquist grid were binned to conform to the Nyquist grid. This corresponds to 28\% of the resultant average sightlines having a smoothed angular coverage with an effective angular diameter of $1\fdg1$ or less, 55\% with $1\fdg2$ or less, and 92\% with $2\fdg0$ or less compared to the $1\arcdeg$ angular resolution of WHAM. We define the effective angular diameter as the diameter of a circle with an area equal to the total area covered by the averaged beams. The further from the main \hi\ structure of the Bridge, the larger the average displacement of the non-Nyquist sampled observations from the Nyquist grid points as these locations were sampled less. The typical effective angular diameter is therefore smallest along the \hi\ Bridge, as illustrated in Figure \ref{figure:num_observations}(b).  

\section{Data Reduction}\label{section:reduce}

Beyond the ring-summing and flat-fielding procedures described in \citet{2003ApJS..149..405H}, the data reduction of the \ha\ map included velocity calibration of the emission, subtraction of the atmospheric emission, stitching together of the spectra taken along the same sightline taken at different velocity intervals, and applying an extinction correction to the \ha\ emission from both foreground dust and dust within the Bridge. 

\subsection{Velocity Calibration}\label{section:v_calibration}

\begin{figure*}
\begin{center}
\includegraphics[scale=0.5,angle=90]{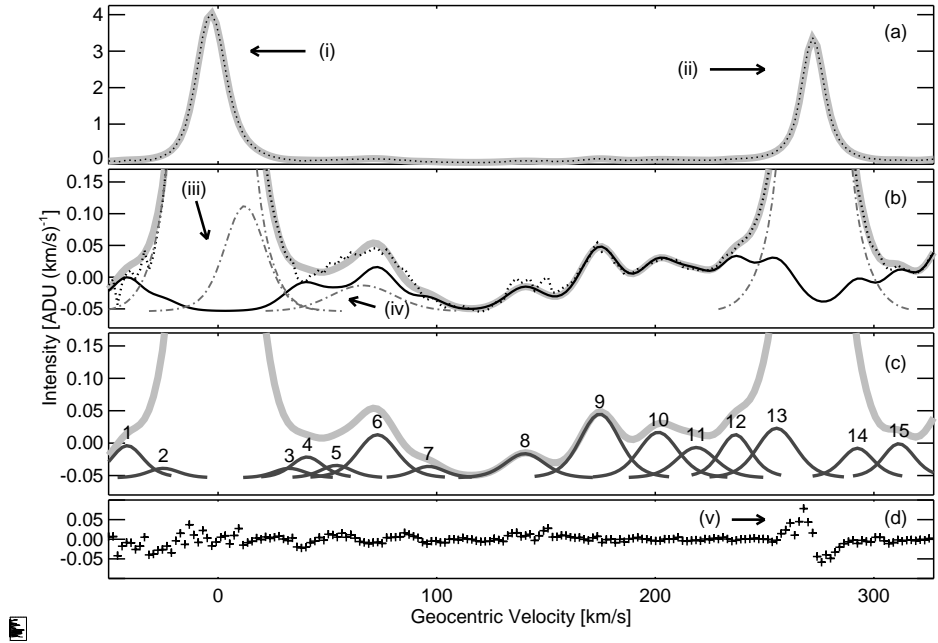} 
\end{center}
\figcaption{
The average \ha\ emission towards (\textit{l,b}) = $(60\fdg0, -67\fdg0)$ and $(89\fdg0, -71\fdg0)$. The (a) and (b) panels show this average spectra as a dotted line and the corresponding fit in gray. The (b) panel emphasizes the faint emission in panel (a) and displays the constructed atmospheric template as a black solid line. Panel (c) illustrates the faint atmospheric lines near \ha\ in dark grey against the fit for the averaged spectra; these faint lines are listed according to the line identification in Table \ref{table:atmlines}. At $+334.4~\kms$, an additional line---not shown here---with a full-width at half-max of $15.0~\kms$ and $2.76$ times the area of line (1), was added to the construction of the average atmospheric template. Panel (d) displays the residuals between the average spectra and the total fit. The \textit{(i)} marker indicates the geocoronal line at $-2.3~\kms$ and the \textit{(ii)} marker denotes a bright OH line at a geocentric velocity of $+272.44~\kms.$ Galactic emission is labeled by markers \textit{(iii)} and \textit{(iv)}. Marker \textit{(v)} indicates residuals from the OH line subtraction caused by a slight mismatch in instrument profile; we decreased these residuals in the final data processing by applying a custom instrument profile for each night (see Section \ref{section:bright}). 
\label{figure:atmosphere}}
\end{figure*}

Once the spectra are pre-processed, ring-summed, and flat-fielded, they are in $2~\kms$ bins over a $200~\kms$ velocity window. These spectra are shifted to the geocentric (geo) velocity frame by adding a constant velocity offset value, determined by identifying bright atmospheric lines with known wavelengths. Both faint and bright atmospheric emission clutter the $-50$ to $+315~\kms$ LSR velocity window of this \ha\ Magellanic Bridge survey. Two bright atmospheric lines dominate the spectra in this survey: the bright geocoronal line at $\vgeo=-2.3~\kms$ and a bright OH line at $\vgeo=+272.44~\kms$ relative to the \ha\ recombination line at $6562.8~\AA$. These two bright lines are labeled \textit{(i)} and \textit{(ii)} in Figure \ref{figure:atmosphere}(a). Because the velocity window of a single observation is only $200~\kms$ wide, multiple exposures are needed to fully sample the spectrum of the Bridge. Each exposure was shifted to include either the geocoronal line or the bright OH line to enable accurate velocity calibration. Although the overlapping emission from the Galactic warm interstellar medium---marked as \textit{(iii)} and \textit{(iv)} in Figure \ref{figure:atmosphere}(b)---with the geocoronal line can add uncertainty in determining positions of emission features below $+50~\kms$, the large contrast in the strengths of these lines generally makes locating the geocoronal-line center easy. Finally, we apply an offset for each observation that shifts them to the LSR frame. 

\subsection{Removing Atmospheric Emission}\label{section:template} 

\begin{deluxetable}{ccccc}
\tablewidth{0pt}
\tablecaption{Faint Atmospheric Lines near \ha\label{table:atmlines}}
\tablehead{
 & \colhead{$v_\mathrm{geo}$} & \colhead{Wavelength} & \colhead{FWHM} & \colhead{Relative} \\
\colhead{Line} & \colhead{[\kms]} & \colhead{[\AA]} & \colhead{[\kms]} & \colhead{Intensity}
}
\startdata
1 & $-41.4$ & 6561.92 & 10 & 1.00  \\
2 & $-24.9$ & 6562.29 & 10 & 0.30  \\
3 & $+32.7$ & 6563.57 & 10 & 0.30  \\
4 & $+40.9$ & 6563.75 & 10 & 0.65  \\
5 & $+54.2$ & 6564.04 & 10 & 0.39  \\
6 & $+73.0$ & 6564.46 & 15 & 1.60  \\
7 & $+96.5$ & 6564.98 & 10 & 0.36  \\
8 & $+140.4$ & 6565.95 & 15 & 0.90  \\
9 & $+174.7$ & 6566.71 & 15 & 2.36  \\
10 & $+201.4$ & 6567.31 & 15 & 1.69  \\
11 & $+218.8$ & 6567.69 & 15 & 1.13  \\
12 & $+236.8$ & 6568.09 & 10 & 1.34  \\
13 & $+255.5$ & 6568.51 & 15 & 1.84  \\ 
14 & $+292.6$ & 6569.33 & 10 & 0.92  \\
15 & $+311.5.$ & 6569.75 & 15 & 1.06  \\
\enddata
\tablecomments{This list excludes the geocoronal line at $-2.3~\kms$ and the bright OH line at $+272.44~\kms$: two lines produced through a different process in a different atmosphere layer.}
\end{deluxetable}

The overlap of the geocoronal line with the \ha\ Magellanic Bridge emission is negligible as the majority of low velocity \hi\ components appear at  $\vgeo >  +30~\kms$. The bright OH line contaminates the spectra at $+260 \gtrsim \vgeo \gtrsim +290~\kms$. As a result, this line does occasionally overlap with the \ha\ emission features close to LMC velocities. Fainter atmospheric emission is present below $\sim0.1~{\rm R}$ at all velocities in this survey. The removal of both the bright and faint atmospheric lines is crucial for detecting the faint \ha\ emission between the Magellanic Clouds. The removal of atmospheric contamination consists of three steps: subtracting the background continuum, subtracting the bright atmospheric lines, and subtracting the faint atmospheric lines.

\subsubsection{Background Subtraction}

We assume an underlying flat background continuum level over all velocities. The baselines are well behaved over the velocity range of this survey, except when contaminated by emission from bright foreground stars. Foreground stars distort the shape of the spectra and create an elevated, non-linear background with absorption lines. Beams that contain stars with $m_V < 6$ $(~\sim9\%$) within a $0\fdg55$ radius are excluded from this survey to minimize this foreground contamination and are replaced with an average of the uncontaminated observations within $1\arcdeg$.

\subsubsection{Bright Atmospheric Line Subtraction}\label{section:bright}

\begin{figure}
\begin{center}
\includegraphics[scale=0.42,angle=0]{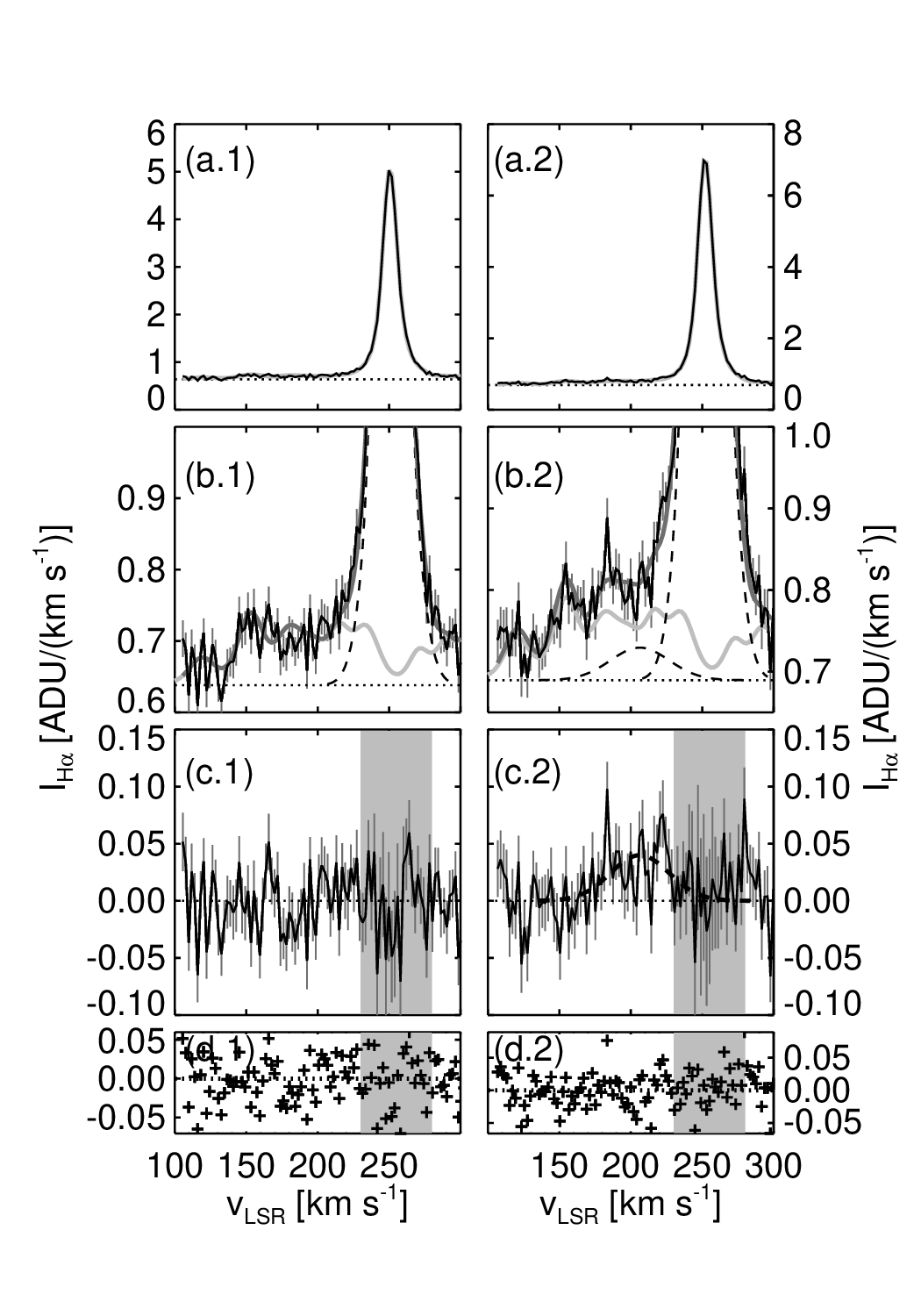}
\end{center}
\figcaption{
Example of atmospheric subtraction. Panels (a.1)--(d.1) correspond to an observation taken towards $(\textit{l,b}) = (89\fdg0, -71\fdg0)$, one of the faint directions used to construct the atmospheric template in Figure~\ref{figure:atmosphere}. Panels \textit{(a.2)--(d.2)} correspond to an observation taken towards the inner region of the Bridge at $(\textit{l,b}) = (289\fdg4, -39\fdg0)$ with a non-extinction corrected $\iha$ of $0.08~\R$. The pre-atmospheric spectra are shown in panels (a.1)--(a.2) and zoomed in to illustrate the faint emission in (b.1)--(b.2). The dotted lines in panels (a.1)--(b.2) indicate the location of the baseline. The dashed lines in lines in panels (b.1--b.2) trace the bright OH line at a geocentric velocity of $+272.44~\kms$ and an \ha\ emission feature produced from the Magellanic Bridge in panel (b.2). The fainter solid gray line in panels (b.1--b.2) indicate the strength and location of the faint atmospheric lines, identified using the atmospheric template in Figure~\ref{figure:atmosphere}. The darker solid gray line in panels (b.1--b.2) indicate the total fit, which includes the baseline, atmospheric template, and contributions from the OH and \ha\ Bridge emission lines. Panels (c.1--c.2) include the spectra after subtracting the atmospheric profile, the bright OH line, and the baseline. The residuals of  the fit are displayed in Panels (d.1)--(d.2). The region highlighted in gray in panels (c.1--2) represent the location that a bright OH line was removed and signifies a velocity range with a lower sensitivity than the surrounding spectra. 
\label{figure:atmosphere_sub}}
\end{figure}

The strength of the bright and faint atmospheric lines vary differently throughout over the course a night and a year. The geocoronal line \citep{2006JASTP..68.1520M} and OH lines \citep{1989JGR....9414629M} are produced from interactions between solar radiation and Earth's upper atmosphere and will, therefore, vary in strength with the direction and the time of the observation. These bright atmospheric lines are displayed in Figure \ref{figure:atmosphere}(a) and are labeled \textit{(i)} and \textit{(ii)}. For this reason, the geocoronal line and the OH line at $\vgeo = +272.44~\kms$ are always individually removed from each spectra before subtracting the faint atmospheric lines. We removed these lines by fitting a single Gaussian profile convolved with the instrument profile. 

Two effects alter the shape bright atmospheric lines: (1) The precision in aligning the dual-etalon transmission functions (spectrometer ``tuning") can result in very slight night-to-night variations in the instrument profile at a level only detectable in narrow, bright lines. (2) A geocoronal ``ghost''---due to an incomplete suppression of a geocoronal line at $\vgeo = -2.3~\kms$ from a neighboring order in the high-resolution etalon (see \citealt{2003ApJS..149..405H}, Figure 2)---lies underneath the OH line at $\vgeo = +272.44~\kms$. Although these affects are minimal, together they can leave residuals that can make detecting the faint \ha\ emission of the diffuse Magellanic Bridge difficult. Each night we constructed a new instrument profile to minimize the residuals associated with the subtraction of the geocoronal and OH lines to account for both of these effects. The P-Cygni shape of the residual in Figure \ref{figure:atmosphere}(d), marked \textit{(v)}, illustrates the result of the line subtraction with the generic WHAM instrument profile instead of using a custom instrument profile each night.

We constructed the instrument profile for each night by modeling the shape of the $+272.44~\kms$ OH line with three gaussians: one to account for the global size and width of the line and one for each wing to account for an asymmetrical shape of the line at the blue wing. The asymmetrical blue wing is due to minor etalon defects \citep{1997PhDT........15T}. We chose to model the instrument profile using the OH line because its intrinsic line width is much narrower than the instrument width and because it is well separated from Galactic emission. We created these profiles from observations towards either (\textit{l,b}) = $(60\fdg0, -67\fdg0)$ or $(89\fdg0, -71\fdg0)$, two sightlines that are observed to have little \ha\ emission and are located far outside of the $(\mathit{l, b}) = (281\fdg5, -30\fdg0)$ to $(302\fdg5, -46\fdg7)$ Bridge survey. 

Due to the high signal strength of the OH line and the subsequent increased noise, the data that overlap with OH line are more noisy than the surrounding spectra. The net result is that the sensitivity of our survey is better at lower velocities between $+100\le \vlsr \le +240~\kms$ with $I_{{\rm H}\alpha}\simeq 30~{\rm mR}$ than at higher velocities between $+240\le \vlsr \le +275~\kms$ with $I_{{\rm H}\alpha}\simeq40~{\rm mR}$. 

Bridge emission at high velocities may be reduced over certain velocities since the some of the emission could be subtracted during the removal of the bright OH line. The intensity of the OH line dominates over this span, hiding \ha\ emission from the Bridge. Removal of some Bridge emission is unavoidable throughout the core of this line. The presence of ionized gas emission over this narrow velocity range may be revealed through other spectral lines, such as \sii\ or \nii. We are undertaking Magellanic Bridge surveys in these lines as well with WHAM.

In the wings of the OH line, the atmospheric and potential Bridge emission become comparable. As mentioned above, determining the instrument profile each night helps to minimize any residuals from subtracting the line. We also observe each sightline multiple times over multiple months so the offset between the geocentric and LSR frames is different for each observation. Combining these multiple-epoch observations minimizes contamination from the OH wing in an individual exposure.

\subsubsection{Faint Atmospheric Line Subtraction}\label{section:faint_atmosphere}

In addition to the bright geocoronal and a OH atmospheric line, faint atmospheric lines litter the spectra. The strength of these faint atmospheric lines changes primarily with airmass. To characterize them, we observed two directions faint in \ha\ emission multiple times over 10 days to create an average spectrum with a high signal-to-noise ratio. This averaged spectrum consists of numerous 30- and 60-second observations, totaling 4.5 hours of integrated exposure time, towards (\textit{l,b}) = $(60\fdg0, -67\fdg0)$ and $(89\fdg0, -71\fdg0)$, which are outside the Bridge survey region of $(\mathit{l, b}) = (281\fdg5, -30\fdg0)$ to $(302\fdg5, -46\fdg7)$. Table \ref{table:atmlines} lists the geocentric velocity, wavelength, width, and relative intensity of these atmospheric lines and Figure \ref{figure:atmosphere}(c) displays their relative size and position. \citet{2002ApJ...565.1060H} and \citet{2003ApJS..149..405H} list similar characteristics for the faint atmospheric lines near \ha\ in the northern hemisphere observed towards the Lockman Window; however, our observations here extend this list to higher positive velocities. We used the averaged atmospheric emission spectrum of the faint lines to construct a synthetic atmospheric template (shown in Figure \ref{figure:atmosphere}(b) as a solid black line). 
 
We removed the faint atmospheric lines from the Magellanic Bridge observations by scaling the synthetic atmospheric template---which accounts for changes in the flux due to airmass and daily fluctuations---to match the atmospheric contamination. This scaled atmospheric template is then subtracted from the observation. The removal of the faint atmospheric lines in this study parallels the reduction method used by \citet{2003ApJS..149..405H}, which provides a more thorough description of this process. Figure \ref{figure:atmosphere_sub} illustrates this process towards (\textit{l,b}) = $(89\fdg0, -71\fdg0)$, a sightline far off the Magellanic Bridge and faint in \ha\ emission, and towards (\textit{l,b}) = $(289\fdg4, -39\fdg0)$, a sightline in the middle of the Magellanic Bridge. Both of these observations were taken on the same night. Panels (b.1--b.2) show the scaled atmospheric template (solid gray line) fit to these spectra and the resultant atmospheric-subtracted spectra in panels (c.1--c.2); the bright OH---centered roughly at $+250~\kms$ in the LSR frame---and the background continuum were fit and subtracted separately. 

\subsection{Co-adding Spectra}

\begin{figure}
\begin{center}
\epsscale{1}\plotone{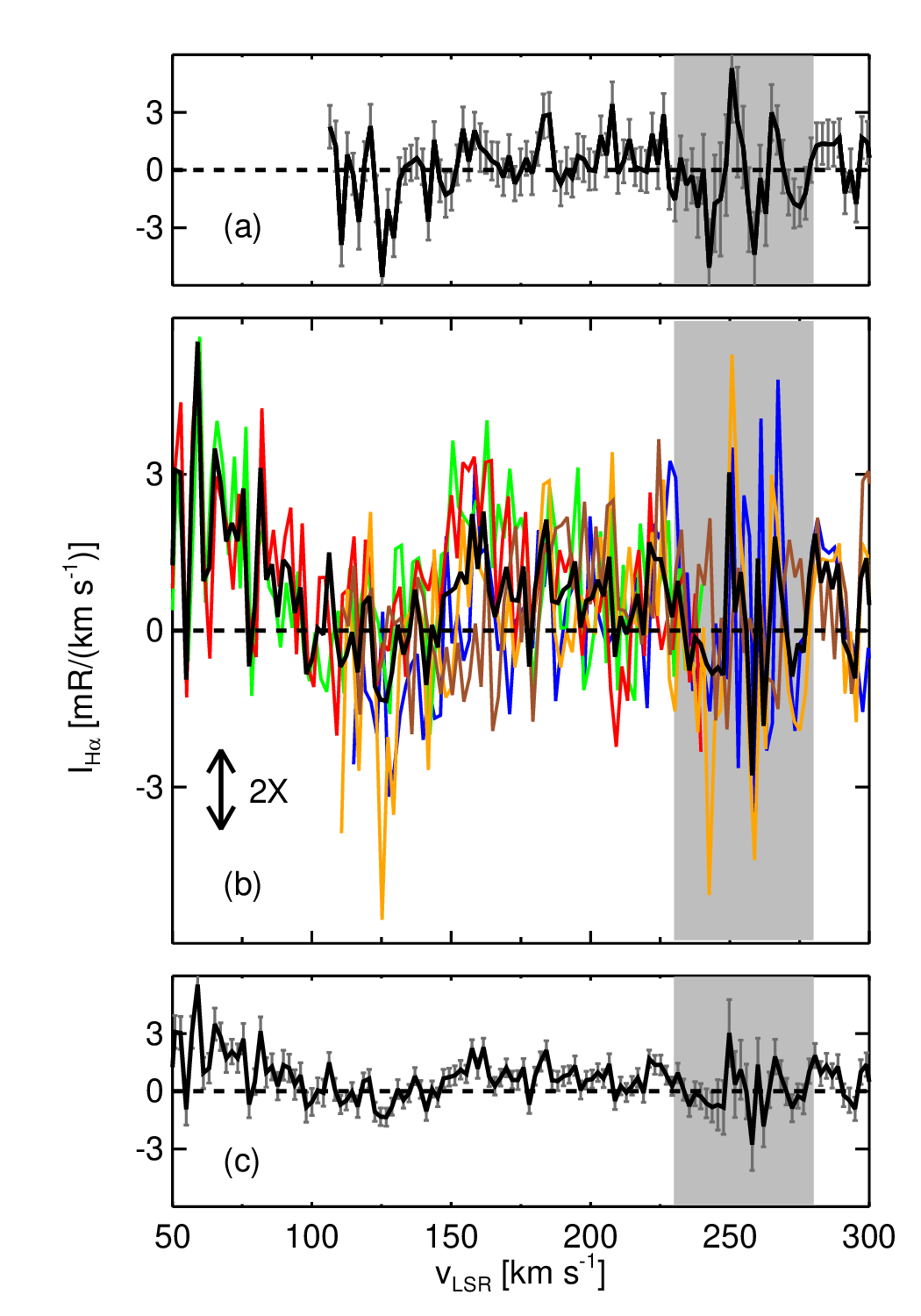} 
\figcaption{
Combining spectra taken over three months towards the direction (\textit{l,b}) = $(291\fdg6, -39\fdg0)$. Panel (a) shows an example of a typical observation and the corresponding error. Panel (b) displays five separate observations, each illustrated with a different color, and their combined average (black) with the y-axis expanded by a factor of two. This average spectra is displayed again in panel (c) with the resulting errors. The elevated intensity between $+50$ and $+100~\kms$ is due to Galactic emission; for this reason, velocities below $+100~\kms$ are excluded in the \ha\ map in Figure \ref{figure:Ha_Bridge}. The region highlighted in gray represent the location that a bright OH line was removed; our sensitivity is lower throughout this velocity region.  
\label{figure:sightline}}
\end{center}
\end{figure}

Each observation produces an average spectrum of the emission within the 1\arcdeg\ beam over a $200~\kms$ wide velocity window at a velocity resolution of $12~\kms$. The dominant \hi\ emission of the Magellanic Bridge spans approximately $+50$ to $+315~\kms$. With this wide velocity range---one and a half the size of the WHAM velocity window---we covered the spectral range with multiple exposures and spliced them together. The spectra were first velocity calibrated and atmospheric subtracted by the methods described in Sections \ref{section:v_calibration} and \ref{section:template}, then combined. Figure \ref{figure:sightline} shows an example of how we combined five separate observations along the same sightline after velocity calibration and atmospheric subtraction. The resultant spectrum in Figure \ref{figure:sightline}(c) is an average of the five multi-color spectra shown in Figure \ref{figure:sightline}(b), with the intensity and uncertainty weighted by the number of observations at each velocity bin. We selected the velocity coverage for each observation to ensure the inclusion of a bright atmospheric line with a stationary position in the geocentric rest frame to ensure accurate alignment.

\begin{figure}
\begin{center}
\epsscale{1}\plotone{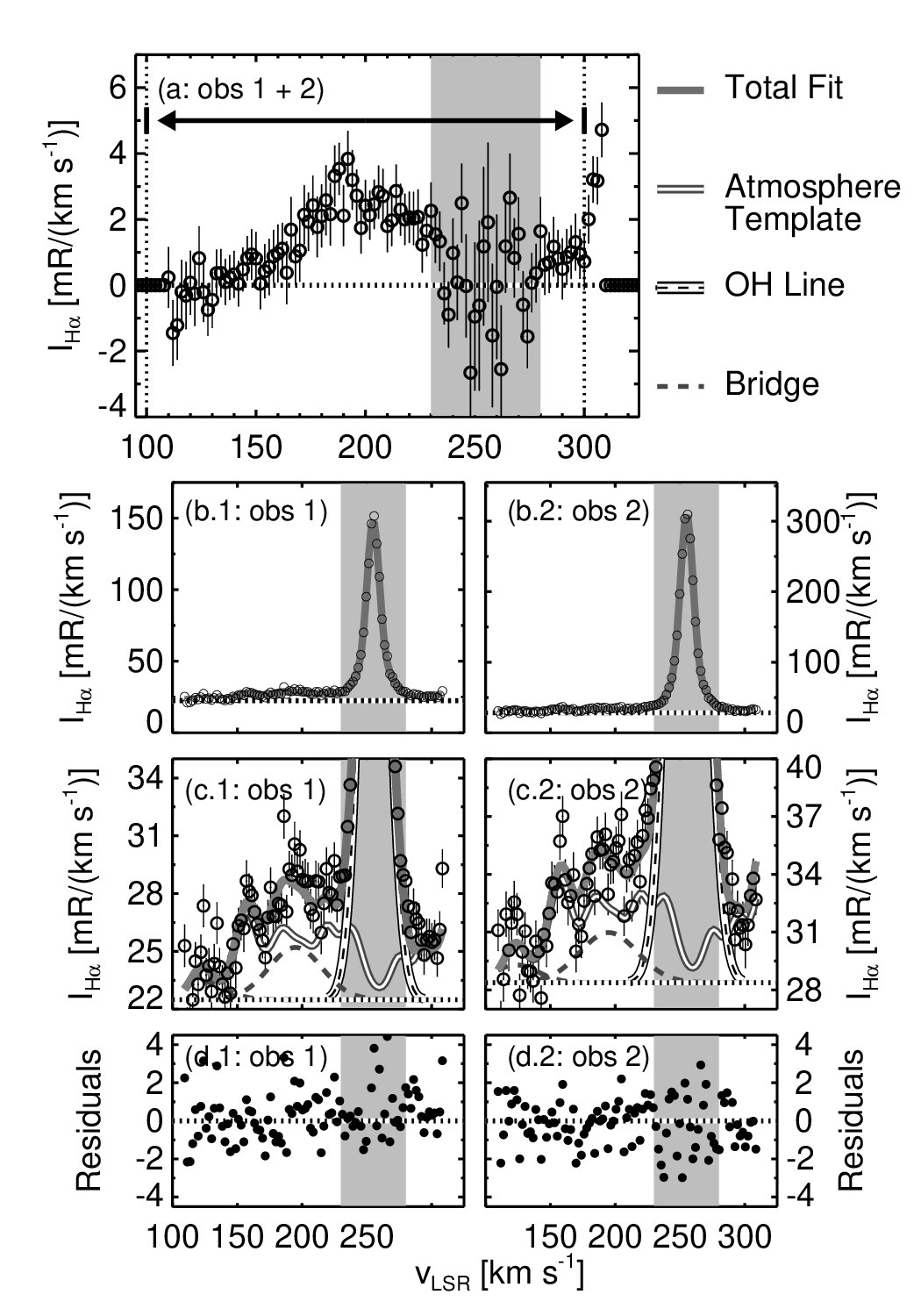}
\end{center}
\figcaption{
The reduction of sightline $(\mathit{l, b}) = (290\fdg9, -41\fdg0)$ in the Magellanic Bridge with two separate observations (obs 1 and obs 2). Panel (a) shows the fully reduced spectra with horizontal arrows and vertical dotted lines labeling the $+100$ to $+300~\kms$ velocity range of the \ha\ survey. This sightline has a non-extinction corrected $\iha$ of $0.18~\R$. The unreduced spectra in panels (b.1--2) are magnified in (c.1--2) to magnify  the faint atmospheric lines and Bridge emission of the non-reduced spectra, where a gray solid line marks the total fit, a hollow gray line signifies the atmospheric template of the faint lines, the outlined gray dashed line labels the bright OH line fit at $\vgeo=+272.44~\kms$, and the gray dashed line traces the Bridge emission. The residuals of the emission from these two observations minus the total fits are included in panels (d.1--2). The region highlighted in gray represents the location of a bright OH line and signifies a velocity range with a lower sensitivity than the surrounding spectra. 
\label{figure:total_reduction}}
\end{figure}

In Figure \ref{figure:total_reduction}, we demonstrate the entire reduction process towards sightline at $(\mathit{l, b}) = (290\fdg9, -41\fdg0)$ in the Magellanic Bridge with two separate observations. The reduced combined spectrum is shown in Panel (a). Panels (b.1--2) include the non-reduced spectra and panels (c.1--2) zoom in on the faint atmospheric lines and Bridge emission with the their corresponding fits. We measure similar Bridge emission from both of these observations before and after we splice the two spectra together.

\subsection{\ha\ Extinction Correction}\label{section:extinction}

\begin{deluxetable*}{lcccccccc}
\tablecolumns{7}
\tabletypesize{\scriptsize}
\tablecaption{Neutral and Ionized Properties\label{table:extinc}} 
\tablewidth{0pt}
\tablehead{
\multicolumn{1}{c}{}  & \multicolumn{3}{c}{Foreground Extinction} &\colhead{} & \multicolumn{3}{c}{Internal Extinction} \\ 
 \cline{2-4} \cline{6-8} 
\colhead{Region}    	& \colhead{$\log\ \langle$\nhi$\rangle$}	&\colhead{A(\mha)\tablenotemark{a}} & \colhead{$\%_{corr}$\tablenotemark{a}} &\colhead{} & \colhead{$\log\ \langle$\nhi$\rangle$} &\colhead{A(\mha)\tablenotemark{a}} & \colhead{$\%_{corr}$\tablenotemark{a}}   \\	
\colhead{} 		& \colhead{[$\cm^{-2}$]}  	&\colhead{\rm [mag]} & \colhead{} 	&\colhead{} & \colhead{[$\cm^{-2}$]}		&\colhead{\rm [mag]} & \colhead{}	  }
\startdata		  		 
Inner Region\tablenotemark{b}		& $20.6$		&\colhead{0.26}	& $27.3\%$	&\colhead{} 	& $19.9$	&\colhead{0.02}	& $1.5\%$	  	 	\\
H{\sc~i} SMC-Tail\tablenotemark{b}	& $20.6$		&\colhead{0.17}	& $16.7\%$	 &\colhead{} 	& $20.5$	&\colhead{0.07}	& $6.8\%$	  	 \\
\ha\ SMC-Tail\tablenotemark{b}		& $20.8$		&\colhead{0.16}	& $15.6\%$	 &\colhead{} 	& $20.4$	&\colhead{0.05}	& $5.2\%$	  	
\enddata
\tablenotetext{a}{Calculated using the average $\log\ \langle$\nhi$\rangle$ of the region.} 
\tablenotetext{b}{Regions defined by a polygon with the following corners: \textit{l}=$(289\fdg0, 283\fdg0, 289\fdg0, 297\fdg0)$ and \textit{b}=$(-30\fdg2, -38\fdg0, -43\fdg0, -35\fdg0)$ for the inner region, \textit{l}=$(301\fdg8, 295\fdg9, 289\fdg1, 295\fdg7)$ and \textit{b}=$(-39\fdg4, -36\fdg2, -43\fdg0, -46\fdg5)$ for the H{\sc~i} SMC-Tail, and \textit{l}=$(300\fdg5, 294\fdg5, 292\fdg0, 297\fdg0)$ and \textit{b}=$(-41\fdg0, -39\fdg5, -42\fdg0, -45\fdg0)$ for the \ha\ SMC-Tail. The boundaries for these regions are displayed in Figure \ref{figure:cartoon_map}.} 
\end{deluxetable*}

The intrinsic \ha\ intensity from the Magellanic Bridge is reduced by foreground dust in the Milky Way and potentially by the dust within the structure itself. In this section, we discuss our prescription for determining the extinction correction due to these sources. 

\subsubsection{Correction for Foreground ISM Extinction}

The position of the Magellanic Bridge below the Galactic plane results in minimal foreground interstellar dust extinction. We expect that most of the extinction comes from local interstellar dust. We use the excess color given in \citet{1994ApJ...427..274D} for a warm diffuse medium:
\begin{equation}
E(B-V)=\frac{\langle N_{\rm H{~\textsc{i}}}\rangle}{4.93\times10^{22}~\mathrm{atoms/(cm^2\cdot mag)}}
\end{equation}
where the average \hi~column density $(\langle N_{\rm H{~\textsc{i}}}\rangle)$ includes only the foreground \hi\ emission \citep{1978ApJ...224..132B}. 
The integrated foreground \hi\ column density is calculated from smoothed Leiden/Argentine/Bonn Galactic H{\sc~i}\ survey (LAB: \citealt{2005A&A...440..775K, 1997agnh.book.....H}) data to match our 1\arcdeg\ angular resolution. To calculate the total foreground extinction along the line-of-sight, we integrated \hi\ column densities over the $-450$ to $+100~\kms$ LSR velocity range. If the extinction follows the $\langle A(\mha)/A(V)\rangle=0.909-0.282/R_v$ 
optical curve presented in \citet{1989ApJ...345..245C} for a diffuse interstellar medium, where R$_v\equiv A(V)/E(B-V)=3.1$, then the expression for the total extinction becomes
\begin{equation}\label{eq:extinction}
A(\mha)=5.14\times10^{-22}~\langle N_{\rm H{~\textsc{i}}}\rangle~\mathrm{cm^{-2}\cdot atoms^{-1} \cdot mag},
\end{equation} so that the foreground extinction correction is $I_{\mha,~corr} = I_{\mha,~obs}~e^{\ A(\mha)/2.5}$. All subsequent mass and ionizing flux calculations are corrected for foreground extinction using the LAB survey \hi\ column densities, unless otherwise specified.

\subsubsection{Correction for Magellanic Bridge ISM Extinction}

\hi\ emission traces most of the dust responsible for the extinction in the inner region of the Bridge. FUSE observations towards the early-type star DI-1388 at $(291\fdg2, -41\fdg3)$ reveal only $\log\left(N_{{\rm H}_2}/\cm^{-2}\right)= 15.45$ \citep{2002ApJ...578..126L} and no ${\rm H_2}$ absorption towards the early-type star DGIK-975 at $(287\fdg2, -36\fdg1)$, which indicates that the faction of ${\rm H}_2$ of the diffuse gas in the central regions of the Bridge is less than $0.004\%$ \citep{2008ApJ...678..219L}.  The non-detections of $^{12}CO(1-0)$ by \citet{2000A&A...363..451S} indicates that this region only contains trace amounts of molecular gas. The lack of molecular gas detections suggests that this region also contains only trace amounts of dust. Therefore we did not apply an extinction correction the central region of the Bridge, which would have only increased the \ha\ intensity by a maximum of $1.5\%$ (see Table \ref{table:extinc}), assuming that this region has a composition similar to the SMC-Tail.

The composition of the SMC-Tail is different than the central regions of the Magellanic Bridge, where dust and molecular gas have been directly observed (see \citealt{2006ApJ...643L.107M} and \citealt{2009ApJ...690L..76G}). \citet{2003ApJ...594..279G} determined the extinction properties of the SMC-Wing by measuring stellar reddening. They found that $E(B-V)=0.263~{\rm mag}$, $R_V\simeq2.05$, and $A_V\simeq1.35\times10^{-22}~\langle N_{\rm H{~\textsc{i}}}\rangle~\mathrm{cm^2\cdot atoms^{-1} \cdot mag}$. To determine the \ha\ extinction in the SMC-Tail, we use the extinction curve presented in their study: $\langle A(\mha)/A(V)\rangle=E(\mha-V)/E(B-V)~R_v^{-1}+1$,  where $E(\mha-V)=0.277~{\rm mag}$ (\citealt{2003ApJ...594..279G}: Equations 1 and 4). This yields the extinction correction for the \ha\ emission of the SMC-Tail: 
\begin{equation}
A(\mha)=2.05\times10^{-22}~\langle N_{\rm H{~\textsc{i}}}\rangle~\mathrm{cm^{-2}\cdot atoms^{-1} \cdot mag}.
\end{equation}
We apply this extinction correction to only the SMC-Tail region, where $\langle N_{\rm H{~\textsc{i}}}\rangle$ is determined using the results from the LAB \hi\ survey smoothed to $1\arcdeg$ to match our angular resolution. Correcting the \ha\ intensities for the dust within the structure results in an average 5.2\% increase for the \ha\ SMC-Tail and 6.8\% for the \hi\ SMC-Tail (see Table \ref{table:extinc} and Figure \ref{figure:cartoon_map}) with \hi\ column densities integrated over the $+100$ to $+300~\kms$ in the LSR velocity range. Because the \ha\ emitting regions lie throughout the SMC-Tail and not behind the structure, this extinction correction represents an upper limit for the \ha\ intensities correction. 

\section{\ha\ Intensity Map}\label{section:intensity_map} 

\begin{figure*}
\begin{center}
\includegraphics[scale=0.5,angle=90]{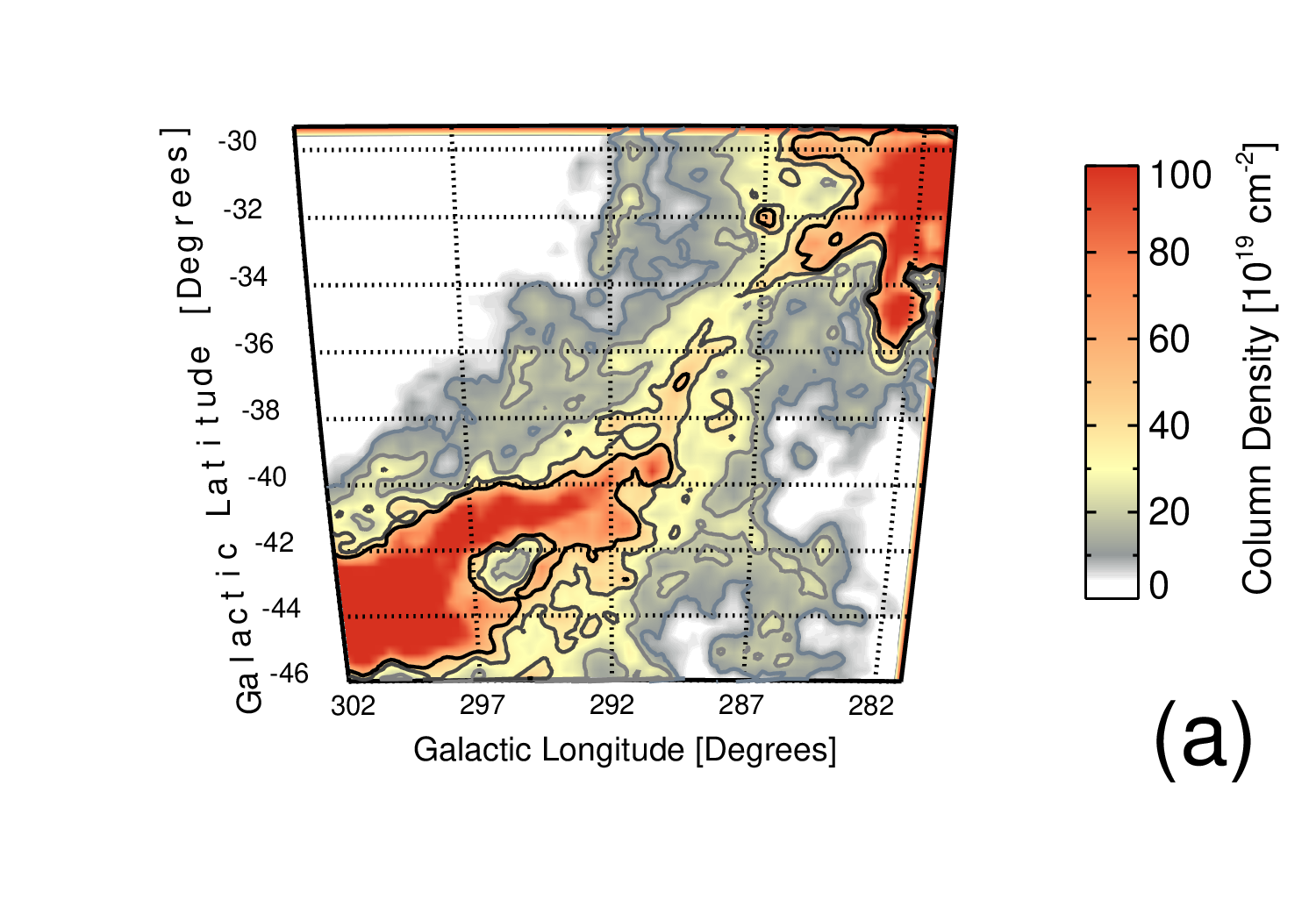} \\
\includegraphics[scale=0.5,angle=90]{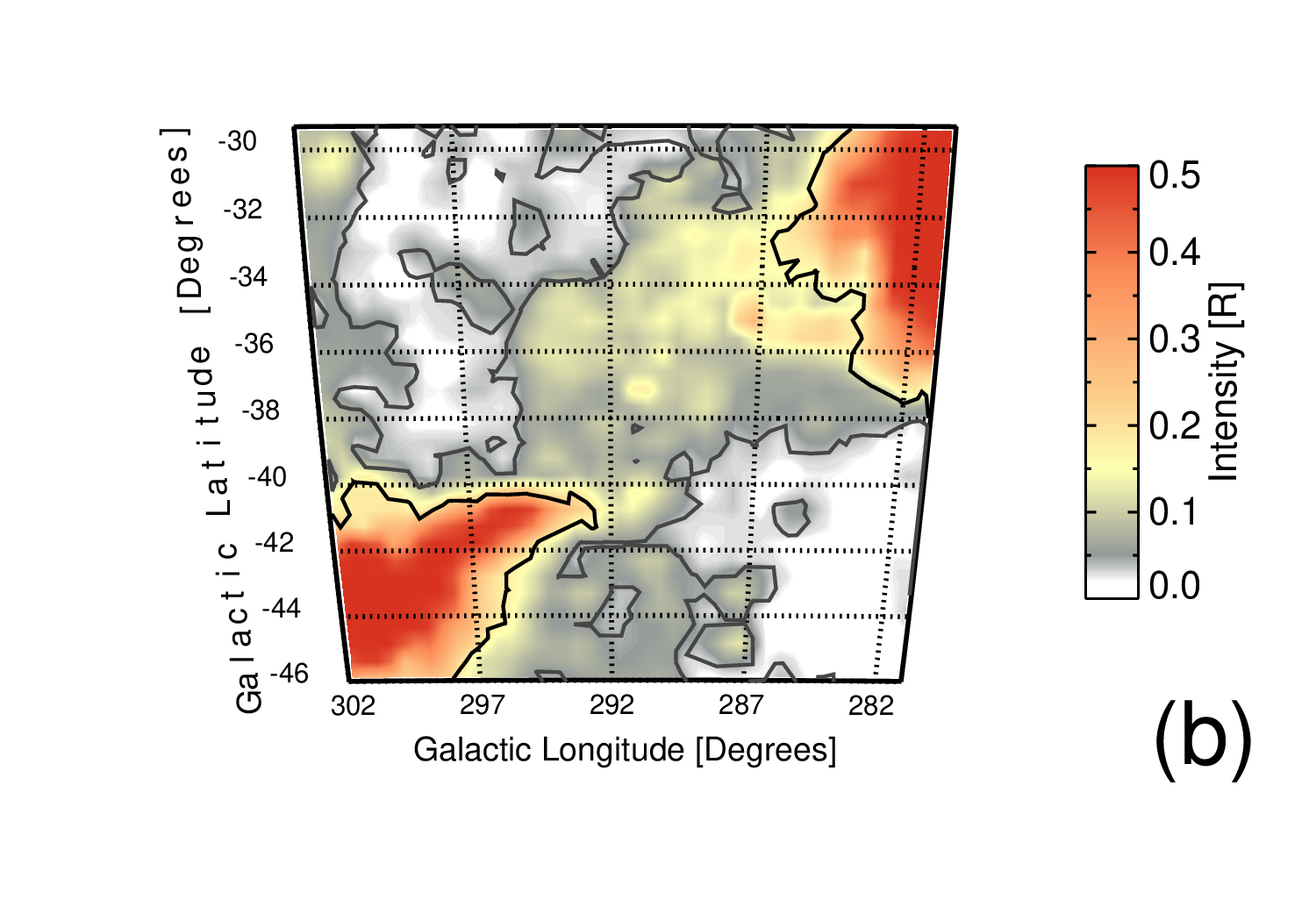}
\end{center}
\figcaption{
The \hi\ \textit{(a)} and \ha\ \textit{(b)} emission maps from the GASS Survey and WHAM observations, respectively. The emission is integrated over the \vlsr\ range of $+100$ to $+300~\kms$. The contour lines in panel \textit{(a)} trace the $10^{19}~\cm^{-2}$ \hi\ column density at increments of 10, 20, 35, and 50. The contour lines in panel \textit{(b)} trace the non-extinction corrected \ha\ intensity at $0.03$ and $0.16~{\rm R}$. The brightest \ha\ emission follows the high column density \hi\ gas in the Small Magellanic Cloud Tail (lower left) and the Large Magellanic Cloud (upper right). 
\label{figure:Ha_Bridge}}
\end{figure*}
 
We surveyed the Magellanic Bridge and SMC-Tail in \ha\ with WHAM from $(\mathit{l, b}) = (281\fdg5, -30\fdg0)$ to $(302\fdg5, -46\fdg7)$ over a velocity range of $0$ to $+315~\kms$ in the LSR frame. Figure \ref{figure:Ha_Bridge} displays both the non-extinction corrected \ha\ intensity and the \hi\ column density over this region, integrated over $+100 \gtrsim \vlsr \gtrsim +300~\kms$. We used Galactic All Sky Survey (GASS) for all the \hi\ spectra, the \hi\ maps, and the \hi\ calculations in this paper \citep{2009ApJS..181..398M, 2010A&A...521A..17K}---except when calculating the extinction correction in Section \ref{section:extinction} where we used the LAB survey---smoothed to 1\arcdeg\ to match the angular resolution of the WHAM observations. To avoid contamination from the Galactic warm interstellar medium, we exclude emission with velocities less than $+100~\kms$. At LSR velocities greater than $+300~\kms$, emission from the LMC contributes in low-longitude and high-latitude regions of the map; for this reason, we chose to also exclude emission at velocities greater than $+300~\kms$. As mentioned in Section \ref{section:bright}, the sensitivity of this survey is decreased from $I_{{\rm H}\alpha}\simeq30~{\rm mR}$  to $I_{{\rm H}\alpha}\simeq40~{\rm mR}$ at $\vlsr\sim+250~\kms$ due to residuals in the bright OH line subtraction at $\vgeo=+272.44~\kms$.        

\ha\ emission, with typical intensities above $0.1~{\rm R}$, spans the entire Magellanic Bridge and SMC-Tail roughly tracking the H I emission. Portions of the \ha\ Bridge exist at  slightly higher latitude than the bright \hi\ Bridge, e.g., the patches of \ha\ emission at $(297\arcdeg, -34\arcdeg)$ and $(295\arcdeg, -32\arcdeg)$ in Figure \ref{figure:Ha_Bridge}(b). Fainter \hi\ emission at column densities of $10^{18}~\cm^{-2}$ does span to these higher latitudes, suggesting that this region could be highly ionized. In Figure \ref{figure:Ha_Bridge}, near the LMC and in the region between $(\mathit{l, b}) = (289\arcdeg, -40\arcdeg)$ and $(283\arcdeg, -45\arcdeg)$ the average \hi\ emission has a higher mean velocity where our measured \ha\ intensity is low.  The loss in \ha\ sensitivity due to the OH line at higher velocities could lead to underrepresentation of emission in certain spatial regions. The general spectroscopic agreement between \hi\ and \ha\ throughout the Bridge combined with our decreased sensitivity limit of $I_{{\rm H}\alpha}\simeq40~{\rm mR}$ over $+240\le \vlsr \le +275~\kms$ does not preclude the existence of diffuse ionized gas with low emission levels associated with the neutral component. Emission maps in other spectral lines (e.g., \sii\ or \nii) may help reveal undetected gas in these regions.

The \ha\ emission decreases radially with distance from both of the Magellanic Clouds away past $0.16~{\rm R}$ contour. The emission becomes constant  in the central $10\arcdeg$ of the Bridge, past the $0.16~{\rm R}$ contour, where the non-extinction corrected intensity ranges from $0.05\le \iha \le0.16~{\rm R}$. The SMC \hi-Tail extends further than the \ha-Tail by a few degrees, where intensities are typically less than $0.4~{\rm R}$.

An elevated region of \ha\ emission exists within the Magellanic Bridge at $(\mathit{l, b}) = (293\arcdeg, -37\arcdeg)$, just off the bright $N_{\rm{H{\textsc{~i}}}}=5\times10^{20}~\cm^{-2}$ contour of the \hi\ SMC-Tail as displayed in Figure \ref{figure:Ha_Bridge}. A bright foreground star, with $M_V=4.7~{\rm mag}$ at $(\mathit{l, b}) = (293\fdg4, -39\fdg8)$, aligns with this line-of-sight. Although we replaced the spectra of the \ha\ map with the average of the nearby spectra for the sightlines within a radius of $1\fdg0$ of the star, this region is still brighter than the neighboring sightlines. The profile of the resultant \ha\ spectra in this region agree with the \hi\ spectra, with \hi\ emission features at $+150 \le \vlsr \le +200~\kms$ and \ha\ features at $+150 \le \vlsr \le +250~\kms$, where the \hi\ column density peaks at $+180~\kms$ compared to $+170~\kms$ for the \ha. The \ha\ spectra also have a less prominent component at $+225~\kms$. The emission feature is only present in the neighboring \hi\ spectra at more negative latitiudes. The similarities between the \ha\ and \hi\ spectral components suggest that this rise in \ha\ intensity at this location is real. 

Although large scale structure of the \ha\ and \hi\ Bridge agree, there are subtle differences at small scales.  In the region above the main bridge, with more positive latitudes, there are small patches of elevated \ha\ emission at roughly $(\mathit{l, b}) = (297\fdg0, -34\fdg0), (295\fdg0, -32\fdg5), (293\fdg0, -30\fdg0), (285\fdg5, -41\fdg0)$, and $(288\fdg0,-45\fdg0)$. The elevation in \ha\ could be explained if these regions are correlated with star forming sites, are more exposed to the Lyman continuum of the LMC and SMC, or are indicators of shock-heated gas that is produced as the Magellanic Bridge travels through Milky Way (MW) halo gas.

\begin{figure}
\begin{center}
\epsscale{1}\plotone{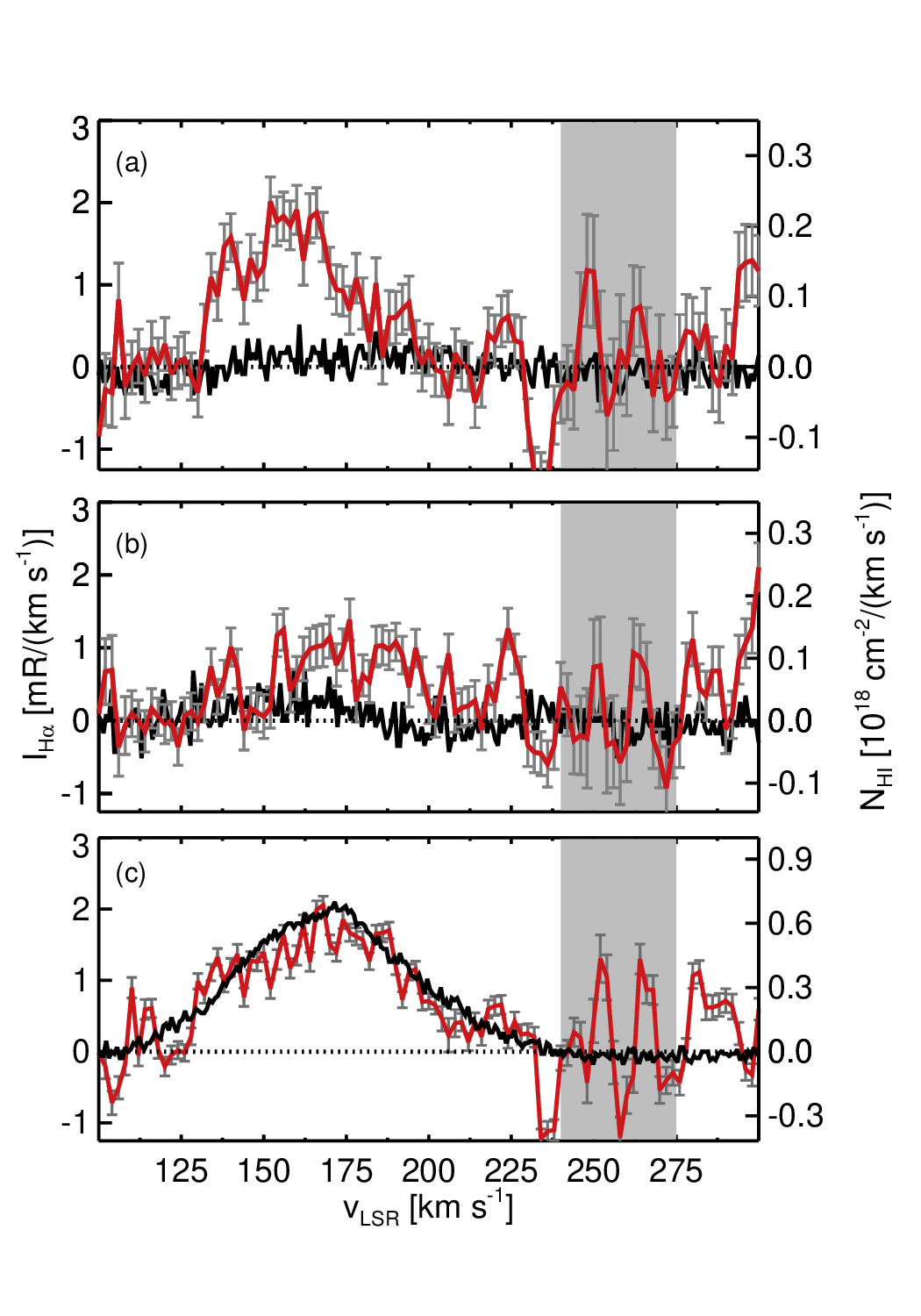}
\end{center}
\figcaption{
Comparison of non-extinction corrected \ha\ intensity (red) and \hi\ column density (black) associated with an elongated, faint \ha\ feature that extends off the \ha\ bridge and spans from $(\textit{l,b})$=$(302\arcdeg, -40\arcdeg)$ to $(302\arcdeg, -30\arcdeg)$. Panel \textit{(a)} includes the emission at $(302\fdg0, -30\fdg7)$, panel \textit{(b)} at $(301\fdg9, -35\fdg1)$, and panel (c) at $(301\fdg8, -40\fdg1)$. The region highlighted in gray represents the location that the bright OH line was removed, marked as \textit{(ii)} in panel \textit{(a)} of Figure \ref{figure:atmosphere} and represents a region where our sensitivity is low. 
\label{figure:LA_spectra}}
\end{figure}

An elongated, faint \ha\ feature exists off the \ha\ bridge that spans a minimum $10\arcdeg$ from $(\mathit{l, b}) = (302\arcdeg, -40\arcdeg)$ and $(\mathit{l, b}) = (302\arcdeg, -30\arcdeg)$ that might be material associated with the Leading Arm or stellar outflows from the SMC. This structure is at the edge of our \ha\ Bridge survey and may extend to higher Galactic longitudes and latitudes. The \ha\ emission component of this structure ranges from $+140\le \vlsr\le+210~\kms$, with typical intensities of roughly $0.03~{\rm R}$. The lack of a complementary \hi\ emission above $N_{\rm H{\textsc{~i}}}=1.6\times10^{18}~\cm^{-2}$, the $3\sigma$ sensitivity of the GASS \hi\ survey at a width of $30~\kms$ \citep{2009ApJS..181..398M}, combined with the faint \ha\ emission could indicate that this gas is low-density or hot $({\rm T }>10^5~{\rm K})$ medium. Although this structure is very faint, it is likely real as the velocity components persist throughout the structure. Figure~\ref{figure:LA_spectra} shows the \hi\ and \ha\ spectra at three locations along this structure. These velocity components are at velocities similar to those observed in the SMC. The small angular distance and velocity difference suggests that this gas is associated with the SMC. With the high star-formation rate of the SMC and the galaxy interactions with the LMC and MW, this structure is probably associated with either SMC stellar feedback or displaced material, removed by galaxy interactions.
       
\section{Comparison of the \ha\ and \hi\ Gas}\label{section: compare}      

The large and small scale similarities and differences between the neutral and ionized gas phases provide clues to the processes affecting the structure. In this section, we compare the differences between the \hi\ and \ha\ velocity distribution and the strength of these emission lines. 
                      
\subsection{\hi\ and \ha\ Velocity Distribution}\label{section:velocity}

\begin{figure}
\begin{center}
\includegraphics[scale=0.35,angle=90]{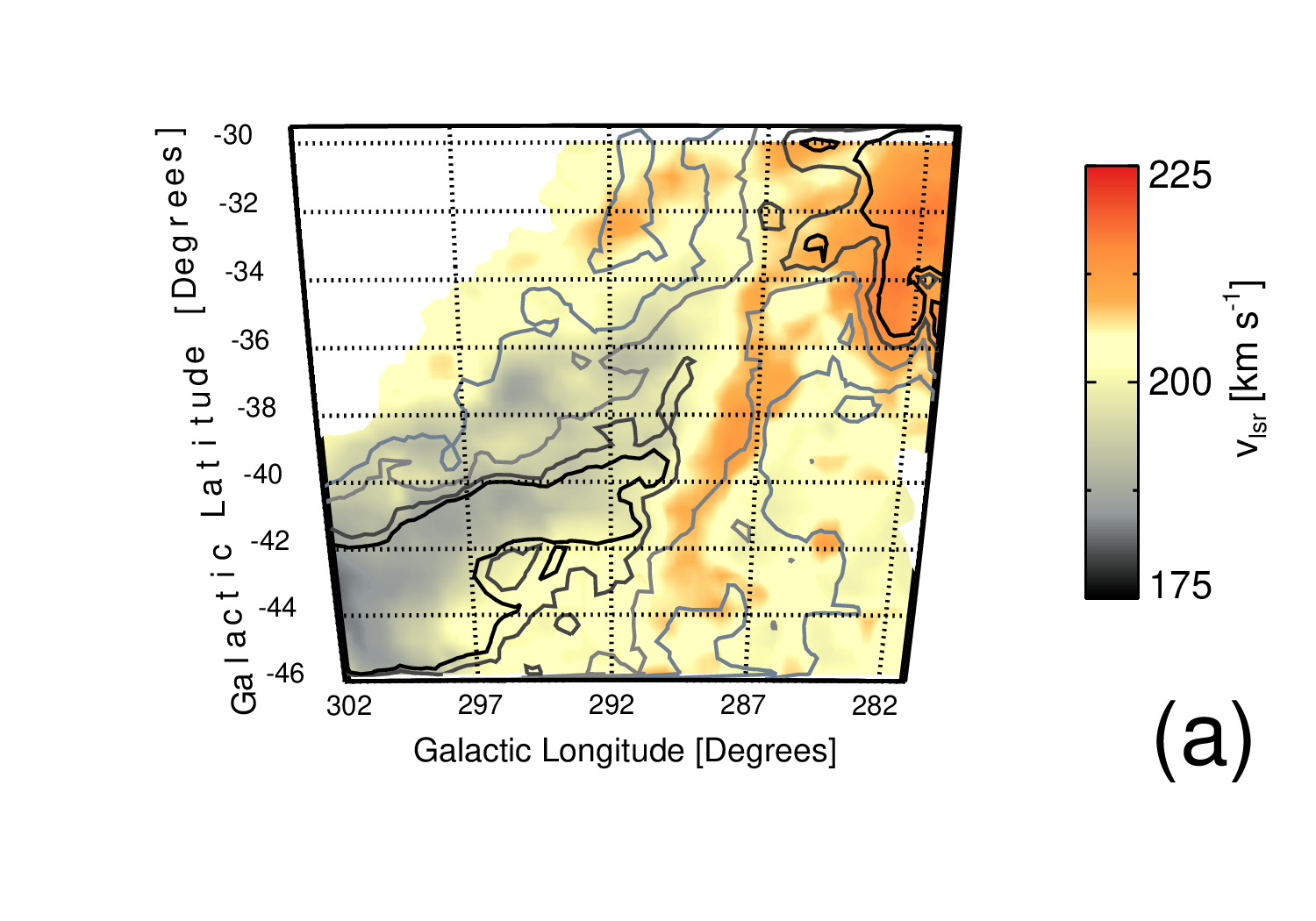} \\
\includegraphics[scale=0.35,angle=90]{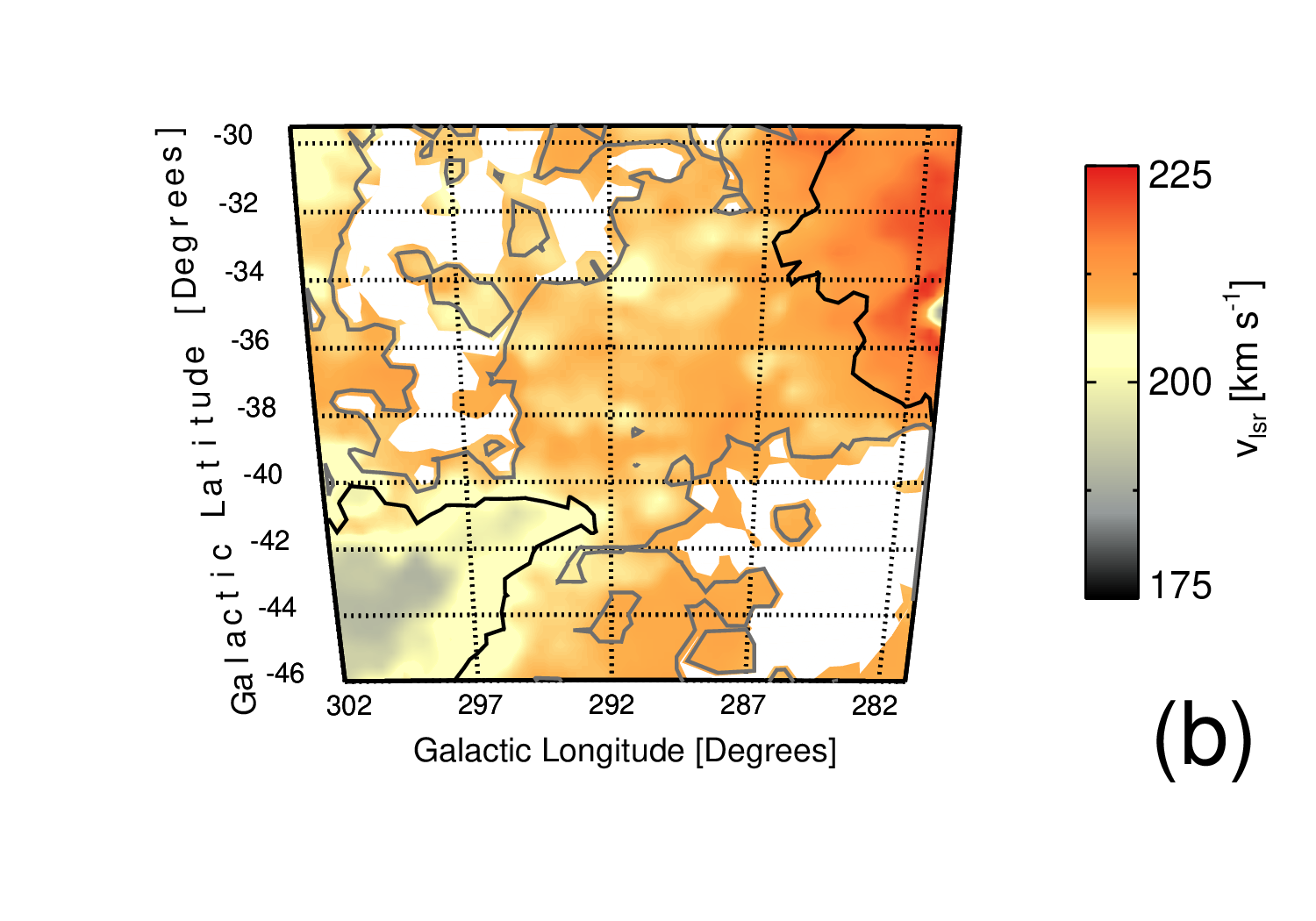}
\end{center}
\figcaption{
The \hi\ \textit{(a)} and \ha\ \textit{(b)} first moment map for the \hi\ column densities greater than $10^{20}~\cm^{-2}$ and the \ha\ intensities greater than the $0.03~\rm{R}$. The contour lines in panel \textit{(a)} trace the $10^{19}~\cm^{-2}$ \hi\ column density at increments of 10, 20, 35, and 50. The contour lines in panel \textit{(b)} trace the non-extinction corrected \ha\ emission at $0.3$ and $0.16~{\rm R}$.
\label{figure:First_Moment}}
\end{figure}

The Magellanic Bridge has a complex velocity distribution. The first moment (also known as the intensity-weighted mean velocity or velocity field: $\bar{v}={\sum v \times I(v)}\div {\sum I(v)}$) of the \ha\ increases from roughly $+175$ to $+225~\kms$\ across the Magellanic Bridge from the SMC-Tail to the LMC. The \hi\ increases from roughly $+125$ to $+250~\kms$\ over the same region. These global velocity trends are shown in Figure \ref{figure:First_Moment}. The smooth \ha\ and \hi\ velocity gradients are a result of blending multiple components in constructing the first moment map. \citet{2005A&A...432...45B} suggest that the \hi\ velocity gradient is largely due to projection effects and indicates that the Magellanic Bridge is likely orbiting parallel with the Magellanic Clouds. 

The \ha\ first-moment map has a much smoother distribution than the corresponding \hi\ map. Three effects cause this difference: (1) The \ha\ emission is much more broad than the \hi, both in width of the individual components and in the overall velocity extent of the multiple components. (2) The angular resolution of the \ha\ survey is much lower than the \hi\ GASS survey at $1\arcdeg$ compared to $16\arcmin$. Each \ha\ observations spans a spatial diameter of $\sim 1~\kpc$, assuming a distance of $55~\kpc$, which causes all the small scale structure to be blended together and diluted in the resultant spectra. (3) The \ha\ survey is less sensitive over the velocity range of $+240 \le \vlsr \le +275~\kms$ due to the residuals from a bight atmospheric OH line (see Section \ref{section:bright}); the average emission of the Bridge shifts to $\vlsr \gtrsim+240 $ at the sightlines closest to the LMC, causing the \ha\ first-map to be less accurate for the faint emission in this region.   

\begin{figure}
\begin{center}
\epsscale{1.25}\plotone{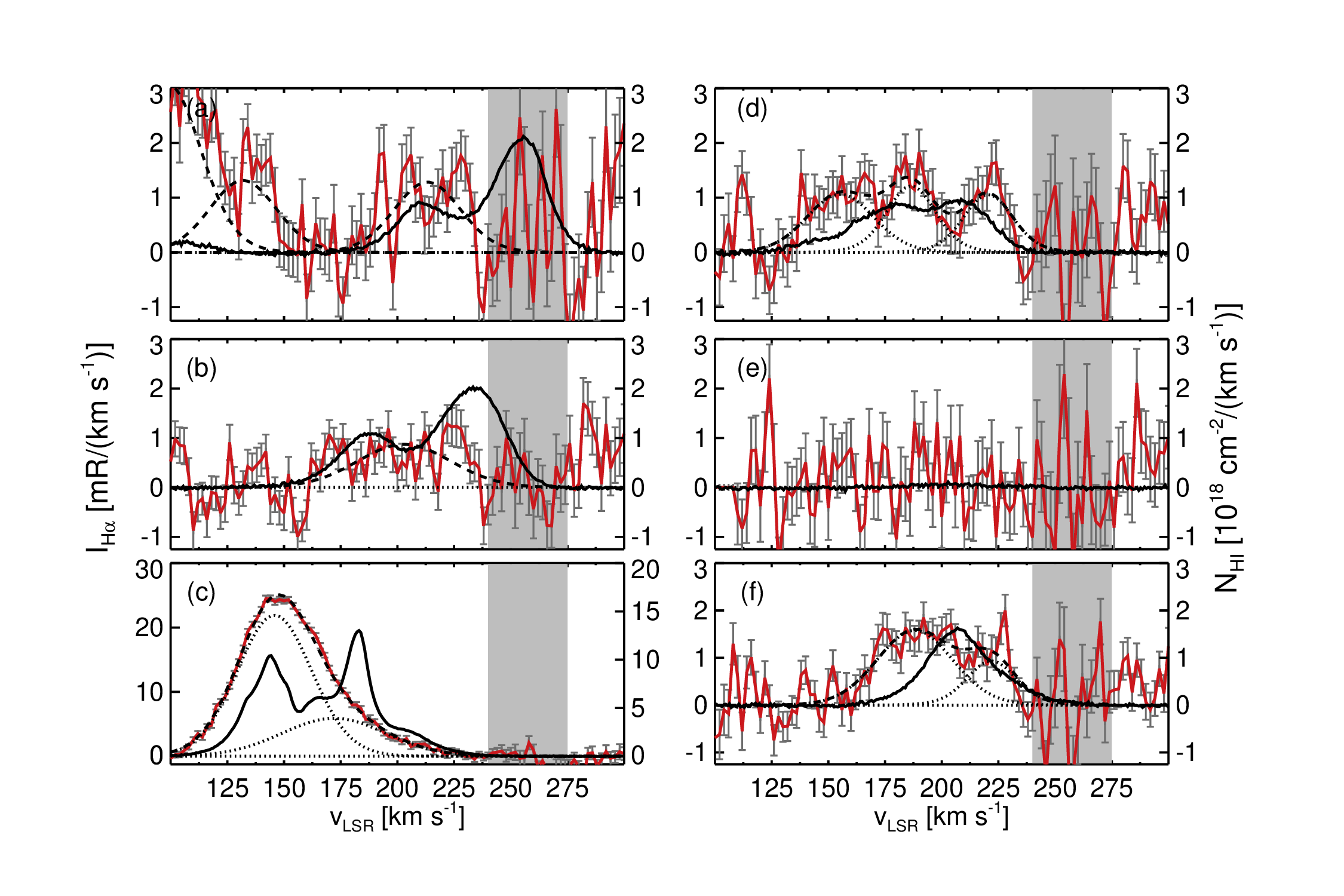}
\end{center}
\figcaption{
Comparison of non-extinction corrected \ha\ intensity (red) and \hi\ column density (black) across the Magellanic Bridge. The dotted Gaussians trace individual \ha\ components; the dashed line is the sum of these components. The region highlighted in gray represents the location that the bright OH line was removed, marked as \textit{(ii)} in panel \textit{(a)} of Figure \ref{figure:atmosphere} and represents a region where our sensitivity is low. Emission from the SMC-Tail is shown in the three left panels: \textit{(a)} at $(290\fdg8, -44\fdg0)$, \textit{(b)} at $(293\fdg2, -46\fdg0)$, and \textit{(c)} at $(297\fdg5, -42\fdg5)$. Panel \textit{(d)} at $(294\fdg5, -36\fdg0)$ displays typical emission features towards the middle of the Magellanic Bridge and panel \textit{(e)} at $(296\fdg2, -33\fdg4)$ illustrates a typical observation off the \ha\ Bridge; this flat spectra indicates that the atmospheric emission has been adequately removed. Panel \textit{(f)} at $(289\fdg5, -32\fdg8)$ shows representative spectra at the Magellanic Bridge and LMC interface.
\label{figure:spectra}}
\end{figure}

Although the global velocity distribution of the \hi\ and \ha\ gradually shift from the SMC to the LMC, the individual spectra have a complex multi-component structure that often have two or more components (see Figures \ref{figure:spectra} and \ref{figure:lehner}). The majority of the brightest \hi\ and \ha\ components peak at roughly the same velocity at most locations; however, there are many places where the dominant \ha\ peak corresponds with a weaker \hi\ peak. This behavior is especially true in the SMC-Tail and near the LMC. Such regions may have a higher ionization fraction, which could indicate active local star formation, more exposure of the gas to ionizing radiation from the galaxies, or an interface between the neutral and ionized gas. 

A statistical investigation of the \hi\ gas components in the SMC-Tail using the Australia Telescope Compact Array (ATCA) and the Parkes telescopes indicates that the multi-component structure might be associated with two kinematically and morphologically distinct arms of gas emanating from the SMC \citep{2004ApJ...616..845M}. Numerical simulations by \citet{1994MNRAS.266..567G} predict that  the lower velocity and more southern arm would extend to the LMC. Figure \ref{figure:spectra} includes a comparison of typical \hi\ and \ha\ spectra towards three locations in the SMC-Tail where the neutral gas exhibits this multi-peaked distribution. In Figure \ref{figure:spectra}(a), the bright \hi\ emission aligns with the \ha\ emission, but in Figure \ref{figure:spectra}(b)--(c) the peak in \ha\ traces the faint \hi\ peak. Many of the \ha\ sightlines towards the SMC-Tail have a two component \ha\ velocity distribution, including the spectrum shown in Figure \ref{figure:spectra}(c).

\begin{figure}
\begin{center}
\includegraphics[scale=0.3,angle=90]{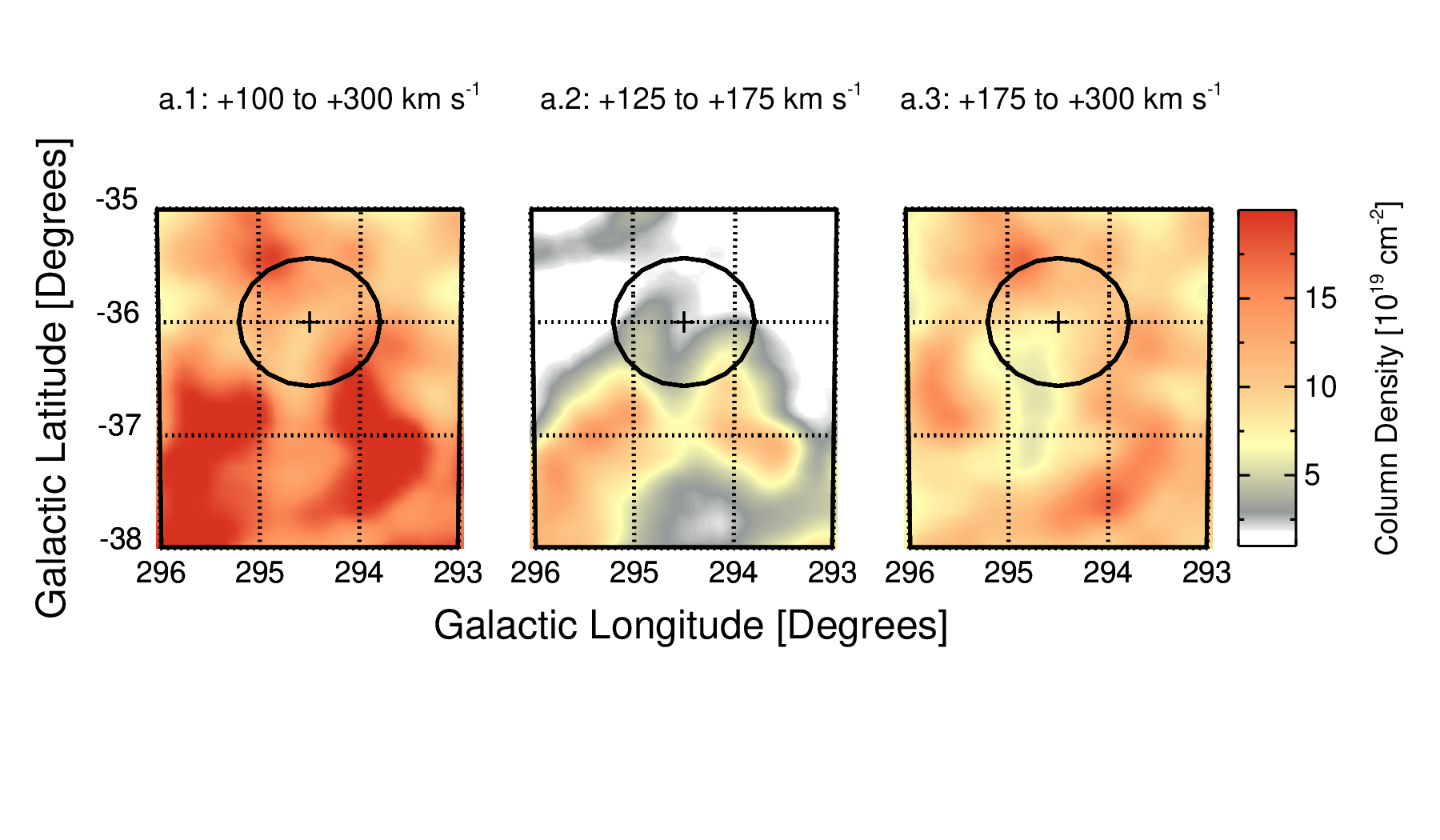} \\
\includegraphics[scale=0.3,angle=90]{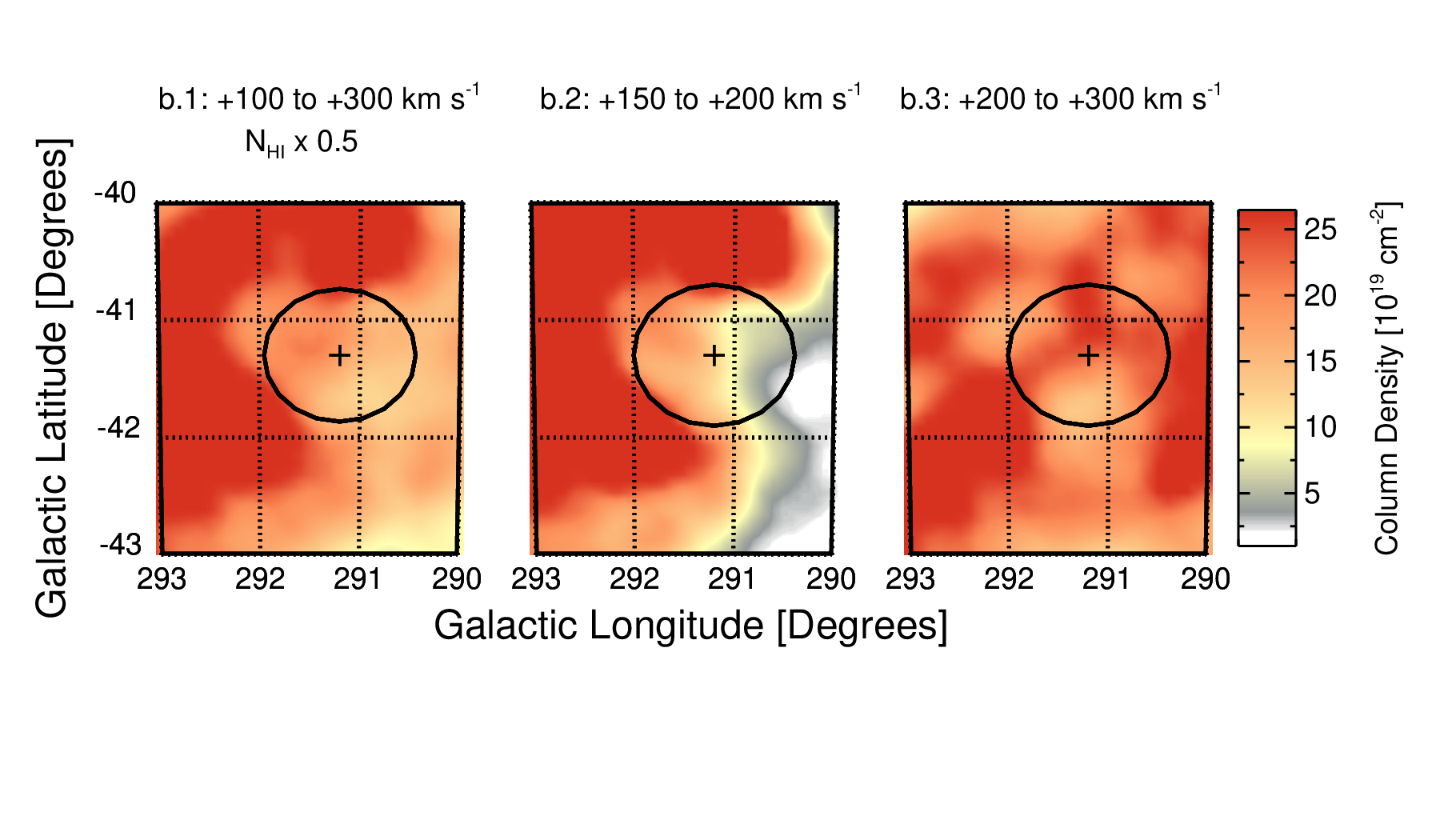} \\
\includegraphics[scale=0.3,angle=90]{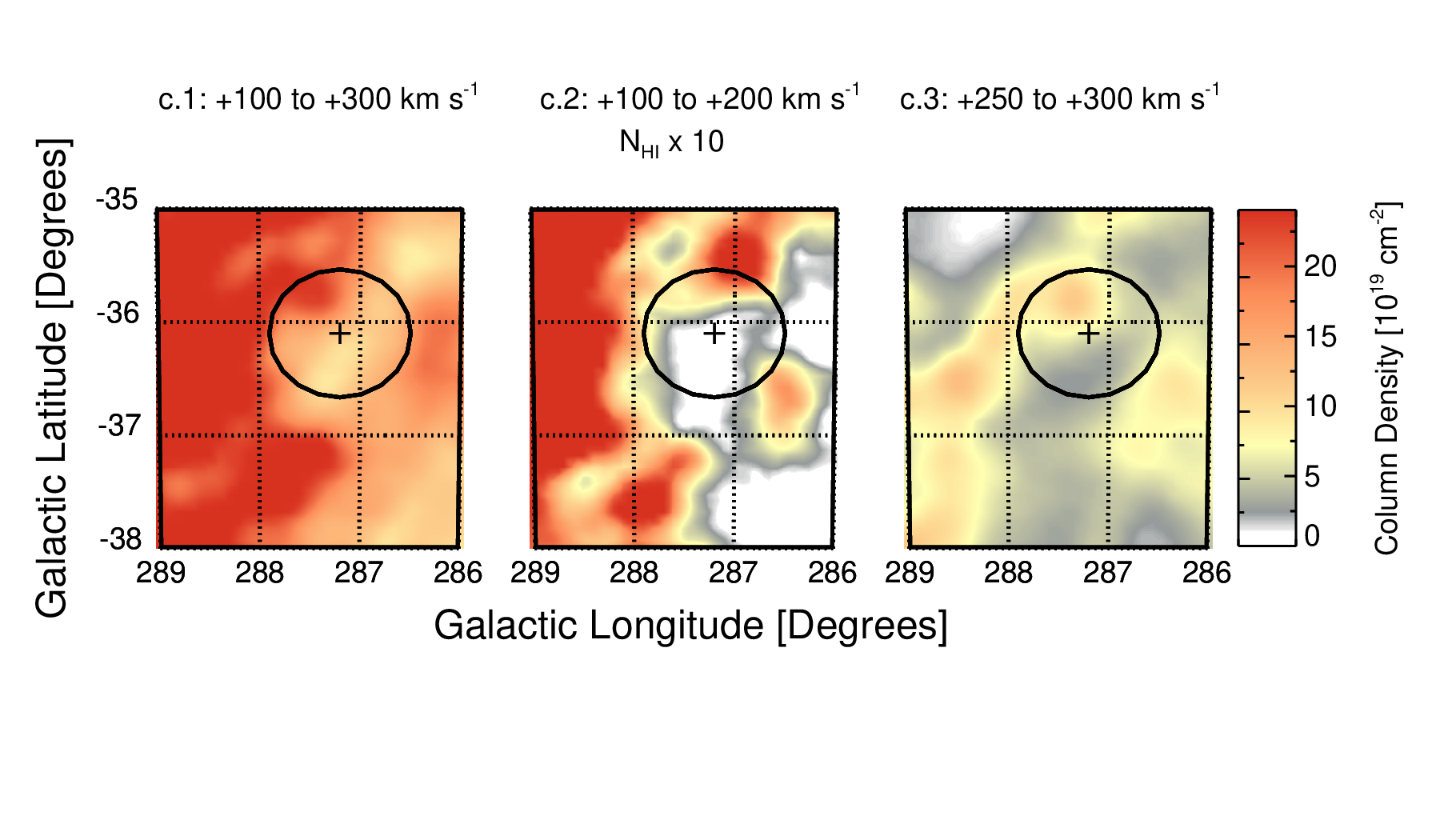} 
\end{center}
\figcaption{
\hi\ column density sub-maps illustrating the substructure of the neutral gas within within one WHAM beam. The emission along these three sightlines produces \ha\ and \hi\ spectra that differ in component structure when the \hi\ is smoothed to the same angular resolution. Panels a.1--3 include the \hi\ Bridge emission at $(294\fdg5, -36\fdg0)$ (see Figure \ref{figure:spectra}d), panels b.1--3 include emission towards early-type star DI~1388 at $(291\fdg2, -41\fdg3)$ (see Figure \ref{figure:lehner}a), and panels c.1--3 include emission towards early-type star DGIK~975 at (287\fdg2, -36\fdg1) (see Figure \ref{figure:lehner}b). Panels a--c.1 integrate the emission $\vlsr$ from $+100$ to $+300~\kms$, panel a.1 from $+100$ to $+175~\kms$, b--c.2 from $+100$ to $+200~\kms$, a.3 from $+175$ to $+300~\kms$, and b--c.3 from $+200$ to $+300~\kms$. The regions used to produce these spectra are outlined within the large black circles, depicting the WHAM beam size.
\label{figure:mapped_pointed}}
\end{figure}

In the Magellanic Bridge, the \ha\ emission has velocity components that lack complementary \hi\ component, possibly revealing a highly ionized region. An example of this behavior is shown in Figure \ref{figure:spectra}(d) towards (\textit{l,b}) = $(294\fdg5, -36\fdg0)$ at $\vlsr\sim165~\kms$. Figure \ref{figure:mapped_pointed}(a.1--3) shows a mini map of the \hi\ emission used to produce the spectra in Figure \ref{figure:spectra}(d), with the 1\arcdeg\ averaged region outlined in black. The velocity range of Figure \ref{figure:mapped_pointed}(a.2) channel map was chosen to highlight the emission where the smoothed \hi\ spectral components differ from the \ha\ components. The small scale structure of the \hi\ gas enclosed within the same angular extent as the WHAM observations reveals a small subregion with a complementary component at $+165~\kms$ towards ($294\fdg3, -36\fdg5$) with ${\textrm \nhi}\sim5\times10^{19}\cm^2$.

\begin{figure}
\begin{center}
\epsscale{1}\plotone{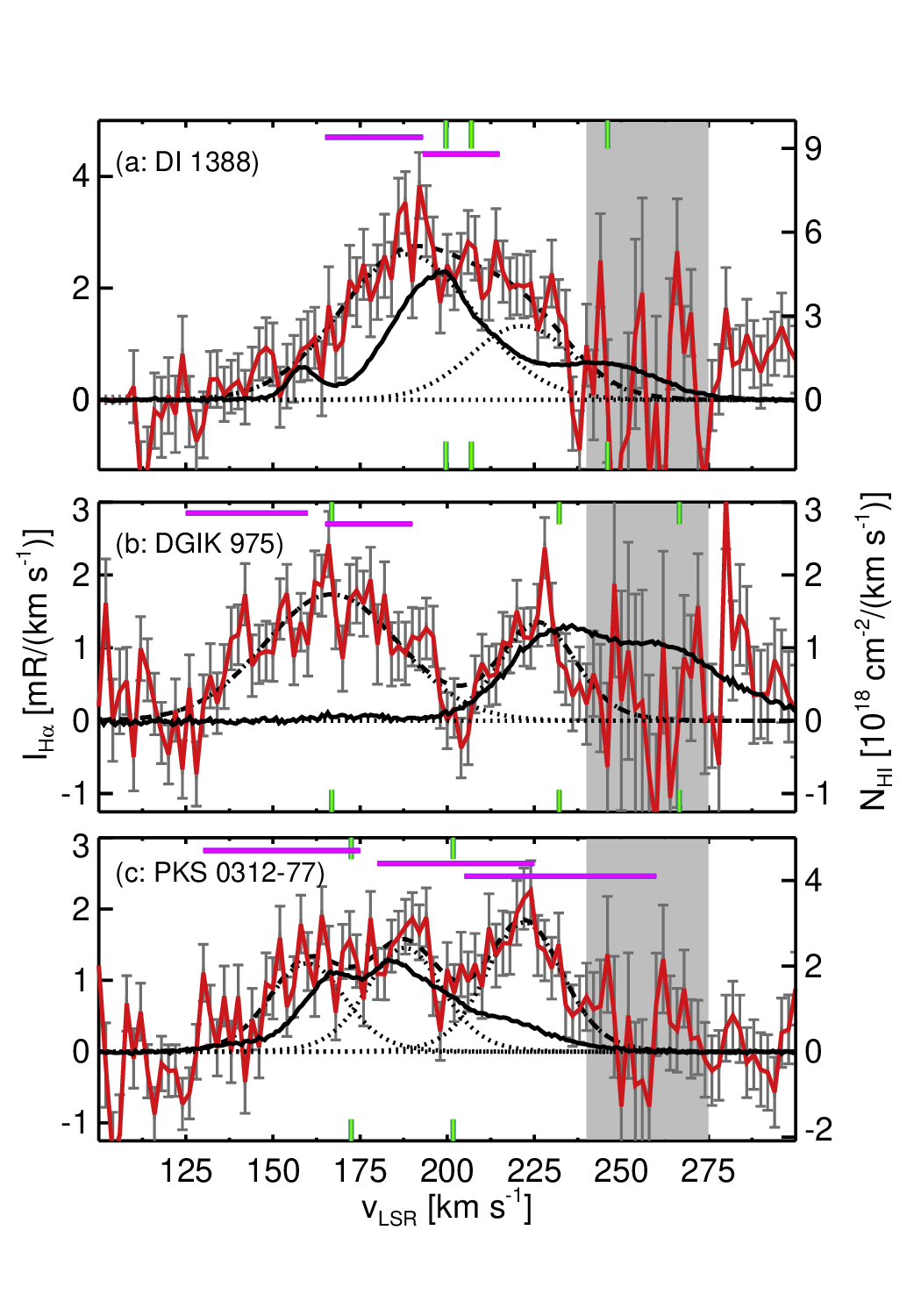}
\end{center}
\figcaption{
Comparison of  non-extinction corrected \ha\ intensity (red), \hi\ column density (black), and UV-absorption species. The dotted Gaussians trace individual \ha\ components; the dashed line is the sum of these components. The green markings indicate the center position of \hi\ emission features and the purple markings indicate the location of known absorption features.  The region highlighted in gray represents the location that the bright OH line was removed, marked as \textit{(ii)} in panel \textit{(a)} of Figure \ref{figure:atmosphere}; unfortunately, this study is insensitive to faint \ha\ emission in this gray region as this emission could easily be subtracted during the OH line removal. Emission towards early-type star DI~1388 is shown in panel \textit{(a)} at $(291\fdg2, -41\fdg3)$, early-type star DGIK~975 in panel \textit{(b)} at $(287\fdg2, -36\fdg1)$, and towards background quasar PKS~0312-77 in panel \textit{(c)} at $(293\fdg5, -37\fdg4)$. \citet{2008ApJ...678..219L} found that the sightline DI~1388 is mostly ionized at $+165\le \vlsr \le+193~\kms$ and partially ionized at $+193\le \vlsr \le+215~\kms$. Sightline DGIK~975 has two absorption features at $+140$, $+175~\kms,$ where the higher velocity component is more neutral; both of these sightlines are towards early-type stars. The PKS~0312-77 sightline, towards a background quasar and shown in panel \textit{(c)}, has absorption features at $+160$, $+200$, $+240$, and $+310~\kms$ (N. Lehner, private communication).
\label{figure:lehner}}
\end{figure}

\citet{2008ApJ...678..219L} also identified highly ionized regions in the inner region of the Magellanic Bridge from the absorption in the spectra of two early-type stars: DI~1388 at (\textit{l,b}) = $(291\fdg2, -41\fdg3)$ and DGIK~975 at (\textit{l,b}) = $(287\fdg2, -36\fdg1)$ marked in Figure \ref{figure:lehner}; FUSE observations from that study revealed gas that is mostly ionized at $+165\le \vlsr \le+193~\kms$ and partially ionized at $+193\le \vlsr \le+215~\kms$ towards DI~1388 and that the gas at $+140~\kms$ has a higher ionization fraction than the gas at $+175~\kms$ towards the DGIK~975 sightline. The \ha\ and \hi\ spectra towards these two early-type stars are displayed in Figure \ref{figure:lehner} where the center of the \hi\ components are marked with green and the positions of the UV-absorption features are marked with purple. The lack correlation of the UV-absorption with the \ha\ and \hi\ emission at $+140~\kms$ in the DGIK~975 sightline suggests this component is highly ionized and possibly influenced by a different process or is exposed to more ionizing flux than the gas at $+175~\kms$, which aligns with a bright \ha\ emission line.   

Both the DI~1388 and DGIK~975 sightlines have \ha\ emission below $+200~\kms$ that is absent in the averaged \hi\ spectra in Figure \ref{figure:lehner}(a--b). Figure \ref{figure:mapped_pointed}(b--c.1--3) includes mini \hi\ emission maps of the region used to produce these \hi\ spectra. Both of the low channel maps contain subregions with bright \hi\ emission within the averaged $1\arcdeg$ region that become dilute when averaged with the surrounding faint emission. These figures illustrate that---although the \ha\ observations excel at mapping the large scale structure of this diffuse Bridge---the small scale details of the ionized gas are unresolvable and might vary greatly within one WHAM beam as seen in the \hi.
            
\subsection{\ha\ Intensity and \hi\ Column Density}\label{section:Intensity_ColumnDensity}

\begin{figure}
\begin{center}
\includegraphics[scale=0.35,angle=90]{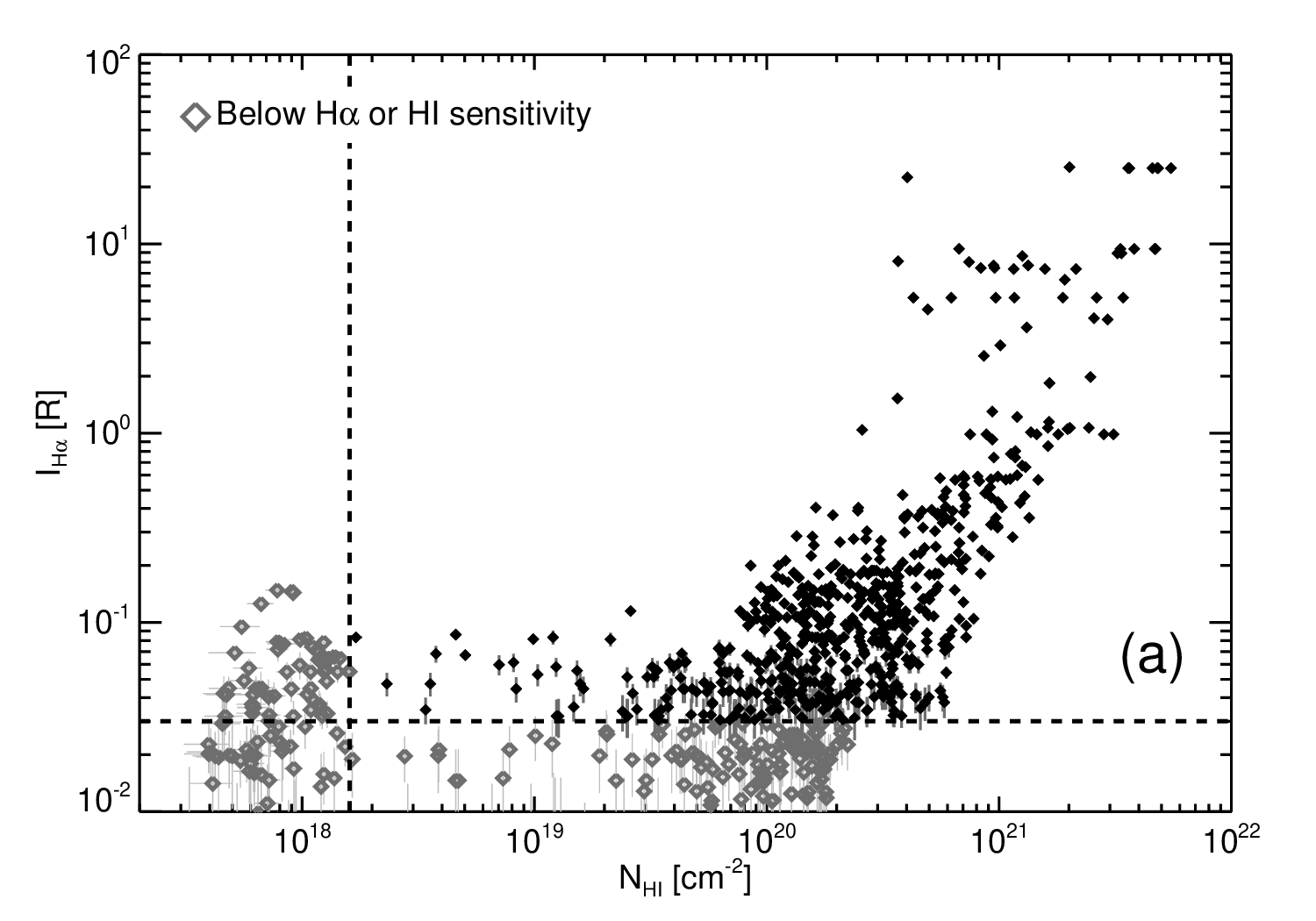} \\
\includegraphics[scale=0.35,angle=90]{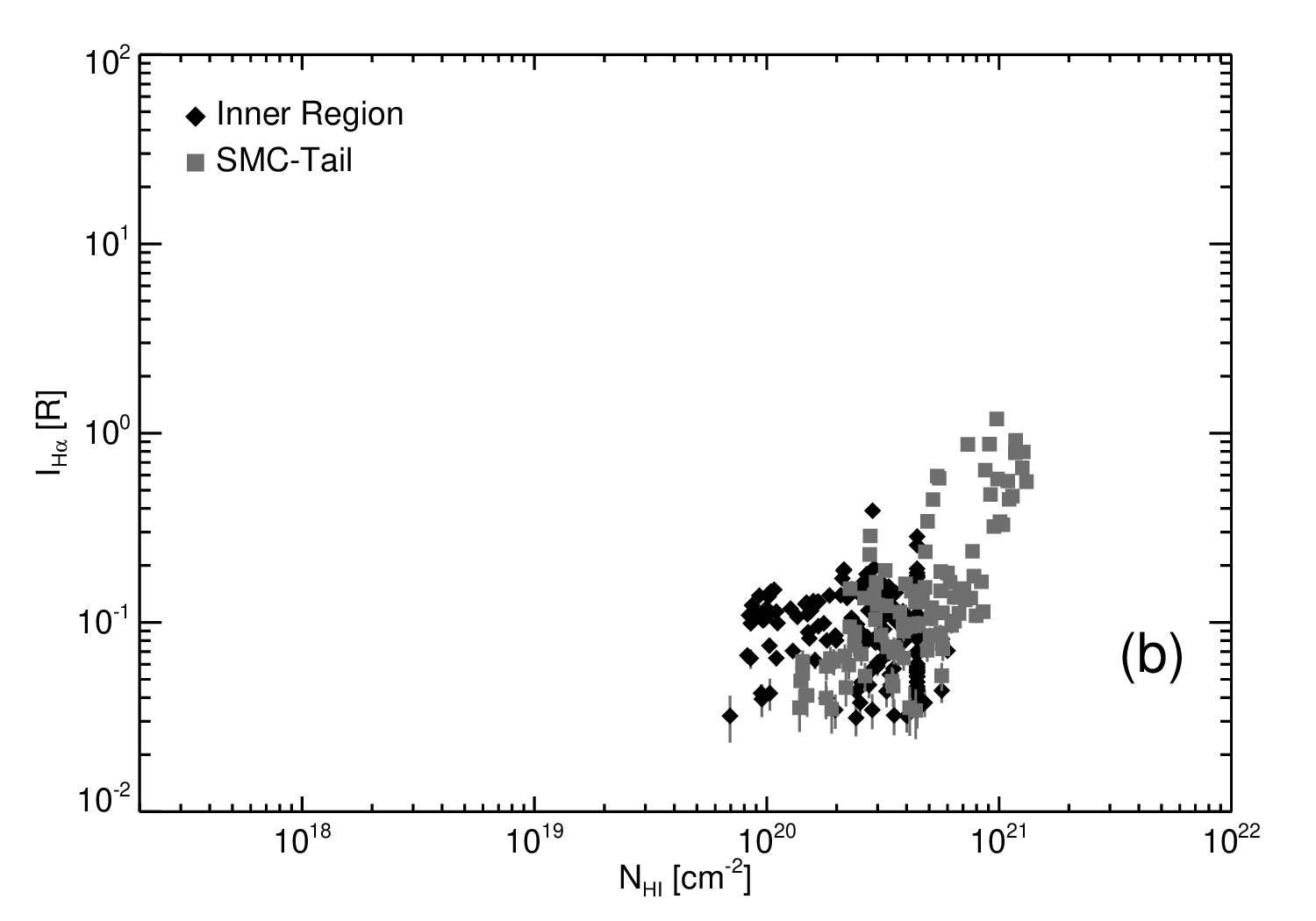} \\
\includegraphics[scale=0.35,angle=90]{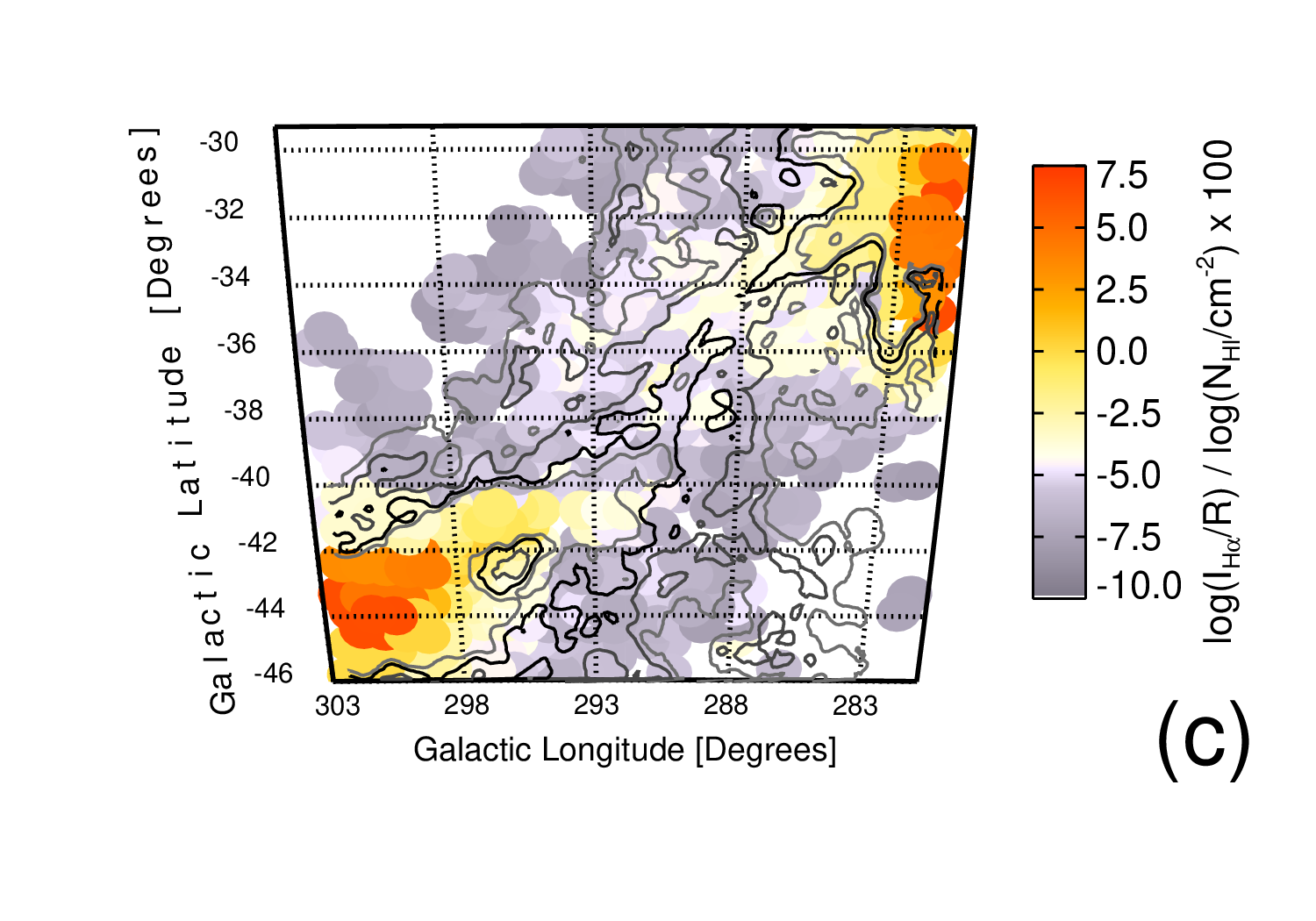} 
\end{center}
\figcaption{
The \ha\ intensity verses the \hi\ column density, integrated over from $+100$ to $+300~\kms$. Panel (a) includes a comparison towards all of the sightlines of the \ha\ map, shown in Figure \ref{figure:Ha_Bridge}b. The open-gray diamonds signify values with $\iha < 0.03~{\rm R}$ and $N_{\rm{H}\textsc{ i}} < 1.6\times10^{18}~\cm^{-2}$, the detection sensitivity as indicated by the dashed lines. Panel (b) is the same as panel (a), but only includes sightlines towards the inner region (black diamonds) and the SMC-Tail (gray squares). Panel (c) maps the $\log I_{\rm H\alpha}/\log N_{\rm H\textsc{~i}}$ line ratio across the Magellanic Bridge.
\label{figure:NHI_Ha}}
\end{figure}

The global morphology of the \ha\ and \hi\ emission agree (see Figure \ref{figure:Ha_Bridge}). The regions near the Magellanic Clouds behave similarly in \hi\ and \ha. Between the galaxies, both the \ha\ intensity and \hi\ column densities decrease substantially. Figure \ref{figure:NHI_Ha}(a) compares the \hi\ column density and non-extinction corrected \ha\ intensity for each sightline in the entire region of this survey over $+100\le \vlsr \le +300~\kms$. The gray dashed lines and the corresponding gray, open diamonds mark where measurements fall below the sensitivity of WHAM or the GASS \hi\ survey. The sensitivity of our survey is $I_{{\rm H}\alpha}\simeq 30~{\rm mR}$ between $+100\le \vlsr \le +240~\kms$ but raises to $I_{{\rm H}\alpha}\simeq40~{\rm mR}$ between $+240\le \vlsr \le +275~\kms$ due to the increased noise from the subtraction of the bright OH line at $\vgeo=+272.44~\kms$. The sensitivity of the GASS \hi\ survey is $N_{\rm H{\textsc{~i}}}=1.6\times10^{18}~\cm^{-2}$ at a width of $30~\kms$ \citep{2009ApJS..181..398M}. 

There is a strong correlation between $\log{N_{\rm H{~\textsc{i}}}}$ and $\log{I_{\rm H\alpha}}$ in Figure \ref{figure:NHI_Ha}(a) at high $N_{\rm H{~\textsc{i}}}$ and $I_{\rm H\alpha}$. Figure \ref{figure:NHI_Ha}(b) separates the SMC-Tail and the Magellanic Bridge into two groupings. The strength of the faint emission from the neutral and ionized gas in the Bridge, represented as black diamonds, show little to no correlation. Here, changes in \ha\ intensity may be due more to changes in the ionization fraction or to the fraction of ionized regions along the line-of-sight and not to the total column of gas. In the Magellanic Bridge, the \ha\ often has an additional emission feature that is unrelated to the \hi\ emission when smoothed to the same angular resolution, as shown in Figure \ref{figure:spectra}(d) and discussed in Section \ref{section:velocity}. This lack of agreement between the strength of the \hi\ and \ha\ emission is also observed in HVCs, even when the number and location of \hi\ and \ha\ components agree (e.g., \citealt{2001ApJ...556L..33H}; \citealt{2003ApJ...597..948P}: Figure 3; \citealt{2005ASPC..331...25H}: Figure 4; \citealt{2012ApJ...761..145B}: Figure 6).

In the SMC-Tail region, $\log{N_{\rm H{~\textsc{i}}}}$ and $\log{\mha}$ track each other and are marked as grey squares in Figure \ref{figure:NHI_Ha}. This behavior suggests that the neutral and ionize gas phases in these regions are affected by similar processes, that these gas phases influence each other, or that they are well mixed. Figure \ref{figure:NHI_Ha}(c) shows that the LMC-Magellanic Bridge interface also behaves similarly with the \ha\ intensity and \hi\ column density both increasing towards the Magellanic Clouds. The increasing line-of-sight depth towards the SMC and LMC could cause the emission from both the neutral and ionized gas to increase even if the gas density stays constant. Such a strong trend is not typical of HVCs suggesting that the presence of either star-formation sites or the adjacent galaxies influence this trend. 

Both the SMC-Tail and the LMC-Magellanic Bridge interfaces have correlated \hi\ and \ha\ emission. These interface regions are undergoing more star formation than the central region of the Bridge. This could mean that either the star-formation rate in the Bridge is either too low to produce a correlation between these lines or that other processes cause this effect. An alternative affect could be related to the ionizing photons that escape from the Magellanic Clouds. These galaxies are only expelling a small fraction of their ionizing radiation into their surrounding ($f_{\rm esc,~LMC} <4.0\%$ and $f_{\rm esc,~SMC}<5.5\%$; see results presented in Section \ref{section:escape}). The close proximity of the SMC-Tail and the LMC-Magellanic Bridge interface with this ionizing source combined with the high \hi\ column density of these regions, compared to typical HVCs ($N_{\rm H{\textsc{~i}},\ HVC}\lesssim10^{18}~\cm^{-2}$), may cause most of the escaping ionizing photons to be absorbed before reaching the inner region of the Bridge. The lower incident ionization from the surrounding galaxies could reduce the \hi\ and \ha\ relationship within the Bridge.   

\section{Distribution and Mass of the Ionized Gas}\label{section:mass}

\begin{deluxetable*}{lcccccccccccccc}
\tablecolumns{14}
\tabletypesize{\scriptsize}
\tablecaption{Neutral and Ionized Properties\label{table:mass}} 
\tablewidth{0pt}
\tablehead{
\multicolumn{1}{c}{}  & \colhead{} & \multicolumn{4}{c}{Neutral Properties} &\colhead{}  & \multicolumn{2}{c}{Ionized Skin (n$_e=n_0$)} & \colhead{} &\multicolumn{2}{c}{Ionized Skin (n$_e=\frac{1}{2}n_0$)} & \colhead{} & \multicolumn{2}{c}{Mixed ($L_{\rm H^+}$=\lhi)}\\ 
 \cline{3-6} \cline{8-9} \cline{11-12}  \cline{14-15}
\colhead{Region}     	& \colhead{}	& \colhead{$\log\ \langle$\nhi$\rangle$}	& \colhead{M$_{\rm H^0}$} 	& \colhead{$\log$~\lhi}		& \colhead{$\log\ \langle{n_0}\rangle$}   & \colhead{$\langle{EM}\rangle$\tablenotemark{a,b}} 		& \colhead{$\log\ \rm{L}_{\rm H^+}$\tablenotemark{a,b}}	& \colhead{M$_{\rm {H^+}}$\tablenotemark{a,b}}   & \colhead{}  &  \colhead{$\log\ \rm{L}_{\rm H^+}$\tablenotemark{a,b}}  & \colhead{M$_{\rm {H^+}}$\tablenotemark{a,b}} & \colhead{}	& \colhead{$\log\ \langle n_e\rangle$\tablenotemark{a,b}}& \colhead{M$_{\rm {H^+}}$\tablenotemark{a,b}}  \\	
\colhead{} 	& \colhead{}	& \colhead{[$\cm^{-2}$]}  	& \colhead{[10$^6$~M$_{\odot}$]} 	& \colhead{[kpc]}		& \colhead{[$\cm^{-3}$]}	& \colhead{[$10^{-3}\ \pc \cm^{-6}$]}  		&\colhead{[kpc]}		& \colhead{[10$^6$~M$_{\odot}$]}	& \colhead{}  & \colhead{[kpc]} & \colhead{[10$^6$~M$_{\odot}$]} & \colhead{}	&\colhead{[$\cm^{-3}$]}	& \colhead{[10$^6$~M$_{\odot}$]}  }
\startdata		  		 
Inner Region\tablenotemark{c}		&& $20.2$		& $123$	& $3.6$	& $-1.9$		& $328$	& $3.1$	& $68$	 && $5.3$		& $135$	&& $-2.1$ & $104$ 	\\
H{\sc~i} SMC-Tail\tablenotemark{c}	&& $20.6$		& $202$	& $3.3$	& $-1.2$		& $487$	& $2.0$	& $16$	 && $2.6$		& $31$	&& $-1.8$ & $63$ \\
\ha\ SMC-Tail\tablenotemark{c}		&& $20.7$		& $125$	& $3.3$	& $-1.0$		& $962$	& $1.8$	& $7.1$	 && $2.5$		& $14$	&& $-1.7$ & $34$
\enddata
\tablenotetext{a}{The average $I_{{\rm H}\alpha}$ extinction correction factor is 1.2, corresponding to $\log\ \langle$\nhi$\rangle=20.7\pm16.8\cm^{-2}$ and $A(\mha)=0.22~{\rm mag}$ (see Equation \ref{eq:extinction}).} 
\tablenotetext{b}{Assumes an electron temperature of $10^4\rm{K}$.} 
\tablenotetext{c}{Regions defined by a polygon with the following corners: \textit{l}=$(289\fdg0, 283\fdg0, 289\fdg0, 297\fdg0)$ and \textit{b}=$(-30\fdg2, -38\fdg0, -43\fdg0, -35\fdg0)$ for the inner region, \textit{l}=$(301\fdg8, 295\fdg9, 289\fdg1, 295\fdg7)$ and \textit{b}=$(-39\fdg4, -36\fdg2, -43\fdg0, -46\fdg5)$ for the H{\sc~i} SMC-Tail, and \textit{l}=$(300\fdg5, 294\fdg5, 292\fdg0, 297\fdg0)$ and \textit{b}=$(-41\fdg0, -39\fdg5, -42\fdg0, -45\fdg0)$ for the \ha\ SMC-Tail. The boundaries for these regions are displayed in Figure \ref{figure:cartoon_map}.} 
\end{deluxetable*}

The velocity distribution of the \hi\ and \ha\ emission from the SMC-Tail and the central region of the Magellanic Bridge suggests the presence of morphologically distinct structures with different ionization fractions. In Section \ref{section:velocity}, we discussed regions in the SMC-Tail with bright \ha\ emission that coincides with the fainter \hi\ velocity component. There are also regions in the central Bridge where the \ha\ emission lacks a complementary \hi\ component (e.g., the slightlines shown in Figure \ref{figure:LA_spectra}). \citet{2002ApJ...578..126L} and \citet{2008ApJ...678..219L} also identified multiple absorption features in the central region of the Magellanic Bridge with different fractions of ionization. Then these components would likely exist at different gas densities and pressures, indicating that the source of the ionization does not uniformly affect the distinct components. 

The Magellanic Bridge has an unknown morphology and distribution along the line-of-sight. The depth of the ionized gas in a distinct structure depends on the distribution of the ionized gas, which could be well-mixed with the neutral gas or separated from the neutral gas. If distributed in an ionized skin, that skin could be either in pressure equilibrium or pressure imbalance with its neutral component. This distribution will depend on the processes influencing the gas. 

If the neutral and ionized gas are well mixed, then the line-of-sight depth of the components are equal. If instead the neutral and ionized components are separated, but in pressure equilibrium, then electron density of an ionized skin would equal half the neutral hydrogen density \citep{2009ApJ...703.1832H}. Because the emission rate of  \ha\ is proportional to the recombination rate, $\iha=\left(4\pi\right)^{-1} \int \alpha_{B}(T)\ \epsilon_{{\rm H}_{\alpha}}(T)\  n_e\ n_p\ dl_{H^+}$, the depth of the ionized of a structure with a constant electron density and temperature over the emitting region can be written as
\begin{equation}\label{eq:L_ion}
L_{H^+}=2.75\ \left(\frac{T}{10^4 K}\right)^{0.924} \left(\frac{\iha}{R}\right) \left(\frac{n_e}{cm^{-3}}\right)^{-2}\pc, 
\end{equation}
where $n_p\approx n_e$ and the probability that the recombination will produce \ha\ emission is $\epsilon_{{\rm H}_{\alpha}}(T)\approx 0.46\left(T/10^4~{\rm K}\right)^{-0.118}$. This relationship assumes that the gas is optically thick to ionizing photons such that the recombination rate is $\alpha_B=2.584\times10^{-13}\ \left(T/10^4~{\rm K}\right)^{-0.806} {\cm}^3\ {\rm s}^{-1}$ \citep{1988ApJS...66..125M}.

Determining the total mass of the Magellanic Bridge helps to quantify the amount of baryons that have been stripped from the Magellanic Clouds, to explore the effects that the gas removal has on the evolution of these galaxies, and to provide insight on the future of this tidal remnant. The distance, the morphology along the line-of-sight, and the distribution of ionized and neutral gas dominate the uncertainty of a mass estimate. The uncertainty further increases for the diffuse Bridge over the $+240 \ge \vlsr \ge +275~\kms$ velocity range as the sensitivity of the \ha\ survey decreases due to enhanced residuals associated with a bright OH atmospheric line (see Section \ref{section:bright}). We assume a distance of $55~\kpc$, average physical conditions along the whole line-of-sight, and three different gas distributions. The three gas distributions considered include an ionized skin in pressure equilibrium with its neutral component, an ionized skin in pressure imbalance with its neutral component, and a fully mixed cloud without an ionized skin. For simplicity, when determining the density of the neutral and ionized gas, we assume that the \hi\ line-of-sight depth is similar to the width of the Magellanic Bridge.

These oversimplified assumptions exclude many effects that will cause the calculated mass to differ from the actual mass of the Magellanic Bridge: (1) The distance to the Magellanic Bridge varies from roughly 50 to 60~\kpc\ from the SMC to the LMC. (2) There are multiple components along many of the sightlines that could exist at different densities and ionizations as discussed in Section \ref{section:velocity}. (3) The distribution of the neutral and ionized gas could differ between components. To more accurately determine the mass of the ionized gas, a statistical analysis of the \ha\ velocity components should be done to to identify morphologically distinct structures and to estimate the density and ionization fraction of the gas, but this analysis is beyond the scope of this first study. Upcoming multiline observations will aid in the recovery missing velocity components due to bright atmospheric OH line residuals over the velocity range $+240\le \vlsr +275~\kms$ and in discerning changes in the physical conditions among components. Here we use our first full survey of the ionized gas to provide a rough estimate of the mass. Note that the calculated mass will exclude the mass of extremely ionized structures (e.g., \citealt{2002ApJ...578..126L} and \citealt{2008ApJ...678..219L}) where any \ha\ emission is below our sensitivity. 

We calculated the mass of the ionized gas as $M_{\rm H^+}=1.4 m_{\rm H} n_e D^2 \Omega L_{H^+}$, where $\Omega$ is the solid angle, $D$ is the distance from the Sun,  $m_{\rm H}$ is mass of a hydrogen atom, and the factor of 1.4 accounts for helium. The integral of the square of the electron density over the path length of ionized gas, also known as the emission measure $\left(EM\equiv\int n_e^2 dl\right)$, affects the strength of the \ha\ emission. Using Equation \ref{eq:L_ion} for the line-of-sight depth of the ionized gas, the EM becomes 
 \begin{equation}\label{eq:EM}
 EM=2.75\ \left(\frac{T}{10^4 K}\right)^{0.924} \left(\frac{\iha}{R}\right) \cm^{-6} \pc.   
\end{equation}
Using Equation \ref{eq:EM}, the mass of the ionized gas within a $1\arcdeg$ circular beam---the angular resolution of the \ha\ observations---becomes \citep{2009ApJ...703.1832H}
\begin{equation}\label{eq:mass_hplus}
\left({\frac{M_{\rm H^{+}}}{M_\odot}}\right)_{beam}=8.26\left(\frac{D}{\kpc}\right)^2 \left(\frac{EM}{\mathrm{pc\cdot cm^{-6}}}\right)\left(\frac{n_e}{\cm^{-3}}\right)^{-1}.
\end{equation}
We calculated the mass of the neutral gas as $M_{\rm H^0}=1.4 N_{\rm H\textsc{~i}} \Omega D^2$. As in the equation for the $M_{\rm H^+}$, the factor of 1.4 accounts for helium. 

To determine the mass of the gas connecting the Magellanic Clouds, we partitioned this structure into three regions: the Magellanic Bridge, the \hi\ SMC-Tail, and the \ha\ SMC-Tail (see Figure \ref{figure:cartoon_map}). We chose to separate the SMC-Tail into two regions because they have very different \hi\ and \ha\ distributions and likely different ionization fractions. Table \ref{table:mass} lists the calculated masses for each region and their corresponding properties. Because the \ha\ emission likely emanates from within the SMC-Tail and not solely from behind, the SMC-Tail extinction corrected mass represents an upper limit. Applying the full SMC-Tail extinction correction increases the mass of the ionized gas by up to 7\% (see Table \ref{table:extinc}). 

Combining the results for the inner region and the SMC-Tail, the total ionized gas mass for the Magellanic Bridge ranges from $\left(0.7-1.7\right)\times10^8~{\rm M}_\odot$, using the three gas distributions described above---compared to $3.3\times10^8~{\rm M}_\odot$ for the neutral mass.  Note that the slight difference in neutral mass determined in this study and the $2.5\times10^8~{\rm M}_\odot$ $(M_{H^0}=1.4 M_{\rm H{\textsc{~i}}})$ found by \citet{2005A&A...432...45B}---which also used observations from the Parkes Magellanic System \hi\ survey---is due to a different definition of the spatial region that encompasses the Magellanic Bridge over a slightly different velocity range. We extend our spatial region to incorporate a larger latitude range and more of the SMC-Tail. We selected our velocity coverage, $+100$ to $+300~\kms$, to avoided Galactic and LMC contamination, as opposed to the $+110$ to $+320~\kms$ used in \citet{2005A&A...432...45B}. The large range in the ionized mass estimate is due to the unknown gas distribution as the neutral and ionized gas distribution that affects both the line-of-sight depth and the density. Identifying the processes influencing this gas, including the source of ionization, will help narrow this range. 

\section{Source of the Ionization}\label{section:ionization}

\begin{figure}
\begin{center}
\includegraphics[scale=0.4,angle=90]{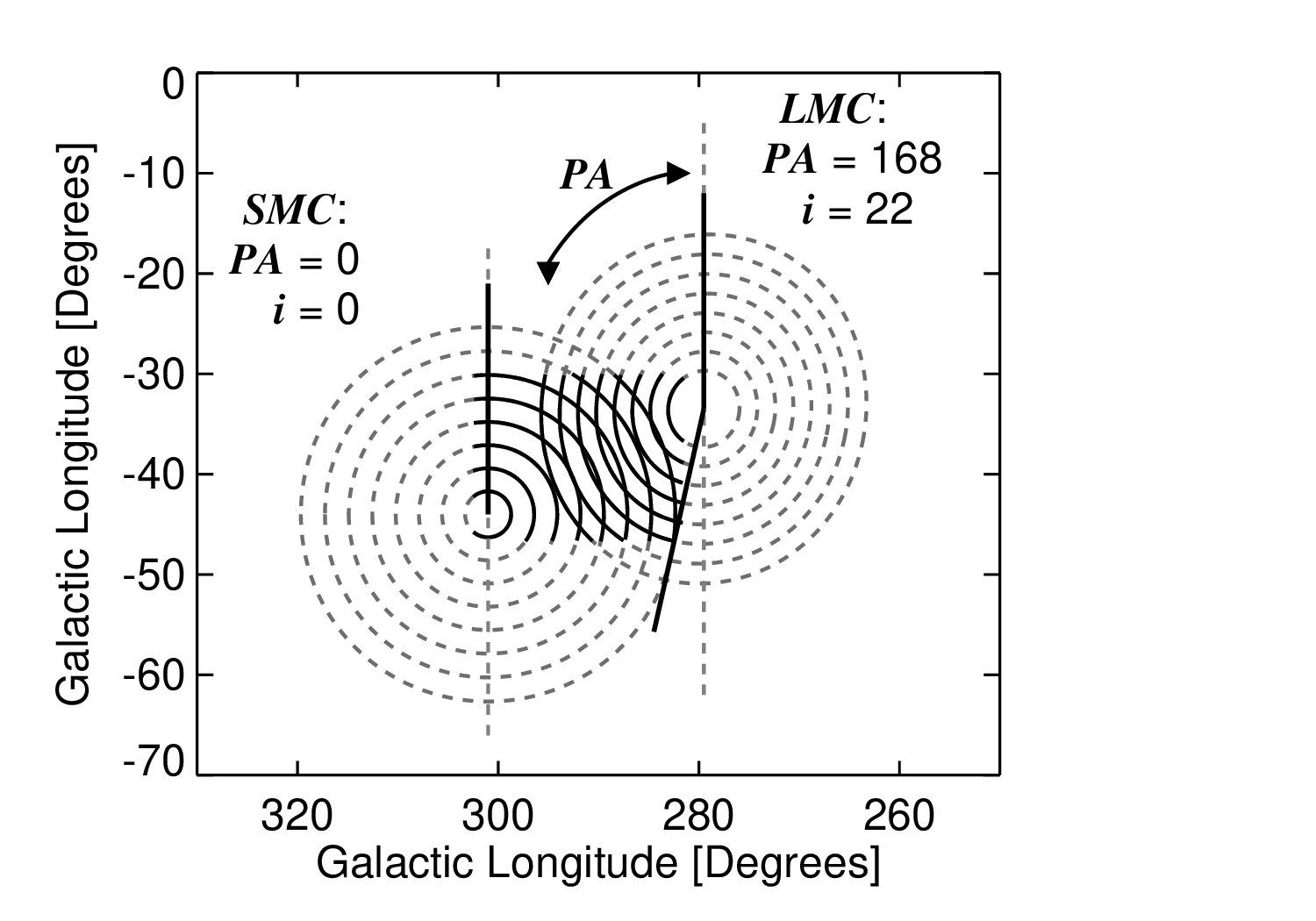} 
\end{center}
\figcaption{
Schematic of the elliptical annuli used to bin the \ha\ intensity radially from the LMC and SMC centers. The spacing depicted between each ring is exaggerated from the actual spacing (0\fdg3) used to determine the mean and median \ha\ intensity values. The LMC annuli is characterized by $PA=168\fdg0$ and $i=22\fdg0$ and expands from the \hi\ disk center at $(l,b)=(279\fdg5, -33\fdg5)$ \citep{1998ApJ...503..674K}. The SMC annuli is circular, where both $PA=0\fdg0$ and $i=0\fdg0$, and expands from $(l,b)=(301\fdg0, -44\fdg0)$; we chose a circular annuli because the \hi\ gas has a random distribution that lacks a disk morphology \citep{2004ApJ...604..176S}. The dashed gray, elliptical lines trace the complete elliptical pattern around each galaxy; the curved, black lines that overlap these gray lines mark the region covered by this survey. The $PA$ of each galaxy is represented by solid black lines that diverge from the vertical axis. 
\label{figure:radial_binning}}
\end{figure}

The Magellanic Bridge contains a substantial amount of ionized gas. Both photoionization and collisional ionization processes might contribute to the ionization of this structure. Sources of photoionization include the extragalactic background, escaping ionizing radiation from the surrounding galaxies (i.e., Milky Way and Magellanic Clouds), and early-type stars within the Bridge. Sources of collisional ionization may include ionization induced by galaxy interactions (e.g., turbulent mixing, ram-pressure stripping, and tidal shocks), strong stellar winds, and supernova explosions. While the \ha\ observations are sensitive to very faint levels of surface brightness, the intrinsic angular resolution is very low. As a result, we are unable to resolve any contribution from compact (e.g., stellar) sources.

The source of the ionization influences the distribution of the neutral and ionized gas. Although the \ha\ emission traces ionized gas and the source of the ionization, definitively identifying which processes affect the Magellanic Bridge requires multiline observations as different ionization sources produce emission and absorption lines of different relative strengths. A future work will use \nii $\lambda6583$ and \sii $\lambda6716$ WHAM observations of the entire  Magellanic Bridge and SMC-Tail to explore the physical conditions of the gas and to discriminate between different sources of ionization.

\begin{figure}
\begin{center}
\includegraphics[scale=0.3,angle=0]{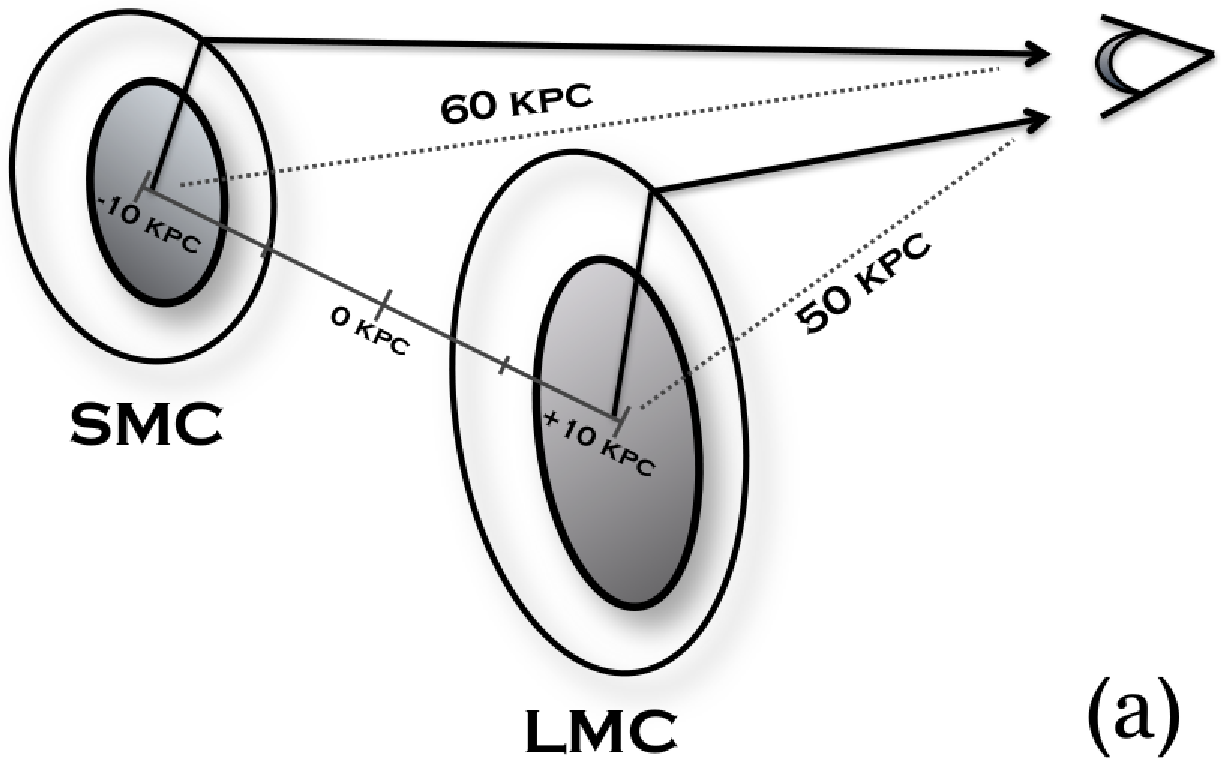} \\
\includegraphics[scale=0.3,angle=0]{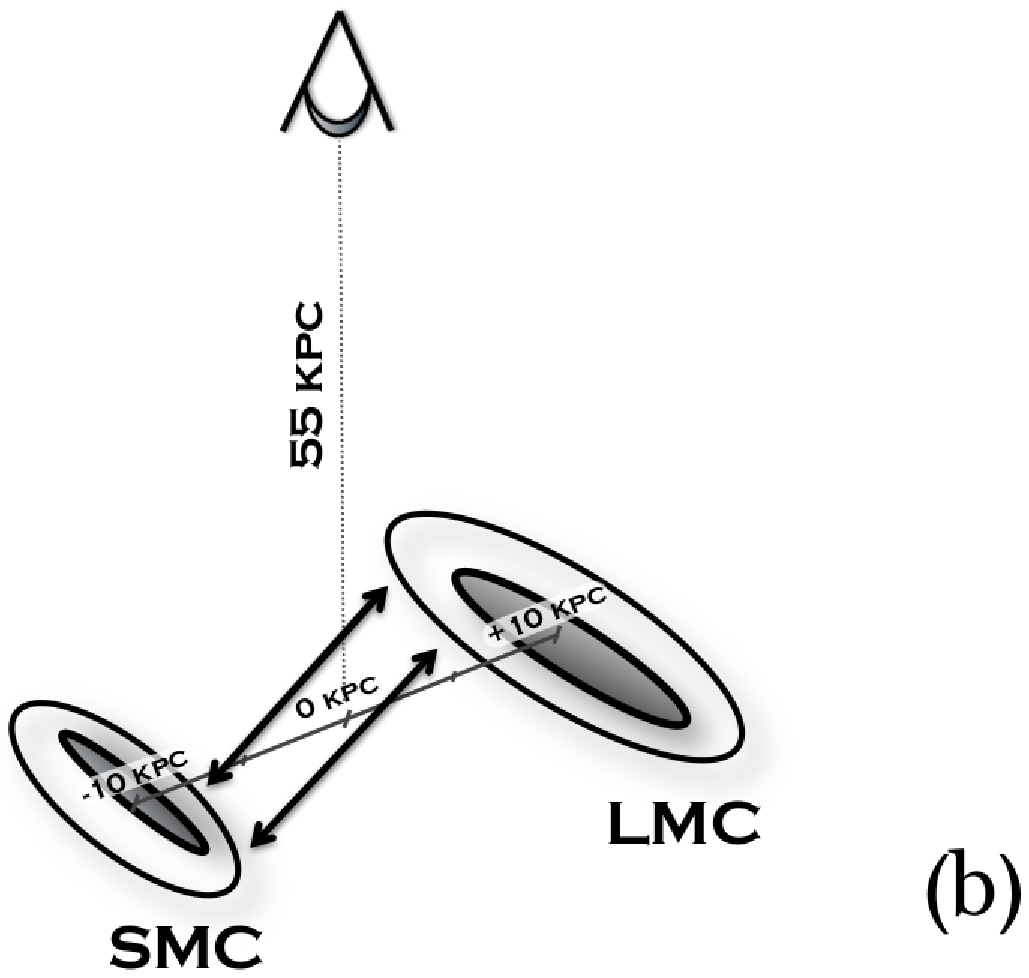}
\end{center}
\figcaption{
Schematics of the LMC and SMC orientation. Panel (a) depicts the geometry of the system, where the SMC and LMC are at a heliocentric distances of $\sim60~\kpc$ and $\sim50~\kpc$, respectively, and are separated by $\sim18.9~\kpc$. Panel (b) illustrates the tilt of the two galaxies with respect to each other---such that the opposite sides of their disks illuminate each other---and the observer. The O and B stellar populations within the disks of these galaxies produce the majority of the Lyman continuum emitted by these galaxies.  
\label{figure:orientation_cartoon}}
\end{figure}

\ha\ emission arrises from the recombination of electrons and protons in an ionized gas. When the ionization is produced through photoionization, the rate of hydrogen recombination will be proportional to the flux of the incident Lyman continuum : $\phi_{\rm{LC}} = \alpha_B\ n_e n_p L_{H^+}$. Using Equation (\ref{eq:L_ion}) for $L_{H^+}$ and the hydrogen recombination coefficient for a gas optically thick to Lyman continuum radiation, $\alpha_B=2.584\times10^{-13}\ (T/10^4~{\rm K})^{-0.806} {\cm}^3\ {\rm s}^{-1}$ \citep{1988ApJS...66..125M} (the \hi\ column density of the Bridge is greater than $10^{18}\ {\cm}^{-2}$), this relationship becomes
\begin{equation}\label{eq:FHalpha}
\phi_{\rm{LC}} = 2.1\times10^5\left(\frac{\iha}{0.1{\rm R}}\right)\left(\frac{T}{10^4 K}\right)^{0.094} {\rm photons}\ {\rm cm}^{-2}\ {\rm s}^{-1}.
\end{equation}
Assuming a constant temperature of $T=10^4~{\rm K}$, the \ha\ emission is then proportional to the strength of the incident ionizing flux. The photoionization sources we explore include the extragalactic background, the Milky Way, the Magellanic Clouds, and the OB stellar population within the Bridge. 

\begin{figure}
\begin{center}
\includegraphics[scale=0.5,angle=0]{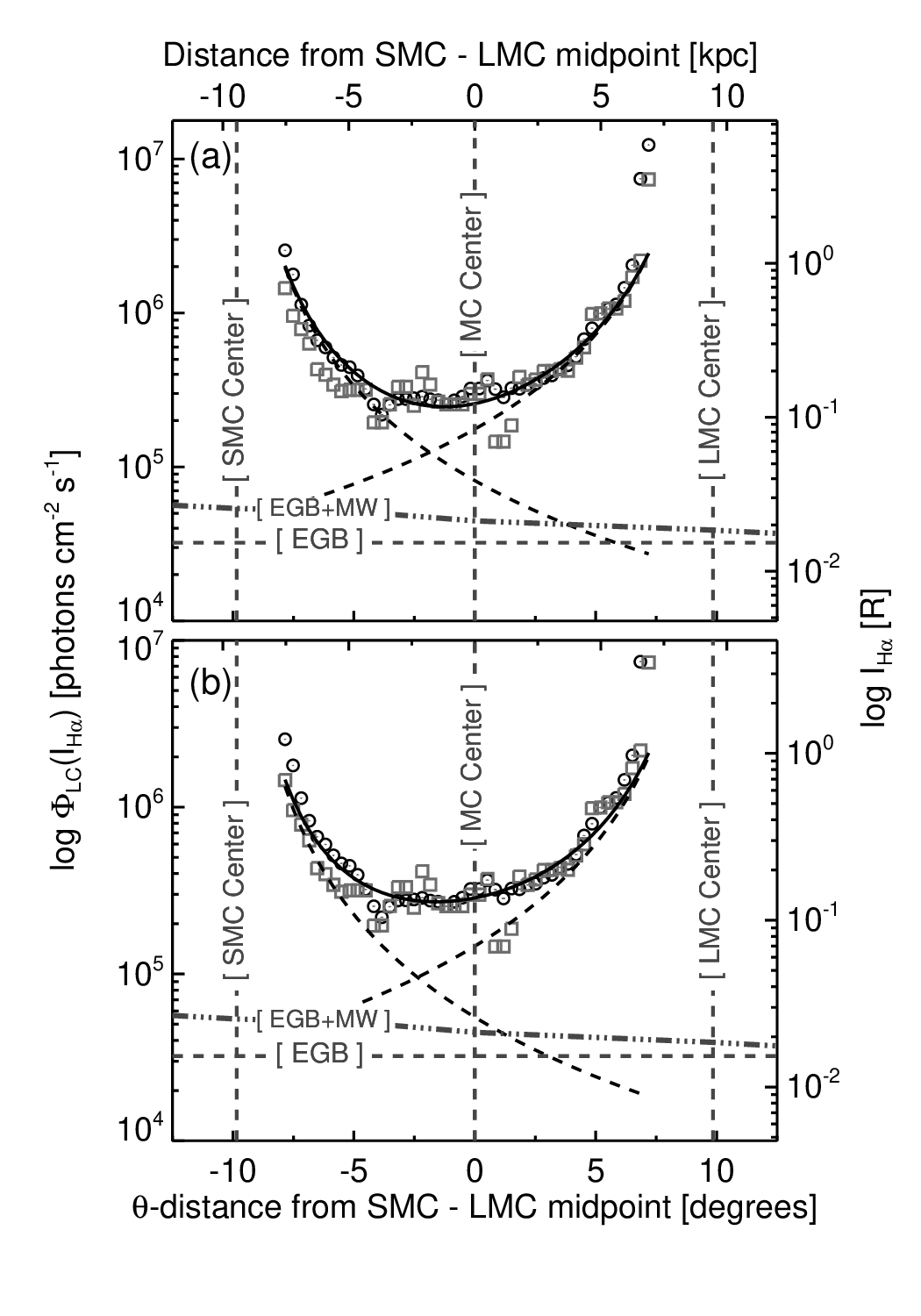}
\end{center}
\figcaption{
Predicted incident ionizing radiation between the Magellanic Clouds based on the observed \ha\ intensity observations of the Magellanic Bridge with $T=10^4~\rm{K}$ (see Equation \ref{eq:FHalpha}). The mean values are displayed with open circles and the median values with open squares. Typical deviations for weighted mean values are less than $0.5\%$ and less than $15\%$ for the median values. The vertical dashes indicate the center location of the SMC (assumed to be $l=302\arcdeg, b=-44\arcdeg,d=60~\kpc$), LMC (assumed to be $l=278\arcdeg, b=-32\arcdeg, d=50~\kpc$), and the Magellanic Clouds. The horizontal dash line labels the Lyman continuum flux from the extragalactic background \citep{2001cghr.confE..64H}. The dash-dot-dot-dot line marks the contribution of the Milky Way \citep{1999ApJ...510L..33B, 2001ApJ...550L.231B} and extragalactic ionizing flux where contribution is $\log\,\Phi_{MW+EGB}=4.73$ and $4.59~{\rm photons}\ {\rm cm}^{-2}\ {\rm s}^{-1}$ at the center of the SMC and LMC, respectively.
\label{figure:Flux_LC}}
\end{figure}

We calculated the strength of an incident ionizing flux that is capable of reproducing the \ha\ observations by determining the mean and median values of the \ha\ intensities along elliptical rings expanding outwards, defined by the position angle ($PA$) and the inclination ($i$) of the SMC and LMC disks (see Figure \ref{figure:radial_binning}).  For the LMC, we defined the shape of the expanding rings by $PA=168\fdg0$ and $i=22\fdg0$ and a center annuli of the \hi\ disk at $(l,b)=(279\fdg5, -33\fdg5)$ \citep{1998ApJ...503..674K}. Because the \hi\ gas distribution in the SMC is randomly orientated with no real disk shape \citep{2004ApJ...604..176S}, we chose circular expanding annuli characterized by $PA=0\fdg0$ and $i=0\fdg0$ with a disk center at $(l,b)=(301\fdg0, -44\fdg0)$. Figure \ref{figure:orientation_cartoon} displays the schematics orientation of the Magellanic Clouds with respect to each other and to the observer, illustrating the different path lengths the emission travels from each galaxy and that the two galaxies illuminate each other. We separated each elliptical ring by $0\fdg3$ to ensure full coverage and only include sightlines with with emission above the sensitivity of this \ha\ survey ($I_{\rm H\alpha}>0.3~{\rm R}$) and of the GASS \hi\ 21-\cm\ survey ($N_{\rm H\textsc{~i}}>1.6\times10^{18}~\cm^{-2}$ at ${\rm FWHM}_{{\rm H}~{\textsc i}}=30~\kms$) to exclude regions beyond the Bridge that might contain only trace amounts of hydrogen. We merged the calculated intensity values along each of the two expanding ellipses at the center of the SMC and LMC with their weighted averages. For warm gas at $10^4~\rm{K}$, Figure \ref{figure:Flux_LC} shows the ionizing flux needed to produce the observed \ha\ emission in the Magellanic Bridge as a function of distance and angular displacement from the center of the Bridge. 

\subsection{Photoionization from the Milky Way and Extragalactic Background}

\citet{1999ApJ...510L..33B, 2001ApJ...550L.231B} modeled the ionizing flux radiating from the Milky Way by assuming that the 90-912 \AA\ radiation is dominated by O-B stars confined to spiral arms and later updated that model in \citet{2005ApJ...630..332F}. This updated model predicts an incident ionizing flux of $\log{(\Phi_{MW}}/[{\rm photons}\ {\rm cm}^{-2}\ {\rm s}^{-1}])=4.33$ at the center of the SMC $(l=301\arcdeg, b=-44\arcdeg,d=60~\kpc$) and $\log(\Phi_{MW}/[{\rm photons}\ {\rm cm}^{-2}\ {\rm s}^{-1}])=3.80$ at the center of the LMC ($l=279.5\arcdeg, b=-33.5\arcdeg, d=50~\kpc$). \citet{2001cghr.confE..64H} predict that the ionizing flux from the extragalactic background radiation is $\log(\Phi_{EGB}/[{\rm photons}\ {\rm cm}^{-2}\ {\rm s}^{-1}])=4.51$ by assuming that the radiation is dominated by quasi-stellar objects, active galaxy nuclei, and active star forming galaxies. 

If photoionization is the only source of the ionization, then the incident ionizing radiation required to produce the typical \ha\ intensities in the central regions of the Magellanic Bridge is roughly $\log(\Phi_{LC}/[{\rm photons}\ {\rm cm}^{-2}\ {\rm s}^{-1}])=5.46,$ assuming the gas is at $T=10^4~{\rm K}$ (see Figure \ref{figure:Flux_LC}). The photoionization from the Milky Way and the extragalactic background, roughly $\log(\Phi_{LC}/[{\rm photons}\ {\rm cm}^{-2}\ {\rm s}^{-1}])=4.63$ at the center of the Magellanic Bridge, are insufficient to produce the \ha\ emission alone (see Figure \ref{figure:Flux_LC}). Additional ionization must therefore come from alternative sources, including the Magellanic Clouds. If the Lyman continuum flux from the Magellanic Clouds dominates the ionization, then the \ha\ intensity traces the fraction of ionizing photons escaping from these galaxies. 

\subsection{Photoionization from the Magellanic Clouds}\label{section:escape}

The Magellanic System provides us with an excellent opportunity to establish $f_{\rm esc}$ for two local dwarf galaxies. If Lyman continuum radiation from the Magellanic Clouds is responsible for producing most of the ionization within the Magellanic Bridge, then the \ha\ emission is a direct measurement of the propagation of the ionizing radiation through this structure, see Equation (\ref{eq:FHalpha}). The LMC and SMC are at distances of roughly $50~\kpc$ and $60~\kpc$ respectively. Their measured star-formation rates are $0.4$ and $0.2~{\rm M}_{\odot}\ {\rm yr}^{-1}$ \citep{2009AJ....138.1243H} and both galaxies have well defined orientations with respect to each other and as observed from Earth (see Figure \ref{figure:orientation_cartoon}). We follow the model procedures of \citet{1999ApJ...510L..33B, 2002ASPC..254..267B} in deriving the ionization levels in the outer disks of the LMC and SMC from their own interval UV sources and from the contribution to the mean field intensity by the other.

In the Galaxy, the total ionizing photon intensity is $2.6\times 10^{53}~{\rm photons} \s^{-1}$ from a disk-averaged star-formation rate of $2\pm1~{\rm M}_{\odot}\ {\rm yr}^{-1}$ \citep{1997ApJ...476..166W}. A similar ratio of star-formation rate to ionizing photon luminosity $(\xi)$ has been measured for M82 \citep{1993ApJ...412..111M}. While $\xi$ may depend on the mean metallicity of the stellar population (e.g., \citealt{2002A&A...382...28S}), we stress that our results are mostly independent of metallicity because our star-formation rates are derived from the Balmer-line intensities.

In Figure \ref{figure:Flux_LC}, we present two models for the \ha\ intensities derived from the outer disks of the LMC and SMC. The data for both galaxies have been binned along annuli centered at each galaxy as described by Section \ref{section:ionization}. We show both the mean and median \ha\ values in each annulus to demonstrate the impact of clumpiness. In our models, we have corrected for the 20\% larger distance for the SMC, which largely accounts for the faster drop-off in its emission compared to the LMC, leading to the well pronounced skewing of the U-shaped profile.

In the first model, we assume that the UV intensity arises from an inner disk of ionizing sources without the need for radiation transfer. Our conversion from ionizing flux
to \ha\ intensities uses Equation \ref{eq:FHalpha}, assuming $T=10^4~{\rm K}$. The first model produces an $f_{\rm esc}$ of approximately $3\%$ for the LMC and $4\%$ for the SMC. A more complicated variation of this model model, one that treats the outer HI disk as a warped structure, works less well (e.g., \citealt{1998ASPC..136..113B}). The fact that the first model works so well may reflect the distributed nature of the young stellar populations and the highly clumpy (fractal) nature of  the ISM in both galaxies (e.g., \citealt{1999MNRAS.302..417S}). 

The second model is a repeat of the first model, but includes the ionizing effect of the other galaxy, i.e., the contribution of the LMC UV intensity on the outer disk of the SMC, and vice versa (see Figure \ref{figure:orientation_cartoon}(b)). We assume the locations in 3-space of both galaxies place them at both ends of an invisible cylinder $10~\kpc$ in length. This reduces $f_{\rm esc,~LMC}$ to $2.5\%$ and $f_{\rm esc,~SMC}$ to $2.8\%$. Since this value is close to the estimate derived for the first model, we conclude that $f_{\rm esc,~LMC} \approx 3.0\pm 1.0\%$ and $f_{\rm esc,~SMC} \approx 4.0\pm 1.5\%$ with a remarkable level of consistency between both methods, which does account for the uncertainty in $\xi$. 

These models assume that photoionization is the dominant source of ionization of the Bridge. If other ionization process also affect the Magellanic Bridge, then $f_{\rm esc}$ would decrease. The dominant source of uncertainty in these $f_{\rm esc}$ estimates is in the uncertainty in the relative positions and orientations of these galaxies. This translates into a less certain galaxy-galaxy ionization effect. We therefore deduce that $f_{\rm esc, ~LMC}<4.0\%$ and that $f_{\rm esc, ~SMC}<5.5\%$. 

\subsection{Photoionization from the O and B stellar population within the Magellanic Bridge}

\begin{figure}
\begin{center}
\includegraphics[scale=0.35,angle=90]{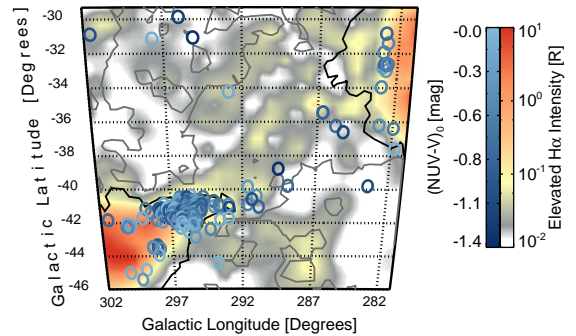}
\end{center}
\figcaption{
Regions with elevated \ha\ emission---compared to the neighboring sightlines---and location of the O and B stars in the Magellanic Bridge, identified through photometry and proper motions \citep{2012ApJ...753..123C}. The shade of the blue circles denote the $(NUV-V)_0$ color of the stars (D.~I. Casetti-Dinescu, private communication). We created the elevated \ha\ emission map using the unsharp mask technique, where the smoothed image is divided by the original image. The contour lines in trace the non-extinction corrected \ha\ intensity at $0.03$ and $0.16~{\rm R}$. 
\label{figure:OB_stars}}
\end{figure}

There are many pockets of elevated \ha\ emission within the Bridge that could be associated with star-formation sites. Multiple studies have identified early-type stars in the direction of the Magellanic Bridge \citep{1971A&A....10....1W, 1990AJ....100..663G, 1990AJ.....99..191I, 1991A&AS...91..171D, 1991AJ....101..911D, 1992AJ....103.1234G, 1998AJ....115.1472B, 1998AJ....115..154D, 1999AJ....118.1700D, 2007ApJ...658..345H, 2012ApJ...753..123C}; however, many of the studies do not distinguish A stars from the O and B stars---which produce the majority of the ionizing photons---or identify the foreground stars in their sample. The \citet{2012ApJ...753..123C} study uses proper motions from the Southern Proper Motion Program 4 (SPM4) to select the stellar population of the Bridge, excluding foreground stars, and uses photometry from the Galaxy Evolution Explore survey (GALEX), the Two Micron All Sky Survey (2MASS), SPM4, and the American Association of Variable Star Observers All Sky Photometric Survey (APASS) to identify O and B stellar candidates. Followup spectroscopy of these candidates is still needed to confirm and place further constraints on their stellar types. The coverage of their survey spans the entire region of this \ha\ study.  

We compare the regions of elevated \ha\ emission with location of the \citet{2012ApJ...753..123C} O and B candidates to test if they are associated with star-formation sites, as illustrated with Figure \ref{figure:OB_stars}. We quantify the elevated \ha\ emission by creating comparing the local emission with the surrounding emission through the unsharp-masked technique---creating smoothed \ha\ map to half the angular resolution and dividing this smoothed map by the original map shown in Figure \ref{figure:Ha_Bridge}(b). Figure \ref{figure:OB_stars} also includes the $(NUV-V)_0$ GALEX color for each of the candidates to indicate which stars emit more ionizing radiation and are therefore earlier-type stars. We find that although some O and B candidates do align with regions of elevated \ha\ emission, many locations---especially those outside $10^{20}~\cm^{-2}$ \hi\ contour---shows no such correlation. The locations lacking in correlation could be explained if the gas is extremely ionized or low density, although these conditions are less conducive of star formation. If the regions of the Bridge at more positive latitudes have a low density, then the cluster of three early-type stars near $(l,b)=(296\arcdeg, -31\arcdeg)$ might be responsible for the latitude offset of the \ha\ emission from the \hi\ emission (see Section \ref{section:intensity_map}), but this offset could also be caused by halo-gas interactions or by shadowing effects associated with the orientation of the Bridge with respect to the Magellanic Clouds. 

\citet{1994ApJ...430..222D} found that OB associations near the solar circle produce roughly $(3-4)\times10^{49}~{\rm ionizing\ photons} \s^{-1}.$ If the OB associations within the Magellanic Bridge produce $\sim10^{49}~{\rm ionizing\ photons} \s^{-1}$ and if the gas is at $10^4~{\rm K}$, then the OB associations within $1~\kpc$ would elevate the \ha\ emission by $0.16~{\rm R}$, using Equation \ref{eq:FHalpha} to convert the ionizing flux to \ha\ intensity. However, some of the ionizing radiation will escape from the Bridge before ionizing the surrounding gas. For example, \citet{1999ApJ...510L..33B, 2001ApJ...550L.231B} estimate that $6\%$ of the ionizing photons of the Milky Way escapes into the halo and Section \ref{section:escape} estimates that $<4\%$ of the Lyman continuum escapes from the LMC into the surroundings and that $<5.5\%$ escapes from the SMC. The low \hi\ column density and dust content of the Bridge would likely make this region more susceptible to losing ionizing radiation. Turbulence can also increase the ionization radiation that escapes from an OB population in the Bridge (e.g., \citealt{2010ApJ...721.1397W}). The sparsity of the O and B stellar populations in the central regions of the Bridge might result in only a marginal increase in \ha\ emission. However, if these OB associations substantially increase the temperature of the gas, then the efficiency of producing \ha\ emission will decrease as $\iha\propto (T/10^4~{\rm K})^{-0.924}$, see Equation (\ref{eq:L_ion}). In the SMC-Tail, the combination of high \hi\ column density and high density of O and B stars will lead to enhanced \ha\ emission. 

\section{Implications of the Ionized Gas}\label{section:implications}

The fraction of UV photons $f_{\rm esc}$  that escape from galaxies has long been recognized as an important physical quantity to measure. At high redshift, the cosmic fog of neutral hydrogen was ionized by either stars or black holes depending on the relative escape fractions of ionizing radiation with the evidence in favor of stars at present \citep{2011ARA&A..49..373B}. At low redshift, the cosmic ionizing intensity conceivably explains the well defined truncation observed in spiral galaxies \citep{1977AZh....54..957B, 1993ApJ...414...41M}. 

In the local universe, there are few reliable measurements of $f_{\rm esc}$ because favorable geometries are called for to make this possible. \citet{1999ApJ...510L..33B} 
initially used the Magellanic Stream as a probe of the Galactic UV field until it was later discovered that in certain locations along the Stream, the H$\alpha$ is too bright. The high-velocity cloud population clearly establish the vertical escape fraction as $\approx 6\%$ or $\approx 1-2\%$ isotropized over a sphere \citep{2002ASPC..254..267B,2003ApJ...597..948P, 2002ApJ...572L.153T}. \citet{1999ApJ...523..575L} determined $f_{\rm esc}$ of order a few percent from a minor axis cloud in the outer halo of M82 (the so-called `cap'), but this was derived from averaging the \ha\ emission from the cloud over its extent. Similar to the early Magellanic Stream estimates, individual clumps within the `cap' are now found to be much too bright to be explained by the central starburst. Interestingly, for both the M82 cap and the Magellanic Stream, the bright clumps can be explained in terms of slow shocks being driven into dense clumps by a fast `wind' of hot gas \citep{2007ApJ...670L.109B, 2012ApJ...761...55M}.

There exists a small sample of dwarf starburst galaxies with estimates for $f_{esc}$: NGC~5253 and Haro~11. NGC~5253 has an advantageous orientation with respect to the Milky Way, allowing for estimates of the $f_{esc}$ along the minor axis of the galaxy. \citet{2011ApJ...741L..17Z} associate this minor axis with an ionization cone where the majority of ionization radiation escapes. If the majority of the escaping ionization radiation funnels through this cone, then $f_{esc}\approx3\%$ (S. Veilleux, private communication). The $f_{esc}$ of blue compact galaxy Haro~11---an extreme starburst dwarf---has been studied by three separate groups, each with considerably varied results. Combining International Ultraviolet Explorer (IUE) spacecraft and FUSE observations \citet{2006A&A...448..513B} predict a $4-10\%$ $f_{esc}$ and conclude that the Lyman continuum radiation escapes through transparent windows of the ISM. Using all three channels on the ACS on HST---covering far-UV, 2200- and U band, and optical wavelengths---the New Technology Telescope at ESO La Silla to collect \ha, \hb, and \oiii\ narrow-band images, and archival X-ray Chandra and XMM-Newton telescopes observations, \citet{2007MNRAS.382.1465H} found a $3\%$ $f_{esc}$. They conclude that $\sim90\%$ of the ionizing photons undergo multiple resonance scattering events, masking their origin. \citet{2007ApJ...668..891G} found no direct connection between outflows and escaping ionizing radiation, placing an upper limit of $f_{esc}\le2\%$ using FUSE and Chandra observations. The contribution of ionizing photons that dwarf galaxies expel into their surroundings has yet to reach a resolution.

The circumgalactic gas structures surrounding the Magellanic Clouds provide a unique avenue for studying the ionizing radiation expelled from the Magellanic Clouds. The ionizing radiation from the Milky Way and the extragalactic background is insufficient for producing the observed \ha\ emission in the Bridge. If the radiation from the Magellanic Clouds dominates the ionization in the Bridge, then the \ha\ emission traces the ionizing photons emitted by these galaxies. This study provides an upper limit on the fraction of escaping ionizing photons from the Magellanic clouds as other process may also contribute to the ionization in the Bridge. These upper limits constrain the contribution of ionizing photons in the extragalactic ionizing background that similar dwarf galaxies generate. In this way, the estimate of the $f_{esc}$ emitted by the LMC and SMC can anchor cosmological simulations working to identify the sources of the Lyman continuum of the extragalactic background from the epoch of reionization to today.

Besides ionizing photons, the disturbed galaxies also expel baryons into their surroundings. The Magellanic systems are surrounded by $6.8\times10^8~{\rm M}_{\odot}$ of neutral gas \citep{2005A&A...432...45B}. The \ha\ observations of the Magellanic Bridge reveal a warm ionized gas phase along the entire structure (see Figure \ref{figure:Ha_Bridge}(b)). In Section \ref{section:mass}, we found that the Magellanic Bridge and the SMC-Tail contain roughly $\left(0.7-1.7\right)\times10^8~{\rm M}_\odot$ in ionized gas and $3.3\times10^8~{\rm M}_\odot$ in neutral gas (see Table \ref{table:mass}). These masses suggest that the Magellanic Bridge and the SMC-Tail have a minimum ionization fraction of $36-52\%$ and $5-24\%$, respectively; however, these mass estimates are insensitive to the faintest and lowest density regions ($N_{\rm H\textsc{~i}}<1.6\times10^{18}~\cm^{-2}$ and $I_{\rm H\alpha}<0.3~{\rm R}$) as well as the extremely ionized gas. FUSE absorption-line observations towards two early-type stars and a background quasar identified components of highly ionized gas, with ionization fractions of $70-90\%$, within the Bridge \citep{2008ApJ...678..219L}. The inclusion of these highly ionized components make this structure more massive and would raise the ionization fraction of the total structure. 

The displacement of $\left(4.0-5.0\right)\times10^8~{\rm M}_{\sun}$ of neutral and ionized gas mass from the Magellanic Clouds, into the Magellanic Bridge, will have repercussions on the future evolution of these galaxies that would, in all likelihood, stunt future stellar production if the material escapes the system. Including the additional neutral mass of the other circumgalactic gas structures of the Magellanic System, as determined by \citet{2005A&A...432...45B}, the Magellanic Clouds have lost at least $\left(7.5-8.5\right)\times10^8~{\rm M_{\sun}}$ in gas. If these other circumgalactic structures were only 25\% ionized, then the total gas currently surrounding the Magellanic Clouds would exceed $10^9~{\rm M_{\sun}}$. \ha\ emission-line observations towards the Magellanic Stream reveals that this structure does have a warm ionized gas phase (\citealt{1996AJ....111.1156W, 2003ApJ...597..948P, 2012ApJ...749L...2Y}; G.J. Madsen, private communication) and \ion{O}{6} absorption-line observations reveal a highly ionized component \citep{2003ApJS..146..165S, 2010ApJ...718.1046F}, which increases ionization fraction of the structure and the total gas mass these galaxies have lost. 

The galaxy interactions could also induce ionization of the Magellanic Bridge. The dissimilar distribution of $N_{\rm H{~\textsc{i}}}$ and $I_{\rm H\alpha}$ distribution in the SMC-Tail and in the Magellanic Bridge (see Figure \ref{figure:NHI_Ha}(c)) suggests either that different sources ionize these structures, that they have extremely different structures, or a combination of both. In the SMC-Tail, 163 \hi\ shells have been observed with 60\% of them lacking an OB associations \citep{2003MNRAS.339..105M}. The presence of the \hi\ shells and active star formation could indicate that this region suffers from a combination of gravitational and pressure instabilities caused by high-velocity cloud impacts or ram pressure effects \citep{2003MNRAS.339..105M}, which would heat the gas and could trigger star formation. This may have created an environment ripe for the propagation of ionizing photons through the creation of large bubbles and chimneys (e.g., \citealt{1989ApJ...345..372N,1994ApJ...430..222D, 2000ApJ...531..846D}). 

The presence of young, $10-40~{\rm Myr}$ old stars, in the central region of the Magellanic Bridge indicates that stars are actively forming \textit{in situ} as they have not had enough time to migrate from the SMC \citep{1998AJ....115..154D}. The scarcity of metals (\citealt{2008ApJ...678..219L}: $Z \approx 0.1\ Z_{\sun}$) and molecular gas  \citep{2000A&A...363..451S, 2002ApJ...578..126L, 2006ApJ...643L.107M} within the Bridge means that the star formation may forgo traditional cooling mechanisms and may instead depend on triggering events. Young stars could spawn out of colliding cloudlets \citep{1983MNRAS.203.1233D}. Numerous simulations of the galaxy interactions provide clues on what type of events may have encourage their formation. The \citet{2012MNRAS.421.2109B} models suggest that a violent collision could have occurred between the LMC and SMC---where the SMC traveled through the disk of the LMC, warping the disk of the LMC and shock heating the gas---and promoted star formation within the Magellanic Bridge. Once stars are formed within the Magellanic Bridge, their ionizing photons, stellar winds, and supernovae would further ionized the surrounding gas. Ram pressure compression could also trigger star formation \citep{2009MNRAS.399.2004M}, but this mechanism would not account for the lack of star formation within the Leading Arm and Magellanic Stream \citep{2012MNRAS.421.2109B}. 

To further distinguish between the sources of the ionization and to further constrain the ionization fraction of the Magellanic Bridge, more observations are needed. Although the \ha\  observations have identified a warm ionized gas phase, they are insensitive to highly ionized and extremely low density gas. The \citet{2002ApJ...578..126L} and  \citet{2008ApJ...678..219L} absorption-line studies reveal that the structure, at minimal, contains pockets of highly ionized gas. At more positive latitudes, the \hi\ column density  of the Magellanic Bridge drops off before the \ha\ intensity decreases, suggesting that this region has as a higher ionization fraction due to internal processes such as star formation, exposure to the ionizing photons from the Magellanic Clouds, or halo-gas interactions as structure travels through the hot halo of the Milky Way; however, this difference could also be due to the decreased sensitivity of this survey at highest velocities due to residuals left behind by a bright OH atmospheric line. To determine if a hot ionized gas phase is rampant throughout this structure, more absorption-line observations are needed towards background objects and not towards early-type stars where the presence highly ionized gas is anticipated. Addition multiline observations will aid in distinguishing between different sources of ionization as each ionization process will produce emission and absorption lines of different relative strengths.

\section{Summary}\label{section:summary}
 
Using WHAM to observe the warm gas phase in the Magellanic Bridge and the SMC-Tail, we mapped the \ha\ emission over $350~{\rm degrees}^2$. These kinematically resolved observations---over the velocity range of $+100$ to $+300$~\kms\ in the LSR frame---include the first full \ha\ intensity map of the Magellanic Bridge. We compare these observations of the warm ionized gas phase with the $21~\cm$ emission in the GASS \hi\ survey. Through these observations, we examined the extent, morphology, velocity gradients, mass of the ionized gas, and the source of the ionization of these structures. This study finishes with the main conclusions from the \ha\ observations of the warm gas component in the Magellanic Bridge and the SMC-Tail: 

\begin{enumerate}

\item{\bf Ionized Gas Mass.} Three quantities dominate the uncertainty in quantifying the ionized-gas mass: the distance, the morphology along the line-of-sight, and the distribution of ionized and neutral gas. Assuming a distance of $55~\kpc$, an \hi\ line-of-sight depth that is similar to the width of the Bridge, a single component structure along each sightline, and three different distributions for the neutral and ionized gas, the mass of the ionized gas in the SMC-Tail and the Magellanic Bridge is between $\left(0.7-1.7\right)\times10^8~{\rm M}_\odot$ compared to $3.3\times10^8~{\rm M}_\odot$ for the neutral mass (see Section \ref{section:mass} and Table \ref{table:mass}). The Magellanic Bridge is significantly more ionized than the SMC-Tail, with the SMC-tail at only $5-24\%$ ionized and the Magellanic Bridge at $36-52\%$ ionized. This survey is insensitive to the faintest ($I_{\rm H\alpha}<30~{\rm mR}$) and lowest density regions. This survey is also less sensitive to \ha\ emission over $+240\ge\vlsr\ge+275~\kms$ velocity range because of increased residuals associated with a bright atmospheric line.

\item{\bf Fraction of Escaping Ionizing Photons.} The Lyman continuum from the extragalactic background and the Milky Way produces a negligible amount of the ionization in the Magellanic Bridge. If the ionizing radiation from the Magellanic Clouds produces the majority of the ionization, then the \ha\ emission traces the $f_{esc}$ of the LMC and SMC. Assuming that the inner disk of the SMC and LMC supply the incident ionizing flux that ionizes the Magellanic Bridge and that the Magellanic Clouds ionize each other, than the LMC and SMC radiate $<4.0\%$ and $<5.5\%$ of their ionizing photons into their surroundings, respectively.

\item{\bf $\boldsymbol N_{\rm H\textsc{~i}}$ and $\boldsymbol I_{\rm H\alpha}$ Distribution.} There is a strong correlation between the $\log{N_{\rm H\textsc{~i}}}$ and the $\log{I_{\rm H\alpha}}$ near the Magellanic Clouds and within the SMC-Tail that is not observed in HVCs (see Figure \ref{figure:NHI_Ha}(c)). This trend indicates that the neutral and ionized gas phases in these regions are related and affected by similar processes and could be influenced by the close proximity to the Magellanic Clouds. The contrast between the $\log N_{\rm H\textsc{~i}}$ and the $\log I_{\rm H\alpha}$ in the central regions on the Magellanic Bridge, shown in Figures \ref{figure:spectra} and \ref{figure:lehner}, suggests that their changes are more related more to the ionization fraction or to the fraction of ionized regions along the line-of-sight. 

\item{\bf Velocity Distribution.}  The global distribution of the \ha\ first moment map agrees with the corresponding \hi\ map and has a relatively smooth velocity gradient of $+175$ to $+225~\kms$ in \ha\ compared to $+125$ to $+250~\kms$ in \hi\ as illustrated in Figure \ref{figure:First_Moment}. This correspondence suggests that both of the neutral and ionized gas phases travel parallel to the LMC and SMC.  Although the global trends of these velocity distributions agree, the intensity-weighted average \ha\ velocity is shifted to lower velocities and is much more smooth than the \hi\ velocity distribution. A combination of broader \ha\ components and an insensitivity at highest velocities may cause these discrepancies. However, even though the velocity gradient is smooth, the spectral velocity distribution of the gas has a complex multiple component structure (see Figures \ref{figure:spectra} \& \ref{figure:lehner}). The strength of the \ha\ intensity compared to the \hi\ column density suggests that many of these structures are predominantly ionized. In the central regions of the Magellanic Bridge, there is often additional \ha\ components at velocities without complementary \hi\ emission when smoothed to the same angular resolution, which indicates the presence of distinct highly ionized structures. 

\end{enumerate}
 
\acknowledgments

The authors acknowledge helpful discussions with Nicolas Lehner, Bart Wakker, and Gurtina Besla. We thank Dana Casetti-Dinescu for providing the locations and properties of the OB stellar population in the Magellanic Bridge and Jay Gallagher for suggesting this project. The authors acknowledge the feedback provided by the referee; comments that lead to an improved and more thorough article. The National Science Foundation supported WHAM  through grants AST~0204973, AST~0607512, and AST~1108911. Barger is further supported through NSF Astronomy and Astrophysical Postdoctoral Fellowship award AST~1203059. 

{\it Facility:} \facility{WHAM}

\bibliographystyle{apj} 
\bibliography{References} 

\begin{thebibliography}{}
\expandafter\ifx\csname natexlab\endcsname\relax\def\natexlab#1{#1}\fi

\bibitem[{{Barger} {et~al.}(2012){Barger}, {Haffner}, {Wakker}, {Hill},
  {Madsen}, \& {Duncan}}]{2012ApJ...761..145B}
{Barger}, K.~A., {Haffner}, L.~M., {Wakker}, B.~P., {et~al.} 2012, \apj, 761,
  145

\bibitem[{{Barkana} \& {Loeb}(1999)}]{1999ApJ...523...54B}
{Barkana}, R., \& {Loeb}, A. 1999, \apj, 523, 54

\bibitem[{{Barnes} \& {Hernquist}(1992)}]{1992ARA&A..30..705B}
{Barnes}, J.~E., \& {Hernquist}, L. 1992, \araa, 30, 705

\bibitem[{{Battinelli} \& {Demers}(1998)}]{1998AJ....115.1472B}
{Battinelli}, P., \& {Demers}, S. 1998, \aj, 115, 1472

\bibitem[{{Bergvall} {et~al.}(2006){Bergvall}, {Zackrisson}, {Andersson},
  {Arnberg}, {Masegosa}, \& {{\"O}stlin}}]{2006A&A...448..513B}
{Bergvall}, N., {Zackrisson}, E., {Andersson}, B.-G., {et~al.} 2006, \aap, 448,
  513

\bibitem[{{Besla} {et~al.}(2012){Besla}, {Kallivayalil}, {Hernquist}, {van der
  Marel}, {Cox}, \& {Kere{\v s}}}]{2012MNRAS.421.2109B}
{Besla}, G., {Kallivayalil}, N., {Hernquist}, L., {et~al.} 2012, \mnras, 421,
  2109

\bibitem[{{Bland-Hawthorn}(1998)}]{1998ASPC..136..113B}
{Bland-Hawthorn}, J. 1998, in Astronomical Society of the Pacific Conference
  Series, Vol. 136, Galactic Halos, ed. {D.~Zaritsky}, 113

\bibitem[{{Bland-Hawthorn} \& {Maloney}(1999)}]{1999ApJ...510L..33B}
{Bland-Hawthorn}, J., \& {Maloney}, P.~R. 1999, \apjl, 510, L33

\bibitem[{{Bland-Hawthorn} \& {Maloney}(2001)}]{2001ApJ...550L.231B}
---. 2001, \apjl, 550, L231

\bibitem[{{Bland-Hawthorn} \& {Maloney}(2002)}]{2002ASPC..254..267B}
{Bland-Hawthorn}, J., \& {Maloney}, P.~R. 2002, in Astronomical Society of the
  Pacific Conference Series, Vol. 254, Extragalactic Gas at Low Redshift, ed.
  {J.~S.~Mulchaey \& J.~T.~Stocke}, 267

\bibitem[{{Bland-Hawthorn} {et~al.}(2007){Bland-Hawthorn}, {Sutherland},
  {Agertz}, \& {Moore}}]{2007ApJ...670L.109B}
{Bland-Hawthorn}, J., {Sutherland}, R., {Agertz}, O., \& {Moore}, B. 2007,
  \apjl, 670, L109

\bibitem[{{Bochkarev} \& {Sunyaev}(1977)}]{1977AZh....54..957B}
{Bochkarev}, N.~G., \& {Sunyaev}, R.~A. 1977, \azh, 54, 957

\bibitem[{{Bohlin} {et~al.}(1978){Bohlin}, {Savage}, \&
  {Drake}}]{1978ApJ...224..132B}
{Bohlin}, R.~C., {Savage}, B.~D., \& {Drake}, J.~F. 1978, \apj, 224, 132

\bibitem[{{Bolton} {et~al.}(2005){Bolton}, {Haehnelt}, {Viel}, \&
  {Springel}}]{2005MNRAS.357.1178B}
{Bolton}, J.~S., {Haehnelt}, M.~G., {Viel}, M., \& {Springel}, V. 2005, \mnras,
  357, 1178

\bibitem[{{Bromm} \& {Yoshida}(2011)}]{2011ARA&A..49..373B}
{Bromm}, V., \& {Yoshida}, N. 2011, \araa, 49, 373

\bibitem[{{Br{\"u}ns} {et~al.}(2005){Br{\"u}ns}, {Kerp}, {Staveley-Smith},
  {Mebold}, {Putman}, {Haynes}, {Kalberla}, {Muller}, \&
  {Filipovic}}]{2005A&A...432...45B}
{Br{\"u}ns}, C., {Kerp}, J., {Staveley-Smith}, L., {et~al.} 2005, \aap, 432, 45

\bibitem[{{Cardelli} {et~al.}(1989){Cardelli}, {Clayton}, \&
  {Mathis}}]{1989ApJ...345..245C}
{Cardelli}, J.~A., {Clayton}, G.~C., \& {Mathis}, J.~S. 1989, \apj, 345, 245

\bibitem[{{Casetti-Dinescu} {et~al.}(2012){Casetti-Dinescu}, {Vieira},
  {Girard}, \& {van Altena}}]{2012ApJ...753..123C}
{Casetti-Dinescu}, D.~I., {Vieira}, K., {Girard}, T.~M., \& {van Altena}, W.~F.
  2012, \apj, 753, 123

\bibitem[{{Chung} {et~al.}(2007){Chung}, {van Gorkom}, {Kenney}, \&
  {Vollmer}}]{2007ApJ...659L.115C}
{Chung}, A., {van Gorkom}, J.~H., {Kenney}, J.~D.~P., \& {Vollmer}, B. 2007,
  \apjl, 659, L115

\bibitem[{{Demers} \& {Battinelli}(1998)}]{1998AJ....115..154D}
{Demers}, S., \& {Battinelli}, P. 1998, \aj, 115, 154

\bibitem[{{Demers} \& {Battinelli}(1999)}]{1999AJ....118.1700D}
---. 1999, \aj, 118, 1700

\bibitem[{{Demers} {et~al.}(1991){Demers}, {Grondin}, {Irwin}, \&
  {Kunkel}}]{1991AJ....101..911D}
{Demers}, S., {Grondin}, L., {Irwin}, M.~J., \& {Kunkel}, W.~E. 1991, \aj, 101,
  911

\bibitem[{{Demers} \& {Irwin}(1991)}]{1991A&AS...91..171D}
{Demers}, S., \& {Irwin}, M.~J. 1991, \aaps, 91, 171

\bibitem[{{Dijkstra} {et~al.}(2004){Dijkstra}, {Haiman}, {Rees}, \&
  {Weinberg}}]{2004ApJ...601..666D}
{Dijkstra}, M., {Haiman}, Z., {Rees}, M.~J., \& {Weinberg}, D.~H. 2004, \apj,
  601, 666

\bibitem[{{Diplas} \& {Savage}(1994)}]{1994ApJ...427..274D}
{Diplas}, A., \& {Savage}, B.~D. 1994, \apj, 427, 274

\bibitem[{{Dove} \& {Shull}(1994)}]{1994ApJ...430..222D}
{Dove}, J.~B., \& {Shull}, J.~M. 1994, \apj, 430, 222

\bibitem[{{Dove} {et~al.}(2000){Dove}, {Shull}, \&
  {Ferrara}}]{2000ApJ...531..846D}
{Dove}, J.~B., {Shull}, J.~M., \& {Ferrara}, A. 2000, \apj, 531, 846

\bibitem[{{Dyson} \& {Hartquist}(1983)}]{1983MNRAS.203.1233D}
{Dyson}, J.~E., \& {Hartquist}, T.~W. 1983, \mnras, 203, 1233

\bibitem[{{Efstathiou}(1992)}]{1992MNRAS.256P..43E}
{Efstathiou}, G. 1992, \mnras, 256, 43P

\bibitem[{{Fernandez} \& {Shull}(2011)}]{2011ApJ...731...20F}
{Fernandez}, E.~R., \& {Shull}, J.~M. 2011, \apj, 731, 20

\bibitem[{{Fox} {et~al.}(2005){Fox}, {Wakker}, {Savage}, {Tripp}, {Sembach}, \&
  {Bland-Hawthorn}}]{2005ApJ...630..332F}
{Fox}, A.~J., {Wakker}, B.~P., {Savage}, B.~D., {et~al.} 2005, \apj, 630, 332

\bibitem[{{Fox} {et~al.}(2010){Fox}, {Wakker}, {Smoker}, {Richter}, {Savage},
  \& {Sembach}}]{2010ApJ...718.1046F}
{Fox}, A.~J., {Wakker}, B.~P., {Smoker}, J.~V., {et~al.} 2010, \apj, 718, 1046

\bibitem[{{Gardiner} \& {Noguchi}(1996)}]{1996MNRAS.278..191G}
{Gardiner}, L.~T., \& {Noguchi}, M. 1996, \mnras, 278, 191

\bibitem[{{Gardiner} {et~al.}(1994){Gardiner}, {Sawa}, \&
  {Fujimoto}}]{1994MNRAS.266..567G}
{Gardiner}, L.~T., {Sawa}, T., \& {Fujimoto}, M. 1994, \mnras, 266, 567

\bibitem[{{Gordon} {et~al.}(2003){Gordon}, {Clayton}, {Misselt}, {Landolt}, \&
  {Wolff}}]{2003ApJ...594..279G}
{Gordon}, K.~D., {Clayton}, G.~C., {Misselt}, K.~A., {Landolt}, A.~U., \&
  {Wolff}, M.~J. 2003, \apj, 594, 279

\bibitem[{{Gordon} {et~al.}(2009){Gordon}, {Bot}, {Muller}, {Misselt},
  {Bolatto}, {Bernard}, {Reach}, {Engelbracht}, {Babler}, {Bracker}, {Block},
  {Clayton}, {Hora}, {Indebetouw}, {Israel}, {Li}, {Madden}, {Meade},
  {Meixner}, {Sewilo}, {Shiao}, {Smith}, {van Loon}, \&
  {Whitney}}]{2009ApJ...690L..76G}
{Gordon}, K.~D., {Bot}, C., {Muller}, E., {et~al.} 2009, \apjl, 690, L76

\bibitem[{{Grimes} {et~al.}(2007){Grimes}, {Heckman}, {Strickland}, {Dixon},
  {Sembach}, {Overzier}, {Hoopes}, {Aloisi}, \& {Ptak}}]{2007ApJ...668..891G}
{Grimes}, J.~P., {Heckman}, T., {Strickland}, D., {et~al.} 2007, \apj, 668, 891

\bibitem[{{Grondin} {et~al.}(1992){Grondin}, {Demers}, \&
  {Kunkel}}]{1992AJ....103.1234G}
{Grondin}, L., {Demers}, S., \& {Kunkel}, W.~E. 1992, \aj, 103, 1234

\bibitem[{{Grondin} {et~al.}(1990){Grondin}, {Demers}, {Kunkel}, \&
  {Irwin}}]{1990AJ....100..663G}
{Grondin}, L., {Demers}, S., {Kunkel}, W.~E., \& {Irwin}, M.~J. 1990, \aj, 100,
  663

\bibitem[{{Haardt} \& {Madau}(2001)}]{2001cghr.confE..64H}
{Haardt}, F., \& {Madau}, P. 2001, in CGHR, ed. {D.~M.~Neumann \&
  J.~T.~V.~Tran}

\bibitem[{{Haffner}(2005)}]{2005ASPC..331...25H}
{Haffner}, L.~M. 2005, in ASP Conf. Ser., Vol. 331, Extra-Planar Gas, ed.
  {R.~Braun}, 25

\bibitem[{{Haffner} {et~al.}(2001){Haffner}, {Reynolds}, \&
  {Tufte}}]{2001ApJ...556L..33H}
{Haffner}, L.~M., {Reynolds}, R.~J., \& {Tufte}, S.~L. 2001, \apjl, 556, L33

\bibitem[{{Haffner} {et~al.}(2003){Haffner}, {Reynolds}, {Tufte}, {Madsen},
  {Jaehnig}, \& {Percival}}]{2003ApJS..149..405H}
{Haffner}, L.~M., {Reynolds}, R.~J., {Tufte}, S.~L., {et~al.} 2003, \apjs, 149,
  405

\bibitem[{{Harris}(2007)}]{2007ApJ...658..345H}
{Harris}, J. 2007, \apj, 658, 345

\bibitem[{{Harris} \& {Zaritsky}(2009)}]{2009AJ....138.1243H}
{Harris}, J., \& {Zaritsky}, D. 2009, \aj, 138, 1243

\bibitem[{{Hartmann, D.~\& Burton, W.~B.}(1997)}]{1997agnh.book.....H}
{Hartmann, D.~\& Burton, W.~B.}, ed. 1997, {Atlas of Galactic Neutral Hydrogen
  (Cambridge: Cambridge University Press)}

\bibitem[{{Hausen} {et~al.}(2002){Hausen}, {Reynolds}, {Haffner}, \&
  {Tufte}}]{2002ApJ...565.1060H}
{Hausen}, N.~R., {Reynolds}, R.~J., {Haffner}, L.~M., \& {Tufte}, S.~L. 2002,
  \apj, 565, 1060

\bibitem[{{Hayes} {et~al.}(2007){Hayes}, {{\"O}stlin}, {Atek}, {Kunth},
  {Mas-Hesse}, {Leitherer}, {Jim{\'e}nez-Bail{\'o}n}, \&
  {Adamo}}]{2007MNRAS.382.1465H}
{Hayes}, M., {{\"O}stlin}, G., {Atek}, H., {et~al.} 2007, \mnras, 382, 1465

\bibitem[{{Hill} {et~al.}(2009){Hill}, {Haffner}, \&
  {Reynolds}}]{2009ApJ...703.1832H}
{Hill}, A.~S., {Haffner}, L.~M., \& {Reynolds}, R.~J. 2009, \apj, 703, 1832

\bibitem[{{Irwin} {et~al.}(1990){Irwin}, {Demers}, \&
  {Kunkel}}]{1990AJ.....99..191I}
{Irwin}, M.~J., {Demers}, S., \& {Kunkel}, W.~E. 1990, \aj, 99, 191

\bibitem[{{Johnson} {et~al.}(1982){Johnson}, {Meaburn}, \&
  {Osman}}]{1982MNRAS.198..985J}
{Johnson}, P.~G., {Meaburn}, J., \& {Osman}, A.~M.~I. 1982, \mnras, 198, 985

\bibitem[{{Kalberla} {et~al.}(2005){Kalberla}, {Burton}, {Hartmann}, {Arnal},
  {Bajaja}, {Morras}, \& {P{\"o}ppel}}]{2005A&A...440..775K}
{Kalberla}, P.~M.~W., {Burton}, W.~B., {Hartmann}, D., {et~al.} 2005, \aap,
  440, 775

\bibitem[{{Kalberla} {et~al.}(2010){Kalberla}, {McClure-Griffiths}, {Pisano},
  {Calabretta}, {Ford}, {Lockman}, {Staveley-Smith}, {Kerp}, {Winkel},
  {Murphy}, \& {Newton-McGee}}]{2010A&A...521A..17K}
{Kalberla}, P.~M.~W., {McClure-Griffiths}, N.~M., {Pisano}, D.~J., {et~al.}
  2010, \aap, 521, A17

\bibitem[{{Kenney} {et~al.}(2008){Kenney}, {Tal}, {Crowl}, {Feldmeier}, \&
  {Jacoby}}]{2008ApJ...687L..69K}
{Kenney}, J.~D.~P., {Tal}, T., {Crowl}, H.~H., {Feldmeier}, J., \& {Jacoby},
  G.~H. 2008, \apjl, 687, L69

\bibitem[{{Kim} {et~al.}(1998){Kim}, {Staveley-Smith}, {Dopita}, {Freeman},
  {Sault}, {Kesteven}, \& {McConnell}}]{1998ApJ...503..674K}
{Kim}, S., {Staveley-Smith}, L., {Dopita}, M.~A., {et~al.} 1998, \apj, 503, 674

\bibitem[{{Lehner}(2002)}]{2002ApJ...578..126L}
{Lehner}, N. 2002, \apj, 578, 126

\bibitem[{{Lehner} {et~al.}(2008){Lehner}, {Howk}, {Keenan}, \&
  {Smoker}}]{2008ApJ...678..219L}
{Lehner}, N., {Howk}, J.~C., {Keenan}, F.~P., \& {Smoker}, J.~V. 2008, \apj,
  678, 219

\bibitem[{{Lehnert} {et~al.}(1999){Lehnert}, {Heckman}, \&
  {Weaver}}]{1999ApJ...523..575L}
{Lehnert}, M.~D., {Heckman}, T.~M., \& {Weaver}, K.~A. 1999, \apj, 523, 575

\bibitem[{{Madau} {et~al.}(1999){Madau}, {Haardt}, \&
  {Rees}}]{1999ApJ...514..648M}
{Madau}, P., {Haardt}, F., \& {Rees}, M.~J. 1999, \apj, 514, 648

\bibitem[{{Maloney}(1993)}]{1993ApJ...414...41M}
{Maloney}, P. 1993, \apj, 414, 41

\bibitem[{{Marcelin} {et~al.}(1985){Marcelin}, {Boulesteix}, \&
  {Georgelin}}]{1985Natur.316..705M}
{Marcelin}, M., {Boulesteix}, J., \& {Georgelin}, Y. 1985, \nat, 316, 705

\bibitem[{{Martin}(1988)}]{1988ApJS...66..125M}
{Martin}, P.~G. 1988, \apjs, 66, 125

\bibitem[{{Mastropietro} {et~al.}(2009){Mastropietro}, {Burkert}, \&
  {Moore}}]{2009MNRAS.399.2004M}
{Mastropietro}, C., {Burkert}, A., \& {Moore}, B. 2009, \mnras, 399, 2004

\bibitem[{{Matsubayashi} {et~al.}(2012){Matsubayashi}, {Sugai}, {Shimono},
  {Hattori}, {Ozaki}, {Yoshikawa}, {Taniguchi}, {Nagao}, {Kajisawa}, {Shioya},
  \& {Bland-Hawthorn}}]{2012ApJ...761...55M}
{Matsubayashi}, K., {Sugai}, H., {Shimono}, A., {et~al.} 2012, \apj, 761, 55

\bibitem[{{Mayer} {et~al.}(2007){Mayer}, {Kazantzidis}, {Mastropietro}, \&
  {Wadsley}}]{2007Natur.445..738M}
{Mayer}, L., {Kazantzidis}, S., {Mastropietro}, C., \& {Wadsley}, J. 2007,
  \nat, 445, 738

\bibitem[{{McClure-Griffiths} {et~al.}(2009){McClure-Griffiths}, {Pisano},
  {Calabretta}, {Ford}, {Lockman}, {Staveley-Smith}, {Kalberla}, {Bailin},
  {Dedes}, {Janowiecki}, {Gibson}, {Murphy}, {Nakanishi}, \&
  {Newton-McGee}}]{2009ApJS..181..398M}
{McClure-Griffiths}, N.~M., {Pisano}, D.~J., {Calabretta}, M.~R., {et~al.}
  2009, \apjs, 181, 398

\bibitem[{{McLeod} {et~al.}(1993){McLeod}, {Rieke}, {Rieke}, \&
  {Kelly}}]{1993ApJ...412..111M}
{McLeod}, K.~K., {Rieke}, G.~H., {Rieke}, M.~J., \& {Kelly}, D.~M. 1993, \apj,
  412, 111

\bibitem[{{Meriwether}(1989)}]{1989JGR....9414629M}
{Meriwether}, Jr., J.~W. 1989, \jgr, 94, 14629

\bibitem[{{Mierkiewicz} {et~al.}(2006){Mierkiewicz}, {Roesler}, {Nossal}, \&
  {Reynolds}}]{2006JASTP..68.1520M}
{Mierkiewicz}, E.~J., {Roesler}, F.~L., {Nossal}, S.~M., \& {Reynolds}, R.~J.
  2006, Journal of Atmospheric and Solar-Terrestrial Physics, 68, 1520

\bibitem[{{Mizuno} {et~al.}(2006){Mizuno}, {Muller}, {Maeda}, {Kawamura},
  {Minamidani}, {Onishi}, {Mizuno}, \& {Fukui}}]{2006ApJ...643L.107M}
{Mizuno}, N., {Muller}, E., {Maeda}, H., {et~al.} 2006, \apjl, 643, L107

\bibitem[{{Muller} \& {Parker}(2007)}]{2007PASA...24...69M}
{Muller}, E., \& {Parker}, Q.~A. 2007, \apj, 24, 69

\bibitem[{{Muller} {et~al.}(2004){Muller}, {Stanimirovi{\'c}}, {Rosolowsky}, \&
  {Staveley-Smith}}]{2004ApJ...616..845M}
{Muller}, E., {Stanimirovi{\'c}}, S., {Rosolowsky}, E., \& {Staveley-Smith}, L.
  2004, \apj, 616, 845

\bibitem[{{Muller} {et~al.}(2003){Muller}, {Staveley-Smith}, {Zealey}, \&
  {Stanimirovi{\'c}}}]{2003MNRAS.339..105M}
{Muller}, E., {Staveley-Smith}, L., {Zealey}, W., \& {Stanimirovi{\'c}}, S.
  2003, \mnras, 339, 105

\bibitem[{{Norman} \& {Ikeuchi}(1989)}]{1989ApJ...345..372N}
{Norman}, C.~A., \& {Ikeuchi}, S. 1989, \apj, 345, 372

\bibitem[{{Putman} {et~al.}(2003{\natexlab{a}}){Putman}, {Bland-Hawthorn},
  {Veilleux}, {Gibson}, {Freeman}, \& {Maloney}}]{2003ApJ...597..948P}
{Putman}, M.~E., {Bland-Hawthorn}, J., {Veilleux}, S., {et~al.}
  2003{\natexlab{a}}, \apj, 597, 948

\bibitem[{{Putman} {et~al.}(2003{\natexlab{b}}){Putman}, {Staveley-Smith},
  {Freeman}, {Gibson}, \& {Barnes}}]{2003ApJ...586..170P}
{Putman}, M.~E., {Staveley-Smith}, L., {Freeman}, K.~C., {Gibson}, B.~K., \&
  {Barnes}, D.~G. 2003{\natexlab{b}}, \apj, 586, 170

\bibitem[{{Sancisi} {et~al.}(2008){Sancisi}, {Fraternali}, {Oosterloo}, \& {van
  der Hulst}}]{2008A&ARv..15..189S}
{Sancisi}, R., {Fraternali}, F., {Oosterloo}, T., \& {van der Hulst}, T. 2008,
  \aapr, 15, 189

\bibitem[{{Schaerer}(2002)}]{2002A&A...382...28S}
{Schaerer}, D. 2002, \aap, 382, 28

\bibitem[{{Sembach} {et~al.}(2003){Sembach}, {Wakker}, {Savage}, {Richter},
  {Meade}, {Shull}, {Jenkins}, {Sonneborn}, \& {Moos}}]{2003ApJS..146..165S}
{Sembach}, K.~R., {Wakker}, B.~P., {Savage}, B.~D., {et~al.} 2003, \apjs, 146,
  165

\bibitem[{{Shapiro} {et~al.}(2004){Shapiro}, {Iliev}, \&
  {Raga}}]{2004MNRAS.348..753S}
{Shapiro}, P.~R., {Iliev}, I.~T., \& {Raga}, A.~C. 2004, \mnras, 348, 753

\bibitem[{{Shaviv} \& {Dekel}(2003)}]{2003astro.ph..5527S}
{Shaviv}, N.~J., \& {Dekel}, A. 2003, ArXiv Astrophysics e-prints,
  arXiv:astro-ph/0305527

\bibitem[{{Smoker} {et~al.}(2000){Smoker}, {Keenan}, {Polatidis}, {Mooney},
  {Lehner}, \& {Rolleston}}]{2000A&A...363..451S}
{Smoker}, J.~V., {Keenan}, F.~P., {Polatidis}, A.~G., {et~al.} 2000, \aap, 363,
  451

\bibitem[{{Stanimirovic} {et~al.}(1999){Stanimirovic}, {Staveley-Smith},
  {Dickey}, {Sault}, \& {Snowden}}]{1999MNRAS.302..417S}
{Stanimirovic}, S., {Staveley-Smith}, L., {Dickey}, J.~M., {Sault}, R.~J., \&
  {Snowden}, S.~L. 1999, \mnras, 302, 417

\bibitem[{{Stanimirovi{\'c}} {et~al.}(2004){Stanimirovi{\'c}},
  {Staveley-Smith}, \& {Jones}}]{2004ApJ...604..176S}
{Stanimirovi{\'c}}, S., {Staveley-Smith}, L., \& {Jones}, P.~A. 2004, \apj,
  604, 176

\bibitem[{{Thoul} \& {Weinberg}(1996)}]{1996ApJ...465..608T}
{Thoul}, A.~A., \& {Weinberg}, D.~H. 1996, \apj, 465, 608

\bibitem[{{Tufte}(1997)}]{1997PhDT........15T}
{Tufte}, S.~L. 1997, PhD thesis, THE UNIVERSITY OF WISCONSIN - MADISON

\bibitem[{{Tufte} {et~al.}(2002){Tufte}, {Wilson}, {Madsen}, {Haffner}, \&
  {Reynolds}}]{2002ApJ...572L.153T}
{Tufte}, S.~L., {Wilson}, J.~D., {Madsen}, G.~J., {Haffner}, L.~M., \&
  {Reynolds}, R.~J. 2002, \apjl, 572, L153

\bibitem[{{van Woerden} \& {Wakker}(2004)}]{2004ASSL..312..195V}
{van Woerden}, H., \& {Wakker}, B.~P. 2004, in ASSL, Vol. 312, High Velocity
  Clouds, ed. {H.~van Woerden, B.~P.~Wakker, U.~J.~Schwarz, \& K.~S.~de Boer },
  195--226

\bibitem[{{Walker}(1999)}]{1999ASSL..237..125W}
{Walker}, A. 1999, Post-Hipparcos Cosmic Candles, 237, 125

\bibitem[{{Weiner} \& {Williams}(1996)}]{1996AJ....111.1156W}
{Weiner}, B.~J., \& {Williams}, T.~B. 1996, \aj, 111, 1156

\bibitem[{{Westerlund} \& {Glaspey}(1971)}]{1971A&A....10....1W}
{Westerlund}, B.~E., \& {Glaspey}, J. 1971, \aap, 10, 1

\bibitem[{{Wilcots} \& {Prescott}(2004)}]{2004AJ....127.1900W}
{Wilcots}, E.~M., \& {Prescott}, M.~K.~M. 2004, \aj, 127, 1900

\bibitem[{{Williams} \& {McKee}(1997)}]{1997ApJ...476..166W}
{Williams}, J.~P., \& {McKee}, C.~F. 1997, \apj, 476, 166

\bibitem[{{Wood} {et~al.}(2010){Wood}, {Hill}, {Joung}, {Mac Low}, {Benjamin},
  {Haffner}, {Reynolds}, \& {Madsen}}]{2010ApJ...721.1397W}
{Wood}, K., {Hill}, A.~S., {Joung}, M.~R., {et~al.} 2010, \apj, 721, 1397

\bibitem[{{Yagi} {et~al.}(2012){Yagi}, {Komiyama}, \&
  {Yoshida}}]{2012ApJ...749L...2Y}
{Yagi}, M., {Komiyama}, Y., \& {Yoshida}, M. 2012, \apjl, 749, L2

\bibitem[{{Zastrow} {et~al.}(2011){Zastrow}, {Oey}, {Veilleux}, {McDonald}, \&
  {Martin}}]{2011ApJ...741L..17Z}
{Zastrow}, J., {Oey}, M.~S., {Veilleux}, S., {McDonald}, M., \& {Martin}, C.~L.
  2011, \apjl, 741, L17

\end{thebibliography}

\end{document}